\documentclass[10pt]{article}

\usepackage[a4paper,margin=2.5cm]{geometry}
\usepackage{helvet}

\usepackage[T1]{fontenc}
\usepackage[utf8]{inputenc}
\usepackage{microtype}
\usepackage{setspace}
\usepackage{amsmath,amssymb,mathtools}
\usepackage{bm}

\usepackage{graphicx}
\usepackage{booktabs}
\usepackage{tabularx}
\usepackage{threeparttable}
\usepackage{caption}
\usepackage{subcaption}

\usepackage{enumitem}
\usepackage{csquotes}
\usepackage[hidelinks]{hyperref}
\usepackage[nameinlink,noabbrev]{cleveref}
\usepackage{url}

\usepackage[
  backend=biber,
  style=apa,
  sorting=nyt,
  maxcitenames=2,
  maxbibnames=99,
  natbib=true
]{biblatex}

\DeclareLanguageMapping{english}{english-apa}
\usepackage{xcolor}
\usepackage{authblk}
\usepackage{orcidlink}
\usepackage{placeins}
\usepackage{amsmath}
\usepackage{graphicx}
\usepackage{algorithm}
\usepackage{algpseudocode}
\usepackage{amssymb}
\usepackage{booktabs}
\usepackage{array}
\usepackage{pdflscape}
  \usepackage{array}
  \usepackage{booktabs}
  \usepackage{longtable}
  \usepackage{url}

\usepackage{authblk}

\title{\textbf{Interpreting Random Forests: Comparing Surrogate Models and Rule Extraction Methods}}

\author[1]{Lea L. Mairhöfer}
\author[1]{Jonathan B. Krause}
\author[1,2,3]{Björn-Hergen Laabs\textsuperscript{*}}
\author[1]{Silke Szymczak\textsuperscript{*}\thanks{Corresponding author: silke.szymczak@uni-luebeck.de; \textsuperscript{*}These authors contributed equally to this work.}}

\affil[1]{Institute of Medical Biometry and Statistics, University of Luebeck, Lübeck, Germany}
\affil[2]{Department of Medical Statistics, University Medical Center Göttingen, Göttingen, Germany}
\affil[3]{CAIMed, Lower Saxony Center for Artificial Intelligence and Causal Methods in Medicine, Göttingen, Germany}

\date{}

\begin{document}

\maketitle

\begin{abstract}
Random forests (RFs) often achieve high prediction performance, but their ensemble structure can be difficult to interpret. 
Surrogate models and rule extraction methods offer interpretable approximations, yet guidance on selecting an appropriate method for a specific research question remains limited. 

We compared 12 methods for interpreting RFs, with a decision tree as a prediction-oriented reference, across 11 simulation settings and 67 benchmark data sets for regression and classification. 
Evaluation addressed prediction and structural similarity to the RF, model complexity, prediction performance, and runtime. Simulations additionally assessed explainability using known effect and noise variables.

No method performed best across all objectives. For regression, SLIM provided the strongest general performance for prediction similarity or performance on test data, while a linear model offered a simpler global interpretation of linear effects when performance was comparable.
For classification, inTrees achieved the strongest overall benchmark performance, with CTree and OC-CART providing useful alternatives. ART consistently produced the most compact representations. 
NH and SIRUS offered strong structural similarity for regression, while MRT most closely reproduced RF variable usage in the classification simulations, illustrating that prediction-oriented rankings do not necessarily identify the best structural explanation.
Outcome-type restrictions, runtime limits, and method-specific failures reduced result availability for several approaches and should be considered alongside performance when selecting a method.

Method selection should therefore be guided by the outcome type, anticipated data structure, and intended interpretability objective. Prediction similarity alone is insufficient for explaining an RF and should be considered jointly with structural similarity and model complexity.

\end{abstract}

\section{Introduction}
Random forests (RFs) are widely used for prediction on tabular data because they can represent nonlinear effects and complex interactions while requiring few assumptions about the functional relationship between variables and outcome \citep{breimanClassificationRegressionTrees1984, breimanRandomForests2001}. However, their ensemble structure makes it difficult to understand how predictions arise. 
This limits their use when the variables and decision structures underlying a prediction are of substantive interest or when the results of the model must be transparently communicated.

Therefore, many methods have been proposed to explain complex prediction models \citep{guidottiSurveyMethodsExplaining2018,molnarInterpretableMachineLearning2022}. 
Surrogate models approximate the predictions of a black-box model with an interpretable model, whereas rule extraction methods derive or build a smaller set of decision rules. 
Both approaches can provide an interpretable global representation of an RF, but differ in their structural assumptions and optimization targets. Some methods primarily aim to reproduce RF predictions, others preserve variables or decision structures used by the forest, and others prioritize compactness.

This variety makes selecting the most appropriate one challenging in practice. A method that closely reproduces RF predictions may rely on different variables and consequently provide a misleading explanation of the RF decision process. 
Conversely, a structurally similar representation may have low prediction similarity or be too complex to interpret. Method choice may also be guided by the outcome type, continuous or categorical, and by the form of the prediction function, such as an approximately linear relationship, subgroup effects or interactions, or a more complex nonlinear pattern. 
For empirical data, this structure is generally unknown. 

Existing comparisons of such approaches cover only selected method classes, small numbers of data sets, or a limited range of performance criteria, and therefore provide insufficient guidance for choosing a method for a specific application \citep{ariaComparisonInterpretativeProposals2021, herbingerLeveragingModelBasedTrees2024}. 
\citet{haddouchiSurveyTaxonomyMethods2024} provided a theoretical taxonomy of RF interpretation methods rather than an empirical comparison. \citet{ariaComparisonInterpretativeProposals2021} compared the two rule extraction methods inTrees and NodeHarvest across six binary classification data sets and found a small overall prediction performance advantage of inTrees.
\citet{herbingerLeveragingModelBasedTrees2024} compared four model-based tree methods (SLIM, GUIDE, MOB, and CTree) on three simulated regression scenarios and showed that the preferred surrogate method depends on the balance between prediction similarity and interpretability, with results also affected by hyperparameter choices.

Addressing these limitations requires systematic comparisons across a broad range of practically relevant settings to establish when existing methods perform well. This is the purpose of phase III methodological research in the framework proposed by \citet{heinzePhasesMethodologicalResearch2024b}, to which the present study contributes.
This study extends previous contributions by comparing 12 different surrogate models and rule extraction methods for RFs, with a decision tree fitted directly to the observed outcomes included as a prediction-oriented reference model, hereafter referred to as the decision tree.
The evaluation combines 11 simulation settings with known linear, nonlinear, subgroup, interaction, and noise structures and 67 heterogeneous benchmark data sets for regression and classification. 
The methods are assessed jointly with respect to agreement with the RF predictions and the variables used by the RF, model complexity, prediction of the observed outcome, and runtime. In the simulation settings, explainability is evaluated by checking which effect and noise variables are used in the interpretable models. 

The aim is to determine which methods should be preferred for different data structures and analytical objectives. 
By combining controlled simulations with a broad empirical comparison, the study distinguishes findings tied to known mechanisms from patterns that generalize across applications. 
Hence, it provides practical guidance for selecting an interpretable RF approximation method while explicitly accounting for the trade-off between prediction performance, similarity, and interpretability.

\section{Interpretable methods}
Interpretability refers to the extent to which a model and the reasons underlying its predictions can be understood by an intended user \citep{molnarInterpretableMachineLearning2022}. 
It depends on the purpose of the explanation and may concern either individual predictions or the global prediction function. 
Common post-hoc approaches such as variable importance measures \citep{breimanRandomForests2001}, partial dependence plots \citep{friedmanGreedyFunctionApproximation2001}, individual conditional expectation curves \citep{goldsteinPeekingBlackBox2015a}, and SHAP values \citep{lundbergUnifiedApproachInterpreting2017} describe selected aspects of a model, but do not provide an interpretable prediction model.

In contrast, this study focuses on approaches that construct such an interpretable model representing either the prediction function or the decision structure of an RF. This model supports both global interpretation, e.g., identifying the variables used, and local interpretation, such as explaining how an individual prediction was obtained. The considered approaches are divided into surrogate models, which fit an interpretable model to approximate the RF, and rule extraction methods, which construct an interpretable rule set from decision structures.
The approaches considered here include both methods specifically developed for model interpretation and general prediction methods used as interpretable surrogates.
Their interpretability depends not only on compactness but also on whether they reproduce the predictions and relevant decisions of the RF. The following sections describe their underlying principles, structural assumptions, and main limitations. The criteria and procedure used to select the methods included in this study are described in \Cref{sec:comparison_study_design}.

\subsection{Surrogate models}
One established strategy for explaining complex machine learning models is the use of interpretable surrogate models \citep{banerjeeIdentifyingRepresentativeTrees2012a,guidottiSurveyMethodsExplaining2018,johanssonAccurateInterpretableRegression2014,molnarInterpretableMachineLearning2022}. A surrogate model is an interpretable prediction model that approximates a complex machine learning model \citep{molnarInterpretableMachineLearning2022}. If the surrogate achieves high similarity to the black-box model, it can be interpreted in its place, enabling both local explanations of individual predictions and global insights into the variables and structures underlying the prediction process \citep{johanssonObtainingAccurateComprehensible2012,guidottiSurveyMethodsExplaining2018,molnarInterpretableMachineLearning2022,renardConceptTreeHighLevel2019}.

Surrogate modeling is generally model-agnostic and can therefore be applied to different types of black-box models \citep{guidottiSurveyMethodsExplaining2018,molnarInterpretableMachineLearning2022}. However, some approaches are designed for specific model classes \citep{banerjeeIdentifyingRepresentativeTrees2012a,laabsConstructionArtificialMost2024} or restricted to particular outcome types, such as regression or classification.

In this work, we focus on RFs as the underlying black-box model because of their strong prediction performance on tabular data \citep{grinsztajnWhyTreebasedModels2022}. 
We thus primarily consider tree-based surrogate approaches, whose hierarchical decision structure is intuitive to interpret and structurally similar to that of an RF \citep{meinshausenNodeHarvest2010}. As a complementary comparison, we additionally consider a linear regression model as a common surrogate model \citep{molnarInterpretableMachineLearning2022}, which is likewise interpretable but relies on a fundamentally different model structure and set of assumptions.

The surrogate approaches considered in this study are grouped into two categories: oracle coaching and representative trees. Oracle coaching fits an interpretable model to the RF predictions, whereas representative tree approaches either select an existing tree from the RF (MRT) or construct a new tree to represent the RF (ART).

\subsubsection{Oracle Coaching}
Oracle coaching represents a naive approach to surrogate modeling. It trains an interpretable model on the predictions of an accurate black-box model, referred to as the oracle, instead of on the true observed outcomes \citep{johanssonObtainingAccurateComprehensible2012,johanssonAccurateInterpretableRegression2014}. 
As any model can be used to serve as an oracle, this method is model agnostic.
The resulting model aims to reproduce the oracle's decision function by generating similar predictions while retaining the transparent structure of the chosen model. Similarity is therefore established only at the prediction level. High prediction similarity does not imply that the surrogate and the oracle rely on similar decision structures.

Training decision trees and linear models on black-box predictions is an established approach to construct interpretable surrogates \citep{johanssonAccurateInterpretableRegression2014,molnarInterpretableMachineLearning2022}. Following the same principle, GUIDE, MOB, SLIM, and CTree have also been investigated as surrogate models  \citep{herbingerLeveragingModelBasedTrees2024}.

The methods described below were applied using the oracle coaching principle.

\paragraph*{Linear Model (LM)}
A linear regression model was used as the interpretable model as proposed by \citet{molnarInterpretableMachineLearning2022}.
It approximates the RF predictions through an additive linear combination of the variables and was used only for regression tasks in this study. The number of variables can be reduced through forward selection, which starts with an intercept-only model and sequentially adds the variable that most improves a chosen model selection criterion until no further improvement is achieved.
A limitation of this application is that it captures only linear main effects and cannot represent nonlinear effects or interactions unless corresponding terms are explicitly included.\\\\
The remaining models used for oracle coaching are tree-based. They recursively partition the variable space but differ in the models fitted within their nodes, the criteria for selecting splitting variables and split points, and the rules for stopping tree growth or pruning.

\paragraph*{Classification and regression tree (OC-CART)}
A single decision tree was used as the interpretable model and trained on the RF predictions \citep{johanssonObtainingAccurateComprehensible2012,johanssonAccurateInterpretableRegression2014}. This surrogate is referred to as OC-CART. The tree was built using the classification and regression tree (CART) algorithm, which recursively partitions the variable space through binary splits and assigns a constant prediction to each terminal node \citep{breimanClassificationRegressionTrees1984}.
CART jointly selects the splitting variable and split point by searching for the greatest reduction in node impurity, measured, for example, by the Gini index for classification or the variance of the response for regression. The reduction is calculated as the impurity of the parent node minus the weighted average impurity of the two child nodes, with weights corresponding to their proportions of observations. 
Because variables with more possible split points offer more opportunities to achieve a large impurity reduction, this selection procedure can favor such variables \citep{lohSplitSelectionMethods1999}.
Splitting continues until no further admissible split is available, for example, because a split would leave fewer than the required minimum number of training observations in a child node or because the node is pure, meaning that all observations have the same class in classification or the same response value in regression.
For prediction, an observation follows the splits from the root to a terminal node, which assigns a constant value for regression or a class label for classification.
Although the classical CART algorithm uses pruning, which removes splits from an initially grown tree, to determine the final tree size, the models in this study were fitted without pruning for comparable model complexity.

\paragraph*{Generalized, Unbiased Interaction Detection and Estimation (GUIDE)}
GUIDE approximates the RF predictions with a tree that is constructed differently from conventional trees such as CART \citep{lohClassificationRegressionTrees2011}. 
Originally developed as a regular prediction model, GUIDE has also been used as an interpretable surrogate fitted to model predictions \citep{herbingerLeveragingModelBasedTrees2024}, an approach adopted here using RF predictions as the response.
In the implementation used here, GUIDE fits linear main effects models within the nodes, so that each terminal node provides a local linear approximation of the RF predictions.
At each node, the split variable and the split point are determined in two separate steps to reduce the selection bias of conventional tree algorithms.
First, statistical tests assess the association between the signs of the residuals of the fitted node model and each candidate variable, with additional tests for pairwise interactions.
Numerical splitting variables are grouped into categories for these tests. GUIDE thus identifies variables associated with systematic overprediction or underprediction by the current node model.
The split variable is then selected based on the minimal p value. 
After selecting the variable, GUIDE evaluates potential split points by fitting models in the two resulting child nodes and selects the split point that minimizes their combined residual sum of squares.
Splitting stops when the improvement in model fit falls below the threshold defined by the improvement criterion or when further splitting would violate growth constraints, such as minimum node-size requirements.
For prediction, an observation follows the splits from the root to a terminal node, where the corresponding linear main effects model generates its prediction using the observation's variable values.
Evaluating many potential split points can be computationally demanding. 
Moreover, accurately approximating the RF predictions may require many splits, making the resulting tree difficult to interpret.
Although GUIDE supports both regression and classification in general, the R implementation used in this study is restricted to regression.

\paragraph*{Surrogate Locally-Interpretable Models (SLIM)}
SLIM approximates the RF predictions by combining a tree with interpretable main effects models within its nodes \citep{huSurrogateLocallyInterpretableModels2020}. 
SLIM was specifically developed as an interpretable surrogate model fitted to the predictions of a complex machine learning model \citep{huSurrogateLocallyInterpretableModels2020}. In this study, RF predictions serve as the response as in \citet{herbingerLeveragingModelBasedTrees2024}.
Similar to CART, the tree partitions the variable space through binary splits. 
SLIM jointly selects the splitting variable and split point.
Candidate splits are evaluated by fitting a separate main effects model in each of the two subgroups resulting from a candidate split. 
The residual sums of squares of both models are added, and the variable and split point with the smallest combined value are selected. 
Splitting stops when the reduction in the residual sum of squares achieved by the best candidate split falls below a specified threshold. 
Tree growth can also be restricted by a maximum tree depth, a maximum number of splits, or a minimum number of observations required in each terminal node.
For prediction, an observation follows the splits from the root to a terminal node, where the corresponding main effects model generates its prediction using the observation's variable values.
SLIM may therefore require fewer splits than OC-CART to approximate a complex prediction function or to display a linear relationship. 
However, it assumes that the RF predictions can be represented adequately by a main effects model within each subgroup. Strong local interactions or irregular effects may reduce prediction similarity or require a large tree. Evaluating models for many candidate splits can also result in a high runtime \citep{herbingerLeveragingModelBasedTrees2024}.

\paragraph*{Model-Based Recursive Partitioning (MOB)}
MOB combines recursive partitioning with parametric statistical models \citep{zeileisModelBasedRecursivePartitioning2008}.
MOB was originally developed to model observed outcomes but has also been used as a surrogate model \citep{herbingerLeveragingModelBasedTrees2024}. In this study, it is used as a surrogate model and is fitted to RF predictions.
Each node contained an intercept-only model, which estimates a single constant rather than regression coefficients for the variables.
First, MOB estimates the constant as the mean RF prediction among all training observations in the current node and calculates their deviations from this mean. 
All available variables were considered as candidate splitting variables.
It then tests whether the common mean is appropriate across the values of each candidate splitting variable. These parameter instability tests use model scores, which are proportional to the deviations from the node mean in this configuration.
For numerical variables, observations are ordered by their values, and cumulative scores are examined for systematic patterns. For categorical variables, scores are aggregated within categories, and a chi-squared test assesses whether they differ systematically between categories. 
If no significant instability remains after Bonferroni correction across the candidate splitting variables, the node is not split. Otherwise, MOB selects the variable with the smallest p value.
For the selected variable, MOB then compares binary splits that satisfy the specified constraints, such as the minimum number of observations required in each child node. 
Each candidate split yields two child-node means. The split that minimizes the combined sum of squared deviations from these means is selected.
The procedure is repeated within each child node until no significant instability is detected or further splitting would violate growth constraints, such as minimum node-size requirements.
For prediction, an observation follows the splits from the root to a terminal node and receives the mean RF prediction of the training observations within that node.
As in OC-CART, predictions are therefore constant within each terminal node, which may require many splits to approximate complex RF prediction functions.

\paragraph*{Conditional Inference Tree (CTree)}
CTree constructs a decision tree using statistical hypothesis tests for recursive partitioning \citep{hothornUnbiasedRecursivePartitioning2006}.
In this study, CTree was used as a surrogate model by fitting it to RF predictions rather than the observed outcomes it was originally designed to predict.
It partitions the variable space through binary splits and assigns a constant prediction to each terminal node.
At each node, CTree tests the independence of the RF predictions and each candidate splitting variable using conditional inference tests. In the implementation used here, quadratic test statistics are evaluated using their asymptotic null distributions, and p values are adjusted across candidate variables using Bonferroni correction.
If no significant association remains, the node is not split. Otherwise, the variable with the smallest p value is selected.
For the selected variable, CTree then compares binary splits that satisfy the specified constraints, such as minimum child-node sizes.
The procedure is repeated within each child node until no significant association is detected or further splitting would violate growth constraints.
For prediction, an observation follows the splits from the root to a terminal node. For regression, it receives the mean RF prediction of the training observations within that node. For classification, it receives the most frequent RF-predicted class within that node.
Separating variable selection from split-point selection reduces the preference of CART for variables with many possible split points. However, constant predictions within terminal nodes may require many splits to approximate complex RF prediction functions.

\subsubsection{Representative Trees}

\paragraph*{Most Representative Tree (MRT)}
The MRT approach was developed to provide an interpretable representation of a tree ensemble by selecting one of its existing trees as a surrogate \citep{banerjeeIdentifyingRepresentativeTrees2012a}.
For each tree of the RF, the pairwise distance to all other trees is calculated and the tree with the smallest mean distance is selected as the MRT.
This distance can be based on structural features such as split variables and their positions \citep{banerjeeIdentifyingRepresentativeTrees2012a,laabsIdentificationRepresentativeTrees2024} or on another tree distance measure. 
Because MRT selects an existing tree, it retains that tree's splits and terminal-node predictions. No additional split selection or stopping rule is required.
For prediction, an observation follows the selected tree's splits from the root to a terminal node and receives its prediction.
The selected tree provides a direct but highly compressed view of the RF.
However, it may be deep and difficult to interpret because trees in an RF are commonly grown with few restrictions. Splits based on noise variables may also occur at any tree depth and lead to misleading interpretations.

\paragraph*{Artificial Representative Tree (ART)}
ART was developed as an interpretable surrogate for RFs, extending the MRT approach by constructing a new representative tree through minimization of a tree-based distance measure rather than selecting an existing tree \citep{laabsConstructionArtificialMost2024}. 
Starting at the root, the algorithm evaluates candidate splitting variables and split points by calculating the distance between each resulting candidate tree and the RF. It selects the split that yields the smallest distance. As in the MRT method, the distance can be based on the split variables used, for example.
Tree growth stops when further splitting increases the distance or leaves it unchanged without a sufficient improvement in prediction performance. Growth can also be restricted by requiring a minimum number of training observations in each terminal node.
For prediction, an observation follows the splits from the root to a terminal node. The node predicts the mean observed outcome for regression or the most frequent observed class for classification among its training observations.
The construction is designed to achieve structural similarity and does not directly optimize prediction similarity. In addition, the algorithm evaluates one split at a time rather than combinations of several splits and may consequently fail to identify the globally optimal tree.

\subsection{Rule extraction}
Rule-based models are supervised machine learning methods that generate predictions based on individual rules or sets of rules \citep{friedmanPredictiveLearningRule2008,benardSIRUSStableInterpretable2021}. 
A rule combines several conditions, such as \textit{if age $> 65$ and weight $> 50$}, and assigns a prediction to observations satisfying them. Predictions are typically derived from the training observations satisfying these conditions, such as the majority class in classification or the mean outcome in regression, analogous to terminal nodes in CART. Multiple rules can be combined to form a complete prediction model \citep{friedmanPredictiveLearningRule2008,benardSIRUSStableInterpretable2021}.
Such approaches can serve as stand-alone prediction models or be used to explain complex machine learning models through rule extraction \citep{guidottiSurveyMethodsExplaining2018,molnarInterpretableMachineLearning2022}. In contrast to surrogate modeling, which approximates a black-box model with an interpretable prediction model, rule extraction represents its decision logic through explicit, human-readable rules \citep{guidottiSurveyMethodsExplaining2018}.

Rule extraction is particularly suitable for RFs because each path from the root node to a terminal node in an individual tree represents a decision rule \citep{breimanRandomForests2001,friedmanPredictiveLearningRule2008}. Individual rules may be easy to understand, but their aggregation across many trees makes the overall RF difficult to interpret. Rule extraction methods therefore aim to identify, generate, or select a smaller set of representative rules that capture important decision mechanisms of the RF while remaining interpretable \citep{benardSIRUSStableInterpretable2021}. In this work, we focus exclusively on rule extraction methods applicable to RFs that yield interpretable prediction models.
We distinguish methods that extract rules directly from the RF trees from those that learn rules by fitting models to RF predictions.

\subsubsection{Rules extracted from RF trees}

\paragraph*{Interpretable Trees (inTrees)}
inTrees was developed to interpret tree ensembles by extracting, simplifying, and selecting decision rules \citep{dengInterpretingTreeEnsembles2019}.
It extracts candidate rules from paths in the trees of a fitted RF.
Each rule consists of conditions on variable values and applies to observations that satisfy all of these conditions.
The rules are evaluated according to their prediction error, coverage, and length. Coverage is the proportion of training observations satisfying a rule, and length is the number of conditions it contains. Rule predictions and errors are calculated using the observed outcomes.
inTrees removes redundant conditions and constructs an ordered rule list by iteratively selecting rules with low prediction error, favoring rules that cover a larger proportion of observations and have fewer conditions in case of ties, and re-evaluating the remaining rules on observations not yet covered.
For prediction, its rules are evaluated in sequence, and the first rule satisfied by an observation determines its prediction. A default prediction is assigned if none of the selected rules applies.
Although individual rules may be easy to interpret, long rules or a large rule list can make the complete model difficult to understand. Moreover, extracting rules from the RF does not guarantee that the resulting model reproduces its predictions.

\subsubsection{Rules learned from RF predictions}

\paragraph*{Stable and Interpretable Rule Set (SIRUS)}
SIRUS was originally proposed as an interpretable prediction method based on RFs \citep{benardSIRUSStableInterpretable2021}. In this study, it is used as a surrogate by fitting it to RF predictions rather than observed outcomes.
SIRUS constructs its own forest using the original variables and the predictions of the RF to be explained as training responses. It therefore learns new trees rather than extracting rules from the original RF. Numerical split points in this new forest are restricted to empirical quantiles to increase the recurrence of the same decision paths across trees.
Rules are extracted from paths leading to both intermediate splitting nodes and terminal nodes.
Rules whose frequency across the trees exceeds a specified threshold are retained, and redundant rules that can be expressed as linear combinations of more frequent rules are removed.
The frequency threshold controls the size of the rule set.
Each rule divides the training observations into those that satisfy its conditions and those that do not. 
Each rule returns the mean prediction among training observations that satisfy the conditions if they are met. Otherwise, it returns the mean prediction among training observations that do not satisfy them.
The final SIRUS prediction is obtained by multiplying each rule's output by its weight, summing these weighted outputs across all selected rules, and adding an intercept.
For regression, the weights are estimated jointly by ridge regression with non-negative coefficients.
The focus on stable and recurring rules can improve reproducibility but may omit less frequent structures that contribute to the predictions.

\paragraph*{Node Harvest (NH)}
NH was originally proposed as an interpretable prediction method based on a collection of potentially overlapping nodes \citep{meinshausenNodeHarvest2010}. In this study, it was used as an RF surrogate by training its own trees and estimating node predictions and weights using RF predictions rather than observed outcomes.
NH first grows its own RF using the original variables and the predictions of the RF to be explained as training responses. 
Candidate rules are derived from both internal and terminal nodes of this newly fitted forest. Each rule consists of the conditions along the path from the root to the corresponding node. 
Its prediction is the mean prediction among the training observations satisfying all of its conditions.
The number and complexity of candidate rules can be restricted, for example, by specifying the size of the initial rule collection, a minimum number of observations satisfying each rule, or a maximum number of conditions per rule.
The final predictions are weighted averages of the node means across all selected nodes covering an observation. 
The node weights are estimated by solving a quadratic optimization problem that minimizes squared prediction error, subject to nonnegative weights and the constraint that the weights of nodes covering each training observation sum to one.
Only rules with positive weights contribute to the final prediction model.
For a new observation, NH identifies all selected rules whose conditions are satisfied. It multiplies their predictions by their respective weights, adds these contributions, and divides by the sum of the weights of the matching rules.
The selected rules can overlap and do not form a single hierarchical partition. This allows flexible representations, but a large collection of overlapping rules may make the complete model difficult to interpret.

\paragraph*{RuleFit}
RuleFit combines decision rules with linear variable effects to construct an interpretable prediction model \citep{friedmanPredictiveLearningRule2008}. In this study, it was used as a surrogate by fitting it to RF predictions rather than observed outcomes.
The implementation used here first grows its own collection of conditional inference trees using the original variables and the predictions of the RF to be explained as training responses.
Candidate rules are extracted from paths leading to internal and terminal nodes of these newly fitted trees. Each rule is represented by a binary indicator that equals one when an observation satisfies all its conditions and zero otherwise.
A regression model is then fitted using these binary variables together with linear terms for the original variables. For regression tasks, this is a linear regression model. For classification, it is a logistic or multinomial regression model.
The coefficients are estimated jointly using the least absolute shrinkage and selection operator (LASSO) \citep{tibshiraniRegressionShrinkageSelection1996}.
The penalty shrinks coefficients towards zero and sets some to zero.
The final model therefore consists of an intercept, the retained linear terms, and the retained rule indicators with their estimated coefficients, which can be positive or negative.
Model complexity can be controlled through the number and depth of the trees used to generate candidate rules and through the strength of the regularization penalty.
For prediction, the retained rule indicators are evaluated for a new observation. For regression, their weighted contributions and those of the retained linear terms are added to the intercept. For classification, the resulting linear predictors are transformed into class probabilities.
Although the final model is linear in its coefficients, the rule indicators allow it to represent nonlinear relationships and interactions between the original variables. A large number of retained terms or strongly overlapping rules can nevertheless make the complete model difficult to interpret.

\subsection{Similarities and differences between interpretation methods}
The methods considered here provide interpretable representations of an RF but differ in the information they use and the models they produce. An overview is provided in \Cref{tab:method_comparison}.

\paragraph{RF access and training targets}
Oracle coaching approaches learn from RF predictions without requiring access to the forest's internal structure. The surrogate model approaches LM, OC-CART, GUIDE, SLIM, MOB, and CTree use this approach in this study. SIRUS, NH, and RuleFit follow the same principle here, although their final models are rule-based. These methods therefore focus on achieving high prediction similarity without considering structural similarity.

MRT, ART, and inTrees instead use parts of the original splits from the fitted forest itself. MRT selects an existing tree, ART constructs a representative tree using distances to the RF, and inTrees extracts rules from its decision paths. Accessing the RF structure does not necessarily mean that RF predictions are used as the training response.

\paragraph{Original purpose}
The methods differ in their original purpose. SLIM, MRT, ART, and inTrees were developed to provide interpretable representations of existing prediction models or tree ensembles \citep{huSurrogateLocallyInterpretableModels2020,banerjeeIdentifyingRepresentativeTrees2012a,laabsConstructionArtificialMost2024,dengInterpretingTreeEnsembles2019}.

Linear regression, CART, GUIDE, MOB, and CTree were originally developed to model observed outcomes rather than explain another fitted model.
SIRUS, NH, and RuleFit were also proposed as prediction methods, but explicitly aimed to combine prediction performance with interpretability \citep{benardSIRUSStableInterpretable2021,meinshausenNodeHarvest2010,friedmanPredictiveLearningRule2008}.
In this study, these prediction methods serve as RF surrogates by using RF predictions as their training responses.

\paragraph{Model structure and prediction}
The resulting models range from a global linear model to individual trees, trees with local models, and collections of rules.
LM predicts through an intercept and the sum of variable values multiplied by their regression coefficients.

OC-CART, CTree, MOB, MRT, and ART provide decision trees with constant predictions within each terminal node. GUIDE and SLIM instead evaluate a local regression model in the terminal node reached by an observation, using its variable values and the locally estimated intercept and coefficients.

inTrees uses the prediction of the first matching rule, with a default prediction if none matches.
SIRUS multiplies each selected rule's output, including its else value when the conditions are not satisfied, by its estimated non-negative weight and adds these contributions to an intercept.
NH computes a weighted average of the predictions of matching rules, dividing their weighted sum by the total weight of those rules.
RuleFit adds the coefficients of satisfied rule indicators and the contributions of retained linear predictor terms to an intercept. For classification, its linear variables are transformed into class probabilities.

\paragraph{Tree growth and rule construction}
The tree-building algorithms differ in how they select splitting variables and split points. OC-CART and SLIM jointly search for the combination that most improves model fit. 

GUIDE, MOB, and CTree first select a variable through statistical testing and then determine its split point. Their tests use different information: GUIDE examines residual signs, MOB assesses parameter instability using model scores, and CTree tests associations between the response and candidate splitting variables. After variable selection, GUIDE and MOB minimize the combined residual sum of squares of the child-node models, whereas CTree maximizes a standardized separation statistic. 

Representative-tree approaches use a different objective: MRT selects a tree according to its average distance to the other RF trees, and ART evaluates candidate splits according to the distance between the resulting tree and the RF. 

inTrees extracts rules from the original RF and selects them based on prediction error, coverage, and length. SIRUS, NH, and RuleFit generate rules from newly fitted trees. SIRUS selects frequently recurring rules, NH selects rules through non-negative weight optimization, and RuleFit uses LASSO to select rules and linear terms.

\paragraph{Complexity control}
Model complexity is controlled at different stages. Tree growth can be restricted by minimum node sizes, maximum depth, a maximum number of splits, or a required improvement in the objective. MOB and CTree additionally use significance-based stopping. 

As MRT inherits the complexity of the selected RF tree, it is not possible to control its complexity directly without modifying the algorithm or the underlying RF itself.

For rule-based models, complexity can be reduced by removing conditions, filtering redundant rules, applying frequency thresholds, or estimating sparse rule coefficients or weights. SIRUS primarily selects rules through their occurrence frequency, while RuleFit uses LASSO regularization and NH retains rules with positive optimization weights. The common complexity restrictions imposed in this study are additional implementation choices rather than inherent properties of all methods.

\section{Related work}
\citet{haddouchiSurveyTaxonomyMethods2024} provided a broad overview and taxonomy of methods for interpreting RFs, including surrogate models and rule extraction methods. They organized existing approaches along several dimensions of interpretability and thereby offered conceptual guidance for selecting tools according to the aspect of an RF that should be explained. 
However, their work was a theoretical review rather than an empirical comparison and therefore did not establish how the methods perform across different data structures or objectives. 

\citet{ariaComparisonInterpretativeProposals2021} combined a theoretical overview of methods for interpreting tree ensembles with an empirical comparison of inTrees and NH, which they characterized as the two main approaches to rule extraction. 
Their evaluation was based on six benchmark data sets for binary classification from the University of California, Irvine (UCI) Machine Learning Repository (\url{https://archive.ics.uci.edu/}). 
Performance was assessed using the confusion matrix to calculate accuracy, precision, sensitivity, specificity, the geometric mean, F1 score, Youden's index, balanced accuracy, and Cohen's kappa. A selection of these measures is also used in the present evaluation.
These measures characterize prediction performance, but do not evaluate model complexity, interpretability, or similarity to the original RF. Although the two methods yielded similar results, inTrees showed a small overall performance advantage and was therefore recommended by the authors.

\citet{herbingerLeveragingModelBasedTrees2024} investigated surrogate modeling more directly by comparing SLIM, GUIDE, MOB, and CTree. 
As part of their work, they provided R implementations of SLIM and GUIDE. 
The four methods were evaluated in a simulation study modeling main effects. This included three scenarios with smooth linear, categorical linear, and mixed linear effects. 
The evaluation considered prediction similarity to the underlying model, measured by the coefficient of determination $R^2$, interpretability, represented by the number of terminal nodes, and stability, quantified by the Rand index. A selection of these measures is also used in the present evaluation.
Their results showed that hyperparameter choices substantially influenced the performance for SLIM and GUIDE, while MOB and CTree were less sensitive to the evaluated settings. All four approaches could contribute to model interpretation, while the preferred method depended on the desired balance between similarity and interpretability.

In summary, the existing studies either provide a broad theoretical taxonomy, compare a limited set of rule extraction methods using prediction performance measures, or focus on a specific class of surrogate models in controlled simulation settings. 
To the best of our knowledge, no previous study has conducted a comprehensive empirical comparison of both surrogate modeling and rule extraction methods across multiple data sets while jointly considering prediction performance, similarity, interpretability, complexity, and stability. The present study is designed to address this gap.

\section{Comparison study design}\label{sec:comparison_study_design}
\begin{figure}
    \centering
    \includegraphics[width=\linewidth]{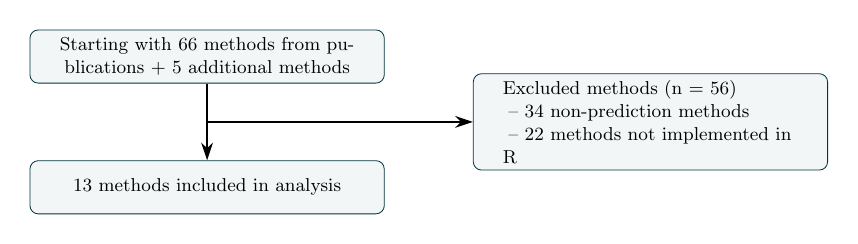}
    \caption{A flowchart showing the selection of methods for the comparison study.}
    \label{fig:flowchart_methods}
\end{figure}

The methods for the comparison study were selected from the three studies discussed above. 
We initially considered all methods mentioned by \citet{ariaComparisonInterpretativeProposals2021}, \citet{haddouchiSurveyTaxonomyMethods2024}, or \citet{herbingerLeveragingModelBasedTrees2024}, as well as subsequent extensions of these methods identified through literature search. This initial set was then restricted according to the selection process shown in \Cref{fig:flowchart_methods}. 
To be eligible, a method had to produce an independent prediction model and had to be applicable to RFs, either as a model agnostic approach or as a method developed specifically for RFs. Since the comparison was conducted in R, an existing R implementation was also required. Furthermore, the RFs in this study were fitted using the R package \emph{ranger}, providing a computationally efficient implementation \citep{wrightRangerFastImplementation2017}. 
Therefore, only methods that supported \emph{ranger} objects directly or could be applied after converting the fitted model were included.
In addition to the selected interpretation methods, a decision tree based on the CART algorithm by \citet{breimanClassificationRegressionTrees1984} was included as a reference model. Its structural similarity to the interpretable models considered in this study makes it a useful baseline, particularly for assessing prediction performance.
An overview of all included methods, their abbreviations, implementations (including the software version), and their applicability to regression and classification tasks is provided in \Cref{tab:method_description}.

\begin{figure}
    \centering
    \includegraphics[width=\linewidth]{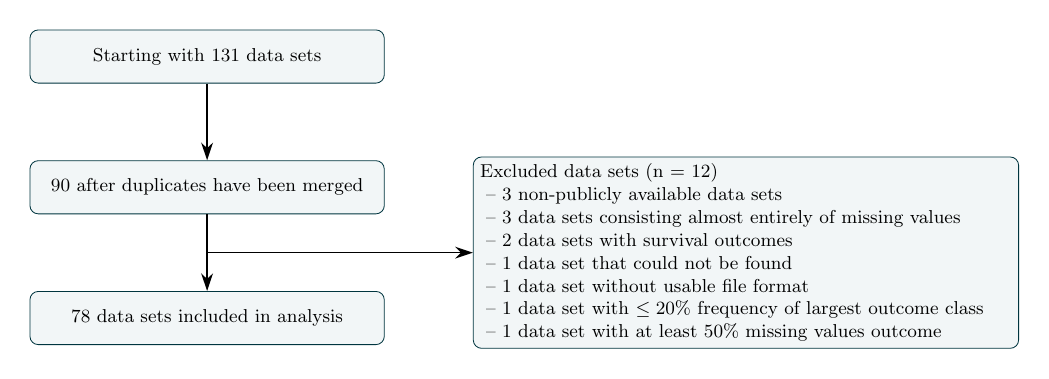}
    \caption{A flowchart showing the selection of data sets for the comparison study.}
    \label{fig:flowchart_datasets}
\end{figure}

The data sets were selected with the aim of ensuring a fair and representative comparison. For each included method, 
we reviewed the data-generating processes and empirical data sets used in its original publication. Where feasible, simulation scenarios were reconstructed, and publicly available benchmark data sets were included. 
Some data sets were identified in more than one publication.
Data sets were excluded if they were not publicly available or could not be identified reliably from the description in the corresponding publication. We also excluded data sets intended for outcomes other than regression or classification. 
Further exclusions concerned data sets that consisted almost entirely of missing values when the original publication did not provide sufficient information about their handling, as well as data sets for which the most frequent outcome category accounted for no more than 20\% of the observations. The complete data set selection process is summarized in \Cref{fig:flowchart_datasets}. 
The final analysis included 78 data sets, comprising 42 classification data sets and 36 regression data sets (see \Cref{tab:dataset_description}). After preprocessing, sample sizes ranged from 55 to 577,601 observations and from 2 to 440 variables.
The comparison study was implemented in \emph{R}, using version 4.3.0.

\section{Comparison using simulated data}
\subsection{Aim}
The simulation study was designed to provide a controlled comparison of surrogate model and rule extraction methods under data-generating mechanisms originally used to introduce or evaluate the respective methods. 
The major advantage of simulated data is that they provide comprehensive knowledge of the characteristics of the variables and their functional relationship with the outcome. 
This enables us to evaluate not only the prediction performance, but also the accuracy with which each method recovers the known signal structure.

\subsection{Data simulation}
We implemented eleven simulation settings, comprising three classification and eight regression settings. The data-generating mechanisms were based on simulation designs reported in the original publications of ART, the CART algorithm, inTrees, NH, and RuleFit \citep{laabsConstructionArtificialMost2024, breimanClassificationRegressionTrees1984, dengInterpretingTreeEnsembles2019, meinshausenNodeHarvest2010, friedmanPredictiveLearningRule2008}.
All simulated data sets could therefore be generated anew in each repetition. 
Two additional synthetic data sets originating from the waveform simulation described by \citet{breimanClassificationRegressionTrees1984} were included as fixed benchmark data sets rather than regenerated in each repetition. Waveform version 1, also used by \citet{johanssonObtainingAccurateComprehensible2012} and \citet{dengInterpretingTreeEnsembles2019}, was obtained from UCI, whereas version 2 was obtained from OpenML. See \Cref{tab:dataset_description} for details regarding the data sets.

\subsubsection*{Simulation Digit recognition / Digit recognition noisy}
The digit-recognition simulation was a multiclass classification problem in which the outcome classes corresponded to the digits 0–9 \citep{breimanClassificationRegressionTrees1984}. 
Each digit was represented by seven binary variables encoding its correct seven-segment display pattern. Variables were drawn from patterns, which were inverted with a probability of $10\%$ in order to add noise to the data.
We considered one scenario without additional noise variables and one scenario with 17 additional independently Bernoulli-distributed noise variables (called digit recognition noisy), resulting in $p=7$ and $p=24$, respectively. Each scenario comprised $200$ training and $5000$ test observations.

\subsubsection*{Simulation XOR-problem}
The XOR-problem was a binary classification problem \citep{dengInterpretingTreeEnsembles2019}. 
For each observation, exactly 10 of the $p=20$ categorical variables were randomly assigned the value $1$, while the remaining variables took the value $0$. The outcome was determined by an exclusive-or rule. Consequently, $X_3,\ldots,X_{20}$ had no direct effect on the outcome. No data split into training and test sets was specified in \citep{dengInterpretingTreeEnsembles2019}. Therefore, we used $100$ observations for training as in \citep{dengInterpretingTreeEnsembles2019} and added $100$ observations for testing.

\subsubsection*{Simulation few large effects}
This simulation represented an additive regression problem with a small number of large effects \citep{laabsConstructionArtificialMost2024}. 
All $p=100$ variables were independent binary variables. The continuous outcome was based on additive effects of five variables, each with an effect size $\beta=2$, and additional standard normally distributed noise. The remaining variables were noise variables. 
The training and test samples each contained $1000$ observations ($n=2000$ in total).

\subsubsection*{Simulation many small effects}
This simulation represented an additive regression problem with many small effects \citep{laabsConstructionArtificialMost2024}. 
All $p=100$ variables were independent binary variables. The continuous outcome was based on additive effects of the first 50 variables, each with an effect size $\beta=0.2$, and additional standard normally distributed noise. The remaining 50 variables were noise variables. The training and test samples each comprised $1,000$ observations ($n=2000$ in total).

\subsubsection*{Simulation correlated variables}
This simulation examined additive effects in the presence of correlated variables \citep{laabsConstructionArtificialMost2024}. 
Ten independent binary variables were generated alongside five blocks of five correlated binary variables. Within each block, each pairwise correlation was set to 0.3. 
The continuous outcome was based on additive effects of five of the ten independent signal variables and the first variable in each of the five correlation blocks, resulting in ten influential variables with an effect size $\beta=2$, plus standard normally distributed noise. 
Independent binary noise variables were added to obtain $p=100$. The training and test samples each comprised $1000$ observations ($n=2000$ in total).

\subsubsection*{Simulation interaction effects}
In this simulation, additive and interaction effects for a continuous regression outcome were both used to simulate the outcome \citep{laabsConstructionArtificialMost2024}. 
All $p=100$ variables were independent binary variables. The outcome was based on additive effects of the first five variables and five pairwise interactions involving the next ten variables. 
All effects had size $\beta=2$, and standard normally distributed noise was added. The training and test samples each comprised $1000$ observations ($n=2000$ in total).

\subsubsection*{Simulation continuous variables}
The last simulation used to introduce ARTs investigated additive effects of binary and continuous variables \citep{laabsConstructionArtificialMost2024}. 
Of the $p=100$ variables, 95 were independent binary variables, and five were independent standard normally distributed variables. The continuous outcome was based on additive effects of five binary and five continuous variables, each with an effect size $\beta=2$, and additional standard normally distributed noise. The other 90 variables were noise variables. The training and test samples each comprised $1000$ observations ($n=2000$ in total).

\subsubsection*{Simulation regression subgroups}
This setting was a regression problem with two subgroups \citep{breimanClassificationRegressionTrees1984}. 
The first variable was sampled from the set $\{-1,1\}$ and determined the relevant subgroup, while the other nine variables were sampled from $\{-1,0,1\}$. 
If $X_1 = 1$, the continuous outcome was based on the additive effects of $X_2$, $X_3$, and $X_4$; if $X_1 = -1$, however, it was based on the additive effects of $X_5$, $X_6$, and $X_7$ instead.
Normally distributed noise was added, and $X_8$, $X_9$, and $X_{10}$ were noise variables. In total, $n=200$ observations with $p=10$ variables were generated for training and $200$ additional observations for testing.

\subsubsection*{Simulation regression nonlinear effects}
This simulation was a nonlinear regression problem \citep{meinshausenNodeHarvest2010}. 
The two independent variables were sampled from $X_1,X_2\sim U(0,1)$, and the continuous outcome was generated according to $Y=\sin(2\pi X_1)\sin(2\pi X_2)+\varepsilon$. 
This nonadditive response surface induced an interaction between the two variables. The error term followed $\varepsilon\sim N(0,0.25)$. A total of $p=2$ variables were generated, without any additional noise variables. 
No data split into training and test sets was specified in \citep{meinshausenNodeHarvest2010}. Therefore, we used $1000$ observations for training as in \citep{meinshausenNodeHarvest2010} and added $1000$ observations for testing.

\subsubsection*{Simulation regression exponential effects}
This setting involved nonlinear regression using exponential distributions \citep{friedmanPredictiveLearningRule2008}. 
Here, $p=100$ independent variables were generated, each sampled uniformly from $\{0,0.1,\ldots,0.9\}$. 
The continuous outcome was given by $Y=F^*(X)+\varepsilon$, where $F^*$ depended on $X_1,\ldots, X_8$ through exponential effects, interactions, a squared sine term, and a linear contrast. The remaining 92 variables, $X_9,\ldots,X_{100}$, were independent noise variables with no effect on the outcome. A total of $5000$ training and $50000$ test observations were generated.

\subsection{Target}
The target was to obtain an interpretable prediction model that approximates the RF.
Its predictions on independent test data were evaluated for agreement with the RF predictions and accuracy with respect to the observed outcomes.

\subsection{Methods}
Despite the structural differences between the methods, model complexity was constrained to a similar extent wherever possible to enable meaningful comparison. 
For tree-based methods, growth was generally restricted by requiring each terminal node to contain at least $10\%$ of the training observations. The name and precise interpretation of this hyperparameter differ across implementations. Since no equivalent hyperparameter was available for inTrees and RuleFit, we subsequently limited the total number of rule conditions and, for RuleFit, active linear terms to approximately match the maximum number of splits allowed by the minimum leaf-size constraint.

Hyperparameters were set according to recommendations in the original publications or, where no recommendation was available, to their software defaults. The implementations, software versions or access dates, and supported outcome types (based on the used implementation) for all applied methods are summarized in \Cref{tab:method_description}.

The RF was fitted to the training data using version 0.17.0 of the R package \emph{ranger} \citep{wrightRangerFastImplementation2017}, with the default of 500 trees and $\mathrm{mtry}=\sqrt{p}$, where $p$ denotes the number of variables. The minimum node size was set to $10\%$ of the number of training observations.

\subsubsection*{ART}
ARTs were fitted using version 3.3.2 of the R package \emph{timbR} from GitHub \url{https://github.com/imbs-hl/timbR} \citep{laabsConstructionArtificialMost2024}. 
We used the variable-selection algorithm Boruta \citep{kursaFeatureSelectionBoruta2010}, controlled by \emph{imp.num.var}, which was proposed as a future improvement in the original publication and has since been implemented. 
Candidate split points were defined using quantiles through \emph{probs\_quantiles}, as recommended for the benchmark analyses in the original publication to reduce computation time. 
Tree growth was restricted by setting \emph{min.bucket} such that each terminal node contained at least $10\%$ of the training observations. All remaining hyperparameters were left at their default values.

\subsubsection*{CTree}
CTree was fitted using version 1.2-26 of the R package \emph{partykit} \citep{hothornPartykitModularToolkit2015}. The RF predictions were used as the outcome for CTree. 
All available variables were included as candidate variables. The \emph{minbucket} parameter required each terminal node to contain at least $10\%$ of the training observations. 
All remaining hyperparameters were left at their default values.

\subsubsection*{GUIDE}
GUIDE was fitted using the R implementation provided by \citet{herbingerLeveragingModelBasedTrees2024} in the GitHub repository \url{https://github.com/slds-lmu/mbt_comparison}. The RF predictions served as the outcome of the surrogate model. 
Since this implementation supports regression only, GUIDE was not applied to the classification data sets. We selected the GUIDE method by setting \emph{split.method} to \texttt{guide}. 
Tree growth was controlled by requiring at least $10\%$ of the training observations in each terminal node through \emph{min.split}. The parameter \emph{n.split} limited the number of tree-growth levels. The default improvement criterion was retained, with \emph{impr.par} set to 0.1 and \emph{pruning} set to \texttt{forward}.

\subsubsection*{inTrees}
inTrees was fitted using version 1.4 of the R package \emph{inTrees} \citep{dengInterpretingTreeEnsembles2019}. 
The package was obtained from the CRAN archive because it was removed from CRAN on January 12, 2026, after package check issues remained unresolved. 
Its original implementation supports forests fitted with the \emph{randomForest} package \citep{liawClassificationRegressionRandomForest2002b}, but not with the \emph{ranger} package \citep{wrightRangerFastImplementation2017}. We therefore adapted the code provided in Issues 3, 15, and 16 of the \emph{inTrees} GitHub repository at \url{https://github.com/softwaredeng/inTrees} to extract rules from the fitted \emph{ranger} forests. 
Since no hyperparameter corresponds to the common leaf-size constraint, we applied the decision budget after rule extraction and pruning. For classification, rules were selected using the original inTrees procedure. For regression, rules were ordered by training error, coverage, and rule length. Rules were retained in this order until adding another rule would exceed the budget, with each rule condition counting as one decision. All other hyperparameters were left at their default values.

\subsubsection*{MOB}
MOB was fitted using version 1.2-26 of the R package \emph{partykit} \citep{hothornPartykitModularToolkit2015}. 
The RF predictions were specified as the outcome. All available variables were included in the model formula. 
Since the implementation supports regression only, MOB was not applied to classification data sets. 
The \emph{minbucket} parameter required each terminal node to contain at least $10\%$ of the training observations. All remaining hyperparameters were left at their default values.

\subsubsection*{MRT}
MRT was fitted using version 3.3.2 of the R package \emph{timbR} from GitHub \url{https://github.com/imbs-hl/timbR} \citep{laabsConstructionArtificialMost2024}. 
The only hyperparameter of this method, \emph{metric}, was set to \texttt{weighted splitting variables}, following the recommendations in \citet{laabsIdentificationRepresentativeTrees2024}.
This metric quantifies pairwise tree distances based on the selected split variables and their positions within each tree. 
A separate constraint on model complexity could not be imposed because MRT selects an existing tree from the RF without pruning it.

\subsubsection*{NH}
NH was fitted using version 0.7-3 of the R package \emph{nodeHarvest} \citep{meinshausenNodeHarvestNodeHarvest2015}. The package was obtained from the CRAN archive because it was removed from CRAN on June 20, 2022, after package check issues remained unresolved. 
The RF predictions served as the outcome of the surrogate model. Since the available implementation supports regression only, NH was not applied to classification data sets. 
Model complexity was constrained through \emph{nodesize}, which required every node to contain at least $10\%$ of the training observations. 
However, because NH combines potentially overlapping nodes, this constraint does not impose the same upper bound on the total number of decisions as the leaf-size constraint for a single binary tree.
All remaining hyperparameters were left at their default values.

\subsubsection*{OC-CART}
OC-CART used a simple model as a surrogate model with the RF predictions as its outcome. Following the decision-tree approach of \citet{johanssonObtainingAccurateComprehensible2012}, we fitted a single tree using version 0.17.0 of the R package \emph{ranger} \citep{wrightRangerFastImplementation2017}. 
For this, we set \emph{num.trees} to 1, \emph{sample.fraction} to 1, and \emph{replace} to \texttt{FALSE}, thereby fitting the tree to the complete training sample without bootstrap resampling. The \emph{mtry} parameter was set to the total number of variables so that every variable was considered at each split.
Tree growth was restricted through \emph{min.bucket}, requiring each terminal node to contain at least $10\%$ of the training observations. 
All remaining hyperparameters were left at their default values.

\subsubsection*{RuleFit}
RuleFit was fitted using version 1.0.8 of the R package \emph{pre} \citep{fokkemaFittingPredictionRule2020}, with the RF predictions serving as the outcome. 
The current implementation cannot extract rules from an existing RF and constructs a separate RF using conditional inference trees. This RF comprised a default of 500 trees, as specified by \texttt{ntrees}. 
No hyperparameter in \emph{pre} corresponds to the common leaf-size constraint. Instead, we constrained the final model by selecting the least penalized model along the existing LASSO regularisation path whose total number of active decisions did not exceed $\lfloor n/\lfloor 0.1n\rfloor\rfloor-1$, where $n$ denotes the number of training observations.
This approximates the maximum number of decisions in a binary tree whose terminal nodes each contain at least $10\%$ of the training observations.
Each active linear term counted as one decision and each condition within an active rule counted as one decision. Predictions and complexity measures were calculated using the selected regularisation parameter. All remaining hyperparameters were left at their default values.

\subsubsection*{SIRUS}
SIRUS was fitted using version 0.3.3 of the R package \emph{sirus}, which was obtained from the CRAN archive after its removal from CRAN on January 15, 2026, after issues were not addressed. 
The available implementation supports regression and classification with a numerically encoded binary outcome. Only three benchmark data sets met the latter requirement. 
SIRUS was therefore not evaluated for classification because restricting its evaluation to these data sets would have prevented a consistent comparison across the broader collection of classification tasks. 
Model complexity was controlled by the default settings of ten rules and a maximum tree depth of two. All remaining hyperparameters were left at their default values.

\subsubsection*{SLIM}
SLIM was fitted using the R implementation provided by \citet{herbingerLeveragingModelBasedTrees2024} in the GitHub repository \url{https://github.com/slds-lmu/mbt_comparison}. 
The RF predictions served as the outcome. Since the implementation supports regression only, SLIM was not applied to classification data sets. 
We selected the SLIM splitting procedure by setting \emph{split.method} to \texttt{slim}. Tree growth was controlled by requiring at least $10\%$ of the training observations in each terminal node through \emph{min.split}. Due to high computation time, we limited \emph{n.split}, the maximum number of splits, to the value of 10. 
All remaining hyperparameters were left at their default values.

\subsubsection*{LM}
Following the oracle coaching principle, a linear regression was included as a surrogate model with the RF predictions as its outcome as in \citet{molnarInterpretableMachineLearning2022}. 
The initial model formula contained all available variables. 
Forward selection was then applied by adjusting the direction accordingly. This was done to obtain a more sparse and interpretable model. 
All remaining hyperparameters were left at their default values.
LM was evaluated only for regression outcomes. We did not use separate logistic and multinomial variants for classification because these models produce class probabilities that require data-set-specific decision rules and may become difficult to interpret when the number of outcome classes is large.

\subsubsection*{Reference model: decision tree}
A decision tree was included as a reference model for prediction performance and was fitted using version 0.17.0 of \emph{ranger} \citep{wrightRangerFastImplementation2017}. 
In contrast to the surrogate models, the decision tree predicted the original outcome. 
Its configuration otherwise followed that of the decision tree used for OC-CART. We set \emph{num.trees} to 1, \emph{sample.fraction} to 1, \emph{replace} to \texttt{FALSE}, and \emph{mtry} to the total number of variables. The resulting tree was fitted to the complete training sample without bootstrap resampling and considered every variable at each split. Tree growth was restricted through \emph{min.bucket}, requiring every terminal node to contain at least $10\%$ of the training observations. 
All remaining hyperparameters were left at their default values.

\subsection{Performance measures}
Each simulation setting was repeated 100 times. The runtime for each method and iteration was limited to 24 hours. Repetitions that exceeded this limit were treated as failures. Timeouts and other problems that prevented a method from producing a result are also reported in the results section.
Variables that occurred directly in the data-generating function of the outcome were defined as effect variables. All remaining variables were considered noise variables. Prediction measures were evaluated on the independent test data, whereas measures of model structure were obtained directly from the fitted models.

Several performance measures required a common definition of the elementary decisions represented by structurally different models. 
For the tree-based methods ART, CTree, GUIDE, MRT, NH, OC-CART, SLIM, and the decision tree, decisions were identified by split variables used in the nodes. 
For the rule-based representations produced by inTrees, MOB, RuleFit, and SIRUS, each condition within a rule was treated as a decision. 
For the linear-model surrogate, the estimated regression coefficients represented the corresponding model components. 
We selected these definitions to ensure that structural comparisons across methods were as consistent as possible.

\subsubsection{Deviation from RF}
Similarity to the RF, and thus the suitability of an interpretable model as a surrogate, was assessed with respect to both its predictions and its decision structure. 

Prediction similarity was evaluated by comparing the predictions of each interpretable model with those of the corresponding RF on the test data. Prediction disagreement was quantified using the root mean squared error (RMSE) for regression and the misclassification rate (MCR) for classification.
Structural similarity was additionally evaluated by comparing the variables used for splitting or rule construction or as regression coefficients of each interpretable model with the split variables used in the individual trees of the corresponding RF.

\subsubsection{Model complexity}
Model complexity as one surrogate for interpretability was characterized by the total number of structural components, defined as splits for tree-based models, rule conditions for rule-based models, and regression coefficients for the linear model.

\subsubsection{Explainability}
To assess the reliability of the surrogate models' decision-making, their decisions were evaluated based on whether they were made using effect or noise variables represented in the fitted models.
We calculated the proportions of true effect variables and noise variables that appeared in splits, rule conditions, or regression coefficients.

\subsubsection{Prediction performance}
Prediction performance on the test data, and thus the potential suitability of the interpretable models as standalone prediction models, was assessed using the F1 score for classification. 
This score is defined as
\[
    \mathrm{Macro\text{-}F1} = \frac{1}{K}\sum_{k=1}^{K}
    \frac{2\,\mathrm{TP}_k}{2\,\mathrm{TP}_k+\mathrm{FP}_k+\mathrm{FN}_k},
\]
where $K$ denotes the number of classes, and $\mathrm{TP}_k$, $\mathrm{FP}_k$, and $\mathrm{FN}_k$ denote the numbers of true positives, false positives, and false negatives for class $k$, respectively, treating each class against all remaining classes. Each class contributes equally to the overall score.
The macro-averaged F1 score combines precision and recall while giving equal weight to each class, preventing performance on frequent classes from dominating the overall assessment.

For regression, prediction performance was evaluated using the coefficient of determination ($R^2$) and an adapted Akaike information criterion (AIC), which balances prediction error on the test data against model complexity:
\[
\mathrm{AIC}=n_{\mathrm{test}}\log(\mathrm{MSE}_{\mathrm{test}})+2D,
\]
where $D$ denotes the number of model decisions and $\mathrm{MSE}_{\mathrm{test}}$ the mean squared error on the test data. Lower values indicate a more favorable trade-off.

\subsubsection{Runtime}
Computations used \emph{R} 4.3.0 and \emph{batchtools} 0.9.17 \cite{langBatchtoolsToolsWork2017b} on a heterogeneous Linux cluster managed by Slurm. Cluster nodes differed in their hardware configurations, with 96 to 738 GB of RAM and 12 to 64 CPUs per node. One CPU was requested per job. All jobs were randomly distributed across the cluster nodes, with no preferential assignment of particular jobs to specific nodes.
We measured user CPU time for surrogate construction with \texttt{proc.time()}, excluding RF training and evaluation.
Because computations were performed on heterogeneous hardware, the reported runtimes indicate approximate computational effort and do not constitute a controlled comparison on identical processors.

\subsection{Results}
Given the large number of methods, data sets, and performance measures, results in the main text are presented as ranks in \Cref{fig:sim_rank_boxplots}.
For each data set and performance measure, the median across repetitions was calculated for each method, and methods were ranked according to these medians, with lower ranks indicating better results. The boxplots summarize the ranks of the methods across the data sets.
Detailed boxplots showing the distributions across all repetitions, without prior aggregation, are provided in the appendix.
 
All figures used for the results displayed in the text or appendix can be recreated in our Shiny app: \url{https://bioweb2.imbs.uksh.de/rsconnect/Comparison_RF_explainability_methods/}. 

GUIDE, LM, SIRUS, SLIM, NH, and MOB were not evaluated for classification because their implementations supported regression only.
In digit recognition, RuleFit completed 95 of 100 repetitions. The five failed fits returned an error indicating that one or no outcome classes were represented, which is consistent with the small sample size relative to the number of classes (\Cref{tab:method_failures}). Ranks and boxplots were based on the available results and should therefore be interpreted with caution.

\subsubsection{Deviation from RF}

\paragraph*{Deviation in predictions}
For regression, GUIDE and LM achieved the highest similarity in the simulation settings with linear effects (see \Cref{fig:sim_rank_boxplots} and \Cref{app:fig:sim_dev_rf_pred}). These results were expected because these methods can represent linear effects directly.
SIRUS performed best for the subgroup structure in the simulation with subgroups, SLIM for the simulation with nonlinear effects, and NH for the simulation with exponential effects. NH had consistently favorable ranks and is the most robust choice when the structure is unknown. inTrees, ART, the decision tree, and MRT generally showed larger deviations and should not be preferred when agreement with RF predictions is the main objective.

For classification, MRT performed best in digit recognition, and the decision tree in the noisy variant. RuleFit was preferable for the interaction structure in the XOR-problem simulation, closely followed by inTrees and ART. 
However, RuleFit and inTrees performed poorly in both multiclass settings. OC-CART and the decision tree had the most favorable average ranks, with OC-CART providing the more direct RF explanation because it was fitted to the RF predictions. 
Five RuleFit repetitions were unavailable in digit recognition, which limits its suitability for small multiclass data sets (\Cref{tab:method_failures}).

\paragraph*{Deviation in decisions}
For regression, NH had the best average structural rank and also provided the most consistent balance between prediction and structural similarity (see \Cref{fig:sim_rank_boxplots} and \Cref{app:fig:sim_dist_sv}). SIRUS was a strong structural alternative in the simulation settings with linear effects. MRT reproduced RF variable usage best in the simulation with subgroups and the simulation with exponential effects, but had the lowest prediction similarity in both settings. In the simulation with nonlinear effects, the distance was zero for ten methods because they used the same variables as the RF. 
GUIDE and LM reproduced predictions well in the linear settings, but structural similarity was low for LM and unavailable for GUIDE.
For GUIDE, the fitted models contained no splits, and the implemented structural measure did not account for variables in the linear node model.

For classification, MRT achieved the highest structural similarity in all settings and was also best in prediction similarity for digit recognition. The decision tree and OC-CART provided the best balance in the noisy variant. In the XOR-problem simulation, RuleFit predicted best but showed low structural similarity, while ART provided the most balanced result. Thus, high prediction similarity alone should not justify interpreting a method's decisions as those of the RF. Such an interpretation may be misleading when the variables used by both models differ.

\subsubsection{Model complexity}
The number of decisions varied substantially across methods despite the common complexity constraint (see \Cref{fig:sim_rank_boxplots} and \Cref{app:fig:sim_number_decisions}).
For regression, ART was the most compact nonempty model, with three to four decisions. The decision tree, CTree, MOB, OC-CART, SLIM, RuleFit, and inTrees generally remained below ten, while SIRUS used approximately 14 to 15. LM and NH reached medians of up to 77 and 103 components. MRT varied most strongly and reached 330 decisions in the simulation with nonlinear effects and 671 in the simulation with exponential effects.
GUIDE often produced no split, but its node regression model still represented the complete data. 
ART is therefore the clearest recommendation for regression when compactness is the primary objective, while NH and MRT should generally be avoided.  
For classification, ART was also the most compact method, with two to five decisions. The decision tree, CTree, OC-CART, RuleFit, and inTrees used between five and nine, compared with 25 to 76 for MRT. ART is therefore also preferable for classification when minimal complexity is required, while the former methods provide moderately sized alternatives.

\subsubsection{Explainability}

\paragraph*{Used effect and noise variables in decisions}
Reliable explanations should cover a large proportion of the true effect variables while including few noise variables. 
For regression, inTrees and NH provided the most favorable balance in the \texttt{Sim\_ART} settings because they covered comparatively many effect variables without using noise variables (see \Cref{fig:sim_rank_boxplots}, \Cref{app:fig:sim_covered_effect_vars}, and \Cref{app:fig:sim_covered_noise_vars}). 
MRT and LM often covered more effects, but also included the largest proportions of noise variables. In the simulation with subgroups, NH provided the best balance by covering 85.7\% of the effects without using noise variables. ART and RuleFit were more conservative alternatives with 57.1\% effect coverage and no noise variables. Most methods recovered both effect variables in the simulation with nonlinear effects, so this setting did not distinguish between them. In the simulation with exponential effects, NH provided the most favorable balance by covering a median of 87.5\% of the effect variables without using noise variables. SIRUS and RuleFit also avoided noise variables but covered only 75\% and 62.5\% of the effects. LM and MRT covered all effect variables, but LM also used 23.9\% and MRT all of the noise variables. 

For classification, MRT covered all effect variables in digit recognition, followed by inTrees. Noise coverage could not be assessed because this setting contained no designated noise variables. In the noisy variant, inTrees provided the best balance by covering most effect variables without using noise variables. The decision tree and OC-CART covered fewer effects and included a small proportion of noise variables. In the XOR-problem simulation, ART and inTrees covered all effect variables without using noise variables and are therefore the strongest recommendations when reliable variable identification is the primary objective. MRT also covered all effects but included most available noise variables in the noisy and interaction settings, which makes its explanations less reliable despite its high effect coverage.

GUIDE produced a single linear regression model without any splits. Because the explainability measures considered only split variables and excluded variables used in the regression model, they were not calculated for GUIDE, preventing comparison with the other methods on these measures.

\subsubsection{Prediction performance}
\begin{figure}
    \centering
    \includegraphics[width=\linewidth]{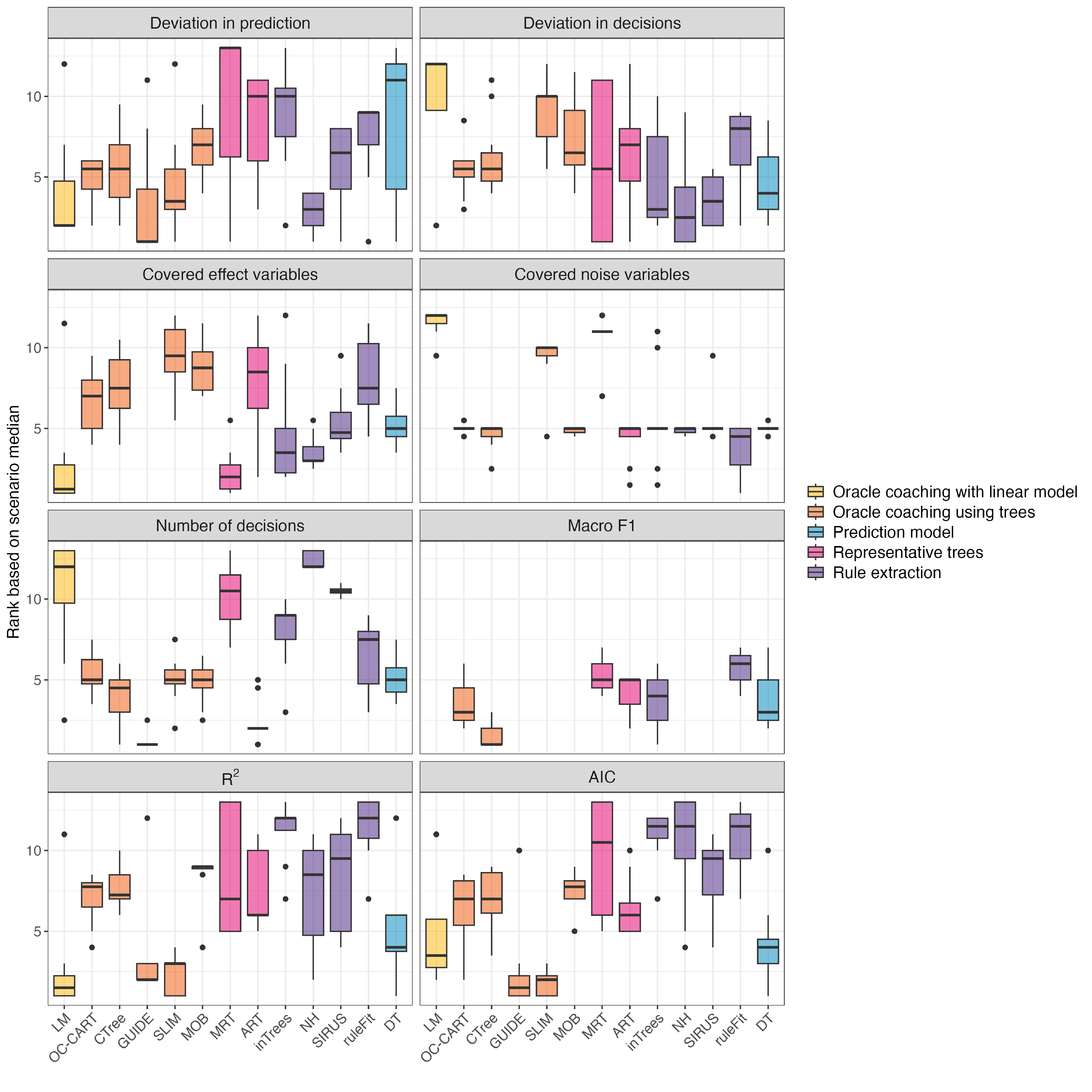}
    \caption{Ranks of methods across simulation scenarios for each performance measure (facets), with lower ranks indicating better performance. Within each scenario, the median performance across 100 repetitions was calculated for each method, and methods were then ranked according to these medians. Boxplots summarize the resulting ranks across scenarios. Methods are displayed on the x-axis and colored by method type: oracle coaching using a linear model, oracle coaching using tree-based methods, representative trees, rule extraction methods, and the decision tree (DT) as a prediction model.}
    \label{fig:sim_rank_boxplots}
\end{figure}

\paragraph{$R^2$ and AIC}
Both $R^2$ and AIC favored the decision tree in the simulation with few large effects (see \Cref{fig:sim_rank_boxplots}, \Cref{app:fig:sim_r_squared}, and \Cref{app:fig:sim_aic}). This advantage is expected because the decision tree was the only method fitted directly to the observed outcome, while the other methods approximated the RF. 
SLIM performed best in the simulations with subgroups, nonlinear effects, and exponential effects. In the simulations with many small effects and continuous variables, LM achieved the highest $R^2$, while GUIDE obtained the lowest AIC. SLIM is therefore the strongest general recommendation, while LM is preferable for prediction and GUIDE for the measured accuracy-complexity trade-off in predominantly linear settings. RuleFit and inTrees ranked lowest overall and sometimes had negative $R^2$ values.

\paragraph{F1 score}
Although the decision tree was the only method fitted directly to the observed outcome, CTree achieved the highest Macro F1 score in both digit-recognition settings and is therefore recommended for multiclass prediction (see \Cref{fig:sim_rank_boxplots} and \Cref{app:fig:sim_macro_f1}). 
In the digit-recognition setting without noise, OC-CART ranked second and the decision tree third, while the decision tree ranked second and OC-CART third in the noisy variant. inTrees attained the best result in the XOR-problem simulation and is preferred for the XOR structure. 
RuleFit had the lowest average rank and is not recommended when classification performance is the primary objective.

\subsubsection{Runtime}
CPU time was measured in minutes (see \Cref{app:fig:sim_runtime}). For regression, all methods had median runtimes below 0.15 minutes except SLIM at 1.12 minutes, RuleFit at 2.28 minutes, and ART at 2.59 minutes. For classification, median runtimes were below 0.16 minutes except for RuleFit at 0.50 minutes. The decision tree and OC-CART were fastest for both outcome types.

\section{Comparison using benchmark data}

\subsection{Aim}
The benchmark comparison aimed to evaluate the methods across a broad range of regression and classification data sets with unknown data-generating mechanisms. 
It examined the similarity to the RF, model complexity, prediction performance, and runtime in order to provide practical guidance for method selection.

\subsection{Data sets}
The empirical comparison included 67 publicly available benchmark data sets, comprising 28 regression and 39 classification tasks. 
The data sets originated from the empirical evaluations reported in the publications identified by the literature review and covered a wide range of application domains, sample sizes, dimensionalities, and variable types. 
The source, persistent identifier where available, dimensions before and after quality control, outcome type, and the publication motivating inclusion are reported in \Cref{tab:dataset_description}. The selection process and reasons for excluding candidate data sets are summarized in \Cref{fig:flowchart_datasets}.

\subsubsection*{Regression data sets}
The regression benchmark comprised 28 data sets with continuous outcomes. 
The collection included 15 data sets obtained from OpenML \citep{bischlOpenMLInsights102025}, four from the UCI Machine Learning Repository \citep{kellyUCIMachineLearning}, six distributed with R packages (\emph{mlbench} V2.1.7 \citep{leischMlbenchMachineLearning2026}, \emph{AER} V1.2.16 \citep{kleiberAppliedEconometrics2008}, \emph{GLMsData} V1.4 \citep{dunnGLMsDataGeneralizedLinear2022}), and three provided through the data archive accompanying \url{https://hastie.su.domains/ElemStatLearn/datasets} \cite{hastieElementsStatisticalLearning2009}.
After preprocessing, the number of observations ranged from 97 to 577,601. The number of recorded variables, including the outcome, ranged from 3 to 361. None of the regression data sets had more variables than observations after preprocessing (see below for details).

\subsubsection*{Classification data sets}
The classification benchmark comprised 39 data sets, of which 34 were obtained from the UCI Machine Learning Repository \citep{kellyUCIMachineLearning}, one from OpenML \citep{bischlOpenMLInsights102025}, and four were distributed with R packages (\emph{mlbench} V2.1.7 \citep{leischMlbenchMachineLearning2026}, \emph{ipred} V0.9.15 \citep{petersIpredImprovedPredictors2024}). 
The collection included both binary and multiclass outcomes.
After preprocessing, the classification data sets contained between 55 and 5,000 observations. Their dimensions ranged from 4 to 441 recorded variables, including the outcome.

\subsubsection*{Preprocessing}
All benchmark data sets were processed using a predefined quality-control procedure before cross-validation. Identifier columns were removed where present. 
Variables with more than 50\% missing values were excluded. 
Observations with any remaining missing values were then removed. 
Rare levels of categorical variables were combined into a common \texttt{other} category when their relative frequency was below 20\%. For classification data sets, this procedure was also applied to the outcome.
Variables that were constant after these steps were also excluded.
To ensure that class labels could be processed consistently by all implementations, hyphens and ampersands in outcome labels were replaced with underscores, and whitespace was removed. The outcome was consistently named \texttt{y}. 
For regression data sets, observations with extreme outcome values were excluded if their outcome lay outside the interval defined by the median plus or minus five interquartile ranges. 
The dimensions before and after the complete preprocessing procedure are reported in \Cref{tab:dataset_description}.

\subsection{Targets}
The targets were the same as those in the simulation study.

\subsection{Methods}
Analogous to the simulation study, the 12 surrogate and rule extraction methods, together with the decision tree, were included in the comparison study.

\subsection{Performance measures}
Performance was evaluated using 10-fold cross-validation. 
The methods were evaluated using the same criteria as in the simulation study, namely deviation from the RF, model complexity, prediction performance, and runtime. Explainability could not be evaluated because the true data-generating process is unknown, making it impossible to distinguish unambiguously between effect and noise variables.

\subsection{Results}
As for the simulated data sets, the benchmark results are presented in a compact, rank-based form in \Cref{fig:bench_rank_boxplots}, while the unaggregated results are provided in the Appendix. The figures can also be reproduced using our Shiny app (\url{https://bioweb2.imbs.uksh.de/rsconnect/Comparison_RF_explainability_methods/}).
Failures and their recorded causes are summarized in \Cref{tab:method_failures}.

NH produced at least one result for only 13 of the 28 regression data sets: complete failures were caused by negative-length-vector or non-positive-definite-quadratic-function errors, while eight of ten folds failed for \texttt{bike\_sharing} because the constraints were inconsistent. ART and RuleFit each produced regression results for 26 data sets; ART exceeded the 24-hour limit in all folds of two data sets and in seven folds of \texttt{nyc-taxi}, whereas RuleFit exceeded the limit in all folds of two data sets. SLIM had individual prediction failures across four regression data sets. For classification, ART returned an empty variable selection in four folds of \texttt{heart\_disease\_s}. These incomplete results should therefore be interpreted with caution, as the displayed summaries are based only on results from successful folds.

\begin{figure}
    \centering
    \includegraphics[width=\linewidth]{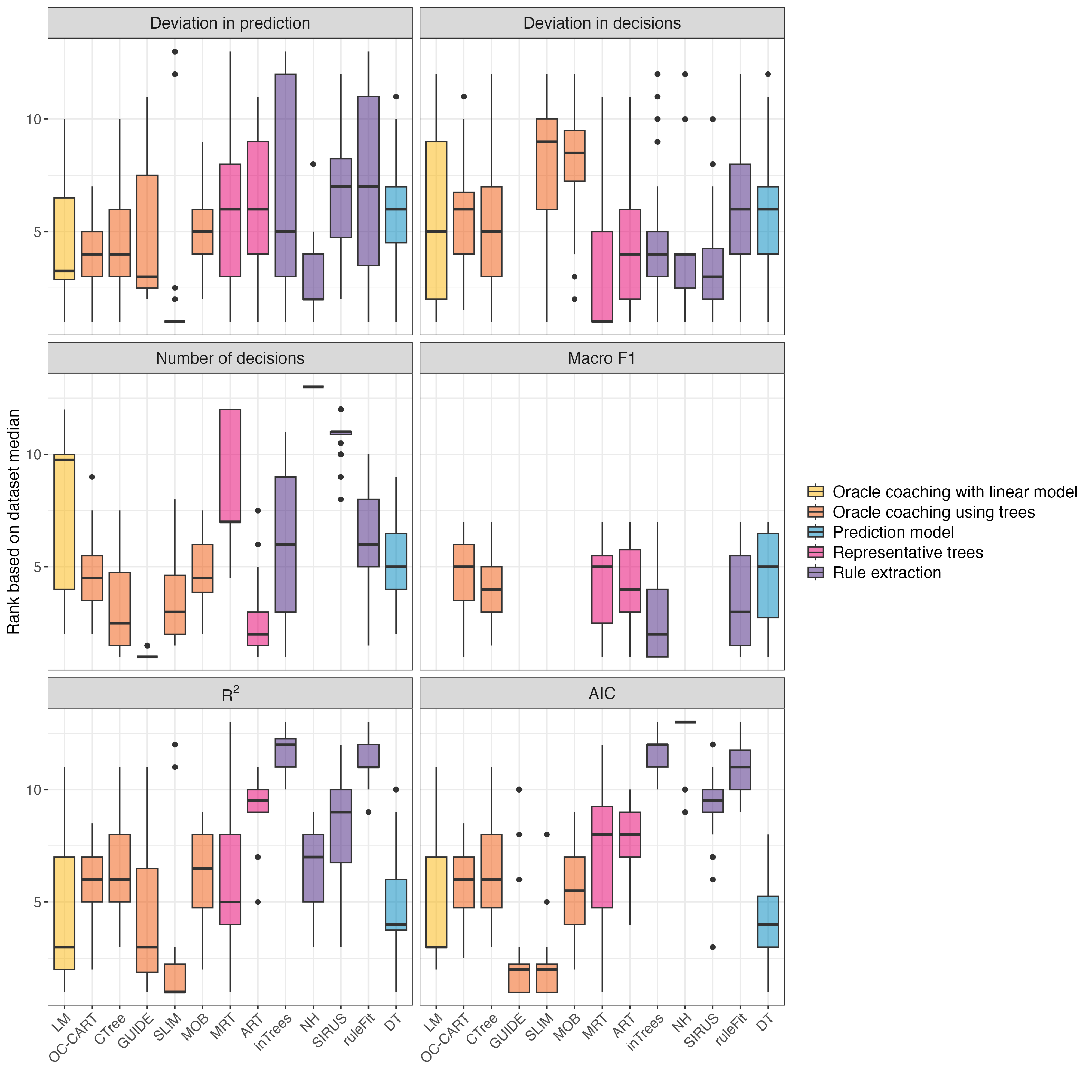}
    \caption{Ranks of methods across benchmark data sets for each performance measure (facets), with lower ranks indicating better performance. Within each data set, the median performance across 10-fold CV was calculated for each method, and methods were then ranked according to these medians. Boxplots summarize the resulting ranks across data sets. Methods are displayed on the x-axis and colored by method type: oracle coaching using a linear model, oracle coaching using tree-based methods, representative trees, rule extraction methods, and the decision tree (DT) as a prediction model.}
    \label{fig:bench_rank_boxplots}
\end{figure}

\subsubsection{Deviation from RF}

\paragraph*{Deviation in predictions}
Across the 28 regression data sets, SLIM achieved the lowest deviation on 23, consistent with the simulation results and its ability to represent both global additive effects and local deviations through separate leaf models (see \Cref{fig:bench_rank_boxplots}, \Cref{app:fig:bench_dev_rf_pred_rmse}, and \Cref{app:fig:bench_dev_rf_pred_mcr}). 
LM ranked among the three best methods on 14 data sets, but outperformed SLIM on only three. 
GUIDE also ranked among the three best methods on 12 of these 14 data sets, suggesting that their RF prediction functions could be approximated reasonably well by a global linear model. This pattern does not establish that the data-generating mechanisms were linear. 
SLIM should therefore be preferred when local heterogeneity is expected. If LM or an unsplit GUIDE model achieves comparably stable similarity, LM may instead be selected because a single set of global coefficients provides a more uniform interpretation than multiple local SLIM models. 
NH ranked among the three best methods on seven of the 13 data sets for which it was available, nine of which were below the median sample size. Its limited coverage prevents a general recommendation.

Across the 39 classification data sets, inTrees achieved the best mean similarity rank, followed by CTree and OC-CART. inTrees performed best on average among the 13 smallest and the 13 highest-dimensional data sets, while CTree led among the 13 data sets with the fewest variables and OC-CART among the 13 largest. This may reflect the ability of inTrees to represent interactions through compact rules, of CTree to identify stable partitions in low dimensions, and of OC-CART to estimate a surrogate tree from larger samples. However, these patterns were descriptive, and neither $p/n$, class number, nor class imbalance consistently explained performance. inTrees is therefore the preferred general choice, with CTree as an alternative for low-dimensional data with limited redundancy and OC-CART for large samples.

\paragraph*{Deviation in decisions}
MRT achieved the highest structural similarity, particularly when few variables and no strongly correlated numeric variables were present (see \Cref{fig:bench_rank_boxplots} and \Cref{app:fig:bench_dist_sv}). 
SIRUS was the strongest regression alternative, and inTrees the strongest classification alternative. Since MRT had comparatively low prediction similarity for regression, it should be selected only when agreement with the RF decision structure is more important than agreement with its predictions. 

\subsubsection{Model complexity}
ART provided the most consistently compact nonempty representation across both outcomes, with CTree similarly compact for classification (see \Cref{fig:bench_rank_boxplots} and \Cref{app:fig:bench_number_decisions}). SLIM used few splits, but the coefficients of its local leaf models were not represented by this measure. The same limitation applies to the unsplit regression model of GUIDE, so its apparent complexity of zero is not directly comparable. NH was complex because it combines overlapping nodes, while MRT inherited the comparatively large structure of an unconstrained RF tree.

\subsubsection{Prediction performance}

\paragraph*{$R^2$ and AIC}
SLIM achieved the best mean rank for both $R^2$ and AIC rank (see \Cref{fig:bench_rank_boxplots}, \Cref{app:fig:bench_aic}, and \Cref{app:fig:bench_r_squared}). SLIM is therefore the strongest general recommendation when prediction performance is prioritized, while GUIDE performed best according to the measured accuracy–complexity trade-off. 
LM and GUIDE were preferable when a global linear model already performed well, whereas SLIM became increasingly favorable as the performance advantage of the RF over LM grew. This agrees with the simulation results and supports selecting LM for a uniform global interpretation when its performance is comparable, and SLIM when nonlinear or locally varying effects provide a relevant performance gain. GUIDE also achieved favorable AIC values, although its unsplit leaf model is not represented by the complexity term. The decision tree ranked below these methods despite being fitted directly to the observed outcome, while RuleFit and inTrees performed worst overall.

\paragraph*{Macro F1 score}
inTrees achieved the best mean Macro F1 rank and is therefore the preferred general choice for classification performance (see \Cref{fig:bench_rank_boxplots} and \Cref{app:fig:bench_macro_f1}). RuleFit ranked second but was unavailable for two data sets and had individual failed folds in four further data sets because of unseen factor levels or insufficiently represented outcome classes (\Cref{tab:method_failures}). CTree provided the strongest alternative with complete coverage. Unlike in the simulation, neither $p/n$, class number, nor class imbalance identified settings in which inTrees or RuleFit was consistently preferable. Method selection should therefore be validated on the data at hand when classification performance is the primary objective.

\subsubsection{Runtime}
The decision tree and OC-CART were fastest across data set sizes and dimensions because they required only a single tree fit (see \Cref{app:fig:bench_runtime}). ART and RuleFit were slowest, with regression medians of approximately 1.0 minute, followed by SLIM at 0.85 minutes. RuleFit and ART were also slowest for classification, with medians of 0.41 and 0.31 minutes.

\section{Discussion}
No single method is universally preferable for interpreting RFs. Method selection should instead reflect the outcome type, anticipated data structure, and whether the objective is prediction similarity, structural similarity, prediction performance, or interpretability.

For regression, SLIM was the strongest general choice when prediction similarity or performance was central, particularly for nonlinear or locally varying effects. When LM performed comparably, its global coefficient vector provided the simpler interpretation. Competitive LM or unsplit GUIDE performance may indicate that a linear approximation is sufficient, but does not establish a linear data-generating mechanism. NH and SIRUS showed favorable structural similarity, while MRT was useful when agreement with RF variable usage outweighed prediction similarity. ART consistently provided the most compact representation, and inTrees and NH often recovered effect variables without including noise variables.

For classification, inTrees was the strongest general choice on the benchmark data for similarity and prediction performance. CTree was a robust alternative with complete coverage, while OC-CART was competitive on larger data sets and computationally efficient. 
In the simulations, CTree achieved the best multiclass prediction, MRT closely reproduced the RF in digit recognition without additional noise variables, and inTrees was well suited to the XOR structure. ART was preferable when compactness or effect-variable recovery was prioritized.

The results demonstrate a trade-off between interpretability and prediction performance. Flexible models may perform well at the cost of greater complexity, whereas compact models may omit relevant structure. 
MRT sometimes combined large models with weak prediction similarity, while ART was compact but did not generally predict best. SLIM combined strong performance with few splits, although its local coefficients add complexity not captured by the split count. Prediction and structural similarity should therefore be considered jointly with complexity.

For data sets with unknown underlying structure, relying on a single surrogate modeling approach may be insufficient. 
Structurally different candidates should be compared using cross-validated similarity, prediction performance, complexity, and stability. 
Comparable LM performance supports a global approximation, while improvements by SLIM or tree- and rule-based methods may suggest local heterogeneity, subgroups, or interactions. The simplest stable model whose performance is not meaningfully worse than that of more flexible alternatives should be selected.

In practice, however, differences between the implementations restricted the comparability of the methods. Some methods did not support classification, and CPU runtimes above 24 hours, numerical optimization errors, incompatible prediction methods, empty variable selections, unseen factor levels, or insufficiently represented outcome classes prevented complete benchmark coverage (\Cref{tab:method_failures}).
Average results therefore refer to different subsets of tasks, although method failure is reported and broadly applicable methods are emphasized. The 67 heterogeneous benchmarks nevertheless provide broader evidence than evaluations based on few selected tasks.
Restricting the number of splits or rules did not make model complexity fully comparable, as compactness was assessed within each model form using measures such as the number of splits in trees or coefficients in linear models.
To improve comparability, the decision tree and OC-CART were fitted without post-pruning, avoiding an additional complexity-reduction step beyond the common minimum leaf-size constraint. However, this choice may have disadvantaged both methods, as pruning could yield a better balance between prediction performance and model complexity.
No data-specific hyperparameter optimization was performed, so alternative settings could change method rankings. Instead, we followed recommendations from the original publications and otherwise used defaults. This reproducible strategy reflects typical use without extensive tuning and avoids favoring methods with larger tuning spaces.

The favorable classification results for inTrees agree with \citet{ariaComparisonInterpretativeProposals2021}, who found slightly better prediction performance for inTrees than for NH on six binary tasks. 
We extend this evidence to a larger benchmark including multiclass outcomes, while showing that inTrees is not superior for every interpretability objective. Direct comparison with their NH result is limited because only regression is implemented.
Our results also support \citet{herbingerLeveragingModelBasedTrees2024}, who concluded that surrogate choice depends on the balance between similarity and interpretability and that hyperparameters influence performance. 
We confirm this trade-off beyond their main-effect simulations and extend it to nonlinear, interaction, subgroup, and multiclass settings and benchmark data. The broader method set further shows the relevance of rule extraction and structural representatives when variable usage rather than prediction is the target.

Future work should evaluate hyperparameter optimization under equal computational budgets, develop complexity measures that jointly represent splits, rules, and coefficients, and examine explanation stability. 
A data-driven selection procedure could then combine cross-validated similarity, prediction performance, stability, and complexity for new data sets.

In conclusion, interpretable RF approximations should be selected for their intended purpose rather than by a universal ranking. SLIM is the strongest general option for regression when prediction similarity is central, while LM is preferable when it performs comparably and a global interpretation is desired. 
inTrees is the strongest general classification option, with CTree and OC-CART as useful alternatives, and ART should be considered when compactness is paramount. Regardless of the selected method, prediction agreement alone is not sufficient for explaining an RF and should be complemented by structural similarity, complexity, and stability.


\section*{Author Contributions}
Lea L. Mairhöfer: conceptualization, data curation, formal analysis, investigation,
methodology, software, visualization, writing---original draft,
writing---review \& editing.\\
Jonathan B. Krause: software, writing---review \& editing.\\
Björn-Hergen Laabs: conceptualization, funding acquisition, supervision, writing---review \& editing.\\
Silke Szymczak: conceptualization, supervision, writing---review \& editing.\\\\
All authors read and approved the final manuscript.


\section*{Acknowledgments}

This study was funded by the Medical Section of the University of Luebeck
(J01--2024, BL).


\section*{Conflict of Interest}

The authors declare no conflicts of interest.


\section*{Data Availability Statement}
The GitHub repository \url{https://github.com/imbs-hl/Comparison_RF_explainability_methods_paper} contains the results from the paper and the full R code, allowing users to reproduce all results and figures or to generate their own results. The code for all simulated data sets is stored there. The benchmark data sets used are all publicly available at the sources stated in \Cref{tab:dataset_description}.

All figures used for the results displayed in the text or appendix can be recreated in our Shiny app: \url{https://bioweb2.imbs.uksh.de/rsconnect/Comparison_RF_explainability_methods/}.


\printbibliography[title={References}]

@article{ariaComparisonInterpretativeProposals2021,
  title = {A comparison among interpretative proposals for random forests},
  author = {Aria, Massimo and Cuccurullo, Corrado and Gnasso, Agostino},
  year = 2021,
  month = dec,
  journal = {Mach. Learn. Appl.},
  volume = {6},
  issn = {2666-8270},
  doi = {10.1016/j.mlwa.2021.100094},
  urldate = {2024-04-12},
  eid = {100094}
}

@article{bachrachBoneMineralAcquisition1999,
  title = {Bone mineral acquisition in healthy Asian, Hispanic, black, and Caucasian youth: a longitudinal study},
  shorttitle = {Bone mineral acquisition in healthy asian, hispanic, black, and caucasian youth},
  author = {Bachrach, L. K. and Hastie, T. and Wang, M. C. and Narasimhan, B. and Marcus, R.},
  year = 1999,
  month = dec,
  journal = {J Clin Endocrinol Metab},
  volume = {84},
  number = {12},
  pages = {4702--4712},
  issn = {0021-972X},
  doi = {10.1210/jcem.84.12.6182},
  langid = {english},
  pmid = {10599739}
}

@article{banerjeeIdentifyingRepresentativeTrees2012a,
  title = {Identifying representative trees from ensembles},
  author = {Banerjee, Mousumi and Ding, Ying and Noone, Anne-Michelle},
  year = 2012,
  journal = {Stat. Med.},
  volume = {31},
  pages = {1601--1616}
}

@article{benardSIRUSStableInterpretable2021,
  title = {SIRUS: stable and interpretable rule set for classification},
  shorttitle = {SIRUS},
  author = {B{\'e}nard, Cl{\'e}ment and Biau, G{\'e}rard and {da Veiga}, S{\'e}bastien and Scornet, Erwan},
  year = 2021,
  month = jan,
  journal = {Electron. J. Stat.},
  volume = {15},
  number = {1},
  pages = {427--505},
  publisher = {{Institute of Mathematical Statistics and Bernoulli Society}},
  issn = {1935-7524, 1935-7524},
  urldate = {2026-10-14},
  langid = {english}
}

@article{bischlOpenMLInsights102025,
  title = {OpenML: insights from 10 years and more than a thousand papers},
  author = {Bischl, Bernd and Casalicchio, Giuseppe and Das, Taniya and Feurer, Matthias and Fischer, Sebastian and Gijsbers, Pieter and Mukherjee, Subhaditya and M{\"u}ller, Andreas C. and N{\'e}meth, L{\'a}szl{\'o} and Oala, Luis and Purucker, Lennart and Ravi, Sahithya and {van Rijn}, Jan N. and Singh, Prabhant and Vanschoren, Joaquin and {van der Velde}, Jos and Wever, Marcel},
  year = 2025,
  journal = {Patterns},
  volume = {6},
  number = {7},
  publisher = {Cell Press},
  doi = {10.1016/j.patter.2025.101317},
  eid = {101317}
}

@book{breimanClassificationRegressionTrees1984,
  title = {Classification and regression trees},
  author = {Breiman, L. and Friedman, J. and Stone, C.J. and Olshen, R.A.},
  year = 1984,
  publisher = {Taylor \& Francis},
  isbn = {978-0-412-04841-8},
  lccn = {83019708}
}

@article{breimanRandomForests2001,
  title = {Random forests},
  author = {Breiman, Leo},
  year = 2001,
  journal = {Mach. Learn.},
  volume = {45},
  pages = {5--32},
  issn = {1573-0565},
  urldate = {2021-05-16},
  langid = {english}
}

@article{dengInterpretingTreeEnsembles2019,
  title = {Interpreting tree ensembles with inTrees},
  author = {Deng, Houtao},
  year = 2019,
  month = jun,
  journal = {Int. J. Data Sci. Anal.},
  volume = {7},
  pages = {277--287},
  issn = {2364-4168},
  urldate = {2026-01-17},
  langid = {english}
}

@misc{dunnGLMsDataGeneralizedLinear2022,
  title = {GLMsData: generalized linear model data sets},
  author = {Dunn, Peter K. and Smyth, Gordon K.},
  year = 2022,
  doi = {10.32614/CRAN.package.GLMsData}
}

@article{fokkemaFittingPredictionRule2020,
  title = {Fitting prediction rule ensembles with R package {{pre}}},
  author = {Fokkema, Marjolein},
  year = 2020,
  journal = {J. Stat. Softw.},
  volume = {92},
  number = {12},
  pages = {1--30},
  doi = {10.18637/jss.v092.i12}
}

@article{friedmanGreedyFunctionApproximation2001,
  title = {Greedy function approximation: a gradient boosting machine},
  shorttitle = {Greedy function approximation},
  author = {Friedman, Jerome H.},
  year = 2001,
  month = oct,
  journal = {Ann. Stat.},
  volume = {29},
  number = {5},
  pages = {1189--1232},
  publisher = {Institute of Mathematical Statistics},
  issn = {0090-5364, 2168-8966},
  urldate = {2025-06-25}
}

@article{friedmanPredictiveLearningRule2008,
  title = {Predictive learning via rule ensembles},
  author = {Friedman, Jerome H. and Popescu, Bogdan E.},
  year = 2008,
  month = sep,
  journal = {Ann. Appl. Stat.},
  volume = {2},
  pages = {916--954},
  publisher = {Institute of Mathematical Statistics},
  issn = {1932-6157, 1941-7330},
  urldate = {2026-03-25},
  langid = {english}
}

@article{goldsteinPeekingBlackBox2015a,
  title = {Peeking inside the black box: visualizing statistical learning with plots of individual conditional expectation},
  shorttitle = {Peeking inside the black box},
  author = {Goldstein, Alex and Kapelner, Adam and Bleich, Justin and Pitkin, Emil},
  year = 2015,
  month = jan,
  journal = {J. Comput. Graph. Stat.},
  volume = {24},
  number = {1},
  pages = {44--65},
  publisher = {Taylor \& Francis},
  issn = {1061-8600},
  doi = {10.1080/10618600.2014.907095},
  urldate = {2026-09-01}
}

@inproceedings{grinsztajnWhyTreebasedModels2022,
  title = {Why do tree-based models still outperform deep learning on typical tabular data?},
  booktitle = {Proceedings of the 36th International Conference on Neural Information Processing Systems},
  author = {Grinsztajn, L{\'e}o and Oyallon, Edouard and Varoquaux, Ga{\"e}l},
  year = 2022,
  month = nov,
  series = {NIPS '22},
  pages = {507--520},
  publisher = {Curran Associates Inc.},
  address = {Red Hook, NY, USA},
  urldate = {2026-11-09},
  isbn = {978-1-7138-7108-8}
}

@article{guidottiSurveyMethodsExplaining2018,
  title = {A survey of methods for explaining black box models},
  author = {Guidotti, Riccardo and Monreale, Anna and Ruggieri, Salvatore and Turini, Franco and Giannotti, Fosca and Pedreschi, Dino},
  year = 2019,
  journal = {ACM Comput. Surv.},
  volume = {51},
  number = {5},
  pages = {1--42},
  issn = {0360-0300},
  doi = {10.1145/3236009},
  urldate = {2025-06-25}
}

@misc{haddouchiSurveyTaxonomyMethods2024,
  title = {A survey and taxonomy of methods interpreting random forest models},
  author = {Haddouchi, Maissae and Berrado, Abdelaziz},
  year = 2024,
  month = jul,
  number = {arXiv:2407.12759},
  eprint = {2407.12759},
  primaryclass = {cs},
  publisher = {arXiv},
  doi = {10.48550/arXiv.2407.12759},
  urldate = {2025-07-28},
  archiveprefix = {arXiv}
}

@book{hastieElementsStatisticalLearning2009,
  title = {The elements of statistical learning},
  author = {Hastie, Trevor and Tibshirani, Robert and Friedman, Jerome},
  year = 2009,
  series = {Springer Series in Statistics},
  edition = {2},
  publisher = {Springer},
  address = {New York, NY},
  doi = {10.1007/978-0-387-84858-7},
  urldate = {2026-08-09},
  copyright = {http://www.springer.com/tdm},
  isbn = {978-0-387-84857-0}
}

@article{heinzePhasesMethodologicalResearch2024b,
  title = {Phases of methodological research in biostatistics---building the evidence base for new methods},
  author = {Heinze, Georg and Boulesteix, Anne-Laure and Kammer, Michael and Morris, Tim P. and White, Ian R.},
  year = 2024,
  month = jan,
  journal = {Biom. J.},
  volume = {66},
  number = {1},
  publisher = {John Wiley \& Sons, Ltd},
  issn = {1521-4036},
  doi = {10.1002/bimj.202200222},
  urldate = {2026-09-07},
  eid = {2200222},
  langid = {english}
}

@inproceedings{herbingerLeveragingModelBasedTrees2024,
  title = {Leveraging model-based trees as interpretable surrogate models for model distillation},
  booktitle = {Artificial Intelligence. ECAI 2023 International Workshops},
  author = {Herbinger, Julia and Dandl, Susanne and Ewald, Fiona K. and Loibl, Sofia and Casalicchio, Giuseppe},
  editor = {Nowaczyk, S{\l}awomir and Biecek, Przemys{\l}aw and Chung, Neo Christopher and Vallati, Mauro and Skruch, Pawe{\l} and {Jaworek-Korjakowska}, Joanna and Parkinson, Simon and Nikitas, Alexandros and Atzm{\"u}ller, Martin and Kliegr, Tom{\'a}{\v s} and Schmid, Ute and Bobek, Szymon and Lavrac, Nada and Peeters, Marieke and {van Dierendonck}, Roland and Robben, Saskia and {Mercier-Laurent}, Eunika and Kayakutlu, G{\"u}lg{\"u}n and Owoc, Mieczyslaw Lech and Mason, Karl and Wahid, Abdul and Bruno, Pierangela and Calimeri, Francesco and Cauteruccio, Francesco and Terracina, Giorgio and Wolter, Diedrich and Leidner, Jochen L. and Kohlhase, Michael and Dimitrova, Vania},
  year = 2024,
  series = {Communications in Computer and Information Science},
  volume = {1947},
  pages = {232--249},
  publisher = {Springer Nature Switzerland},
  address = {Cham},
  doi = {10.1007/978-3-031-50396-2_13},
  isbn = {978-3-031-50396-2}
}

@article{hothornPartykitModularToolkit2015,
  title = {{{partykit}}: a modular toolkit for recursive partytioning in R},
  shorttitle = {partykit},
  author = {Hothorn, Torsten and Zeileis, Achim},
  year = 2015,
  journal = {J. Mach. Learn. Res.},
  volume = {16},
  number = {118},
  pages = {3905--3909},
  issn = {1533-7928},
  urldate = {2026-08-07}
}

@article{hothornUnbiasedRecursivePartitioning2006,
  title = {Unbiased recursive partitioning: a conditional inference framework},
  shorttitle = {Unbiased recursive partitioning},
  author = {Hothorn, Torsten and Hornik, Kurt and Zeileis, Achim},
  year = 2006,
  month = sep,
  journal = {J. Comput. Graph. Stat.},
  volume = {15},
  number = {3},
  pages = {651--674},
  publisher = {Taylor \& Francis},
  issn = {1061-8600},
  doi = {10.1198/106186006X133933},
  urldate = {2026-01-18}
}

@misc{huSurrogateLocallyInterpretableModels2020,
  title = {Surrogate locally-interpretable models with supervised machine learning algorithms},
  author = {Hu, Linwei and Chen, Jie and Nair, Vijayan N. and Sudjianto, Agus},
  year = 2020,
  month = jul,
  number = {arXiv:2007.14528},
  eprint = {2007.14528},
  publisher = {arXiv},
  urldate = {2026-10-14},
  archiveprefix = {arXiv}
}

@inproceedings{johanssonAccurateInterpretableRegression2014,
  title = {Accurate and interpretable regression trees using oracle coaching},
  booktitle = {2014 IEEE Symposium on Computational Intelligence and Data Mining (CIDM)},
  author = {Johansson, Ulf and S{\"o}nstr{\"o}d, Cecilia and K{\"o}nig, Rikard},
  year = 2014,
  pages = {194--201},
  publisher = {IEEE}
}

@article{johanssonObtainingAccurateComprehensible2012,
  title = {Obtaining accurate and comprehensible classifiers using oracle coaching},
  author = {Johansson, Ulf and S{\"o}nstr{\"o}d, Cecilia and L{\"o}fstr{\"o}m, Tuve and Bostr{\"o}m, Henrik},
  year = 2012,
  month = mar,
  journal = {Intell. Data Anal.},
  volume = {16},
  number = {2},
  pages = {247--263},
  publisher = {SAGE Publications},
  issn = {1088-467X},
  doi = {10.3233/IDA-2012-0522},
  urldate = {2026-01-02}
}

@misc{kellyUCIMachineLearning,
  title = {The UCI machine learning repository},
  author = {Kelly, Markelle and Longjohn, Rachel and Nottingham, Kolby},
  urldate = {2026-08-09},
  howpublished = {https://archive.ics.uci.edu}
}

@book{kleiberAppliedEconometrics2008,
  title = {Applied econometrics with R},
  author = {Kleiber, Christian and Zeileis, Achim},
  year = 2008,
  publisher = {Springer-Verlag},
  address = {New York},
  doi = {10.1007/978-0-387-77318-6}
}

@article{kursaFeatureSelectionBoruta2010,
  title = {Feature selection with the Boruta package},
  author = {Kursa, Miron and Rudnicki, Witold},
  year = 2010,
  month = sep,
  journal = {J. Stat. Softw.},
  volume = {36},
  pages = {1--13}
}

@inproceedings{laabsConstructionArtificialMost2024,
  title = {Construction of artificial most representative trees by minimizing tree-based distance measures},
  booktitle = {Explainable Artificial Intelligence},
  author = {Laabs, Bj{\"o}rn-Hergen and Kronziel, Lea L. and K{\"o}nig, Inke R. and Szymczak, Silke},
  editor = {Longo, Luca and Lapuschkin, Sebastian and Seifert, Christin},
  year = 2024,
  series = {Communications in Computer and Information Science},
  volume = {2154},
  pages = {290--310},
  publisher = {Springer Nature Switzerland},
  address = {Cham},
  doi = {10.1007/978-3-031-63797-1_15},
  isbn = {978-3-031-63797-1},
  langid = {english}
}

@article{laabsIdentificationRepresentativeTrees2024,
  title = {Identification of representative trees in random forests based on a new tree-based distance measure},
  author = {Laabs, Bj{\"o}rn-Hergen and Westenberger, Ana and K{\"o}nig, Inke R.},
  year = 2024,
  journal = {Adv. Data Anal. Classif.},
  volume = {18},
  pages = {363--380},
  issn = {1862-5355}
}

@article{langBatchtoolsToolsWork2017b,
  title = {{{batchtools}}: tools for R to work on batch systems},
  author = {Lang, Michel and Bischl, Bernd and Surmann, Dirk},
  year = 2017,
  journal = {J. Open Source Softw.},
  volume = {2},
  eid = {135}
}

@misc{leischMlbenchMachineLearning2026,
  title = {{{mlbench}}: machine learning benchmark problems},
  author = {Leisch, Friedrich and Dimitriadou, Evgenia},
  year = 2026,
  doi = {10.32614/CRAN.package.mlbench}
}

@article{liawClassificationRegressionRandomForest2002b,
  title = {Classification and regression by randomForest},
  author = {Liaw, Andy and Wiener, Matthew},
  year = 2002,
  journal = {R News},
  volume = {2},
  number = {3},
  pages = {18--22}
}

@article{lohClassificationRegressionTrees2011,
  title = {Classification and regression trees},
  author = {Loh, Wei-Yin},
  year = 2011,
  month = jan,
  journal = {Wiley Interdiscip. Rev. Data Min. Knowl. Discov.},
  volume = {1},
  number = {1},
  pages = {14--23},
  issn = {1942-4787, 1942-4795},
  doi = {10.1002/widm.8},
  urldate = {2026-03-14},
  copyright = {http://onlinelibrary.wiley.com/termsAndConditions\#vor},
  langid = {english}
}

@article{lohRegressionTressUnbiased2002,
  title = {Regression trees with unbiased variable selection and interaction detection},
  author = {Loh, Wei-Yin},
  year = 2002,
  journal = {Statistica Sinica},
  volume = {12},
  number = {2},
  eprint = {24306967},
  eprinttype = {jstor},
  pages = {361--386},
  publisher = {Institute of Statistical Science, Academia Sinica},
  issn = {1017-0405},
  urldate = {2026-06-05}
}

@article{lohSplitSelectionMethods1999,
  title = {Split Selection Methods for Classification Trees},
  author = {Loh, Wei-Yin and Shih, Yu-shan},
  year = 1999,
  month = jul,
  journal = {Stat Sinica},
  volume = {7}
}

@inproceedings{lundbergUnifiedApproachInterpreting2017,
  title = {A unified approach to interpreting model predictions},
  booktitle = {Proceedings of the 31st International Conference on Neural Information Processing Systems},
  author = {Lundberg, Scott M. and Lee, Su-In},
  year = 2017,
  month = dec,
  series = {NIPS'17},
  pages = {4768--4777},
  publisher = {Curran Associates Inc.},
  address = {Red Hook, NY, USA},
  urldate = {2025-06-25},
  isbn = {978-1-5108-6096-4}
}

@article{meinshausenNodeHarvest2010,
  title = {Node harvest},
  author = {Meinshausen, Nicolai},
  year = 2010,
  month = dec,
  journal = {Ann. Appl. Stat.},
  volume = {4},
  number = {4},
  pages = {2049--2072},
  publisher = {Institute of Mathematical Statistics},
  issn = {1932-6157, 1941-7330},
  doi = {10.1214/10-AOAS367},
  urldate = {2026-01-02}
}

@misc{meinshausenNodeHarvestNodeHarvest2015,
  title = {nodeHarvest: node harvest for regression and classification},
  author = {Meinshausen, Nicolai},
  year = 2015,
  doi = {10.32614/CRAN.package.nodeHarvest}
}

@book{molnarInterpretableMachineLearning2022,
  title = {Interpretable machine learning: a guide for making black box models explainable},
  author = {Molnar, Christoph},
  year = 2022,
  edition = {2}
}

@misc{petersIpredImprovedPredictors2024,
  title = {{{ipred}}: improved predictors},
  author = {Peters, Andrea and Hothorn, Torsten},
  year = 2024,
  doi = {10.32614/CRAN.package.ipred}
}

@misc{renardConceptTreeHighLevel2019,
  title = {Concept tree: high-level representation of variables for more interpretable surrogate decision trees},
  shorttitle = {Concept tree},
  author = {Renard, Xavier and Woloszko, Nicolas and Aigrain, Jonathan and Detyniecki, Marcin},
  year = 2019,
  month = jun,
  number = {arXiv:1906.01297},
  eprint = {1906.01297},
  primaryclass = {cs, stat},
  publisher = {arXiv},
  urldate = {2024-02-07},
  archiveprefix = {arXiv},
  langid = {english}
}

@article{tibshiraniRegressionShrinkageSelection1996,
  title = {Regression shrinkage and selection via the lasso},
  author = {Tibshirani, Robert},
  year = 1996,
  month = jan,
  journal = {J. R. Stat. Soc. Ser. B Methodol.},
  volume = {58},
  number = {1},
  pages = {267--288},
  issn = {0035-9246},
  doi = {10.1111/j.2517-6161.1996.tb02080.x},
  urldate = {2026-09-09}
}

@article{wrightRangerFastImplementation2017,
  title = {{{ranger}}: a fast implementation of random forests for high dimensional data in C++ and R},
  shorttitle = {ranger},
  author = {Wright, Marvin N. and Ziegler, Andreas},
  year = 2017,
  journal = {J. Stat. Softw.},
  volume = {77},
  pages = {1--17},
  issn = {1548-7660},
  urldate = {2021-05-15},
  copyright = {Copyright (c) 2017 Marvin N. Wright, Andreas Ziegler},
  langid = {english}
}

@article{zeileisModelBasedRecursivePartitioning2008,
  title = {Model-based recursive partitioning},
  author = {Zeileis, Achim and Hothorn, Torsten and Hornik, Kurt},
  year = 2008,
  month = jun,
  journal = {J. Comput. Graph. Stat.},
  volume = {17},
  number = {2},
  pages = {492--514},
  publisher = {Taylor \& Francis},
  issn = {1061-8600},
  doi = {10.1198/106186008X319331},
  urldate = {2026-03-25}
}

\FloatBarrier
\newpage
\section*{Appendix}

\subsection*{Overview of supplementary material}

\begingroup
\setlength{\parindent}{0pt}
\setlength{\parskip}{0.3em}

\newcommand{\suppentry}[3]{%
  \par\noindent
  \hangindent=1.5em
  \hangafter=1
  #1~\ref{#2}: #3\nobreak\dotfill
  \mbox{\pageref{#2}}\par
}

\subsubsection*{Supplementary tables}

\suppentry{Table}{tab:method_comparison}
  {Comparison of interpretation methods}

\suppentry{Table}{tab:method_description}
  {Methods and implementations}

\suppentry{Table}{tab:dataset_description}
  {Overview of data sets}

\suppentry{Table}{tab:method_failures}
  {Method failures and recorded causes}

\subsubsection*{Supplementary figures: Simulation study}

\suppentry{Figure}{app:fig:sim_dev_rf_pred}
  {Deviation from RF predictions}

\suppentry{Figure}{app:fig:sim_dist_sv}
  {Structural similarity to the RF}

\suppentry{Figure}{app:fig:sim_number_decisions}
  {Model complexity}

\suppentry{Figure}{app:fig:sim_covered_effect_vars}
  {Effect variables used}

\suppentry{Figure}{app:fig:sim_covered_noise_vars}
  {Noise variables used}

\suppentry{Figure}{app:fig:sim_r_squared}
  {Prediction performance: $R^2$}

\suppentry{Figure}{app:fig:sim_aic}
  {Prediction performance: AIC}

\suppentry{Figure}{app:fig:sim_macro_f1}
  {Prediction performance: Macro F1 score}

\suppentry{Figure}{app:fig:sim_runtime}
  {Runtime}

\subsubsection*{Supplementary figures: Benchmark experiment}

\suppentry{Figure}{app:fig:bench_dev_rf_pred_rmse}
  {Deviation from RF predictions: regression}

\suppentry{Figure}{app:fig:bench_dev_rf_pred_mcr}
  {Deviation from RF predictions: classification}

\suppentry{Figure}{app:fig:bench_dist_sv}
  {Structural similarity to the RF}

\suppentry{Figure}{app:fig:bench_number_decisions}
  {Model complexity}

\suppentry{Figure}{app:fig:bench_r_squared}
  {Prediction performance: $R^2$}

\suppentry{Figure}{app:fig:bench_aic}
  {Prediction performance: AIC}

\suppentry{Figure}{app:fig:bench_macro_f1}
  {Prediction performance: Macro F1 score}

\suppentry{Figure}{app:fig:bench_runtime}
  {Runtime}

\endgroup
\clearpage

\renewcommand{\thefigure}{A.\arabic{figure}}
\setcounter{figure}{0}

\renewcommand{\thetable}{A.\arabic{table}}
\setcounter{table}{0}

\begin{landscape}
\begingroup
\footnotesize
\setlength{\tabcolsep}{3pt}
\renewcommand{\arraystretch}{1.18}

\begin{longtable}{
    @{}
    >{\raggedright\arraybackslash}p{1.0cm}
    >{\raggedright\arraybackslash}p{2.5cm}
    >{\raggedright\arraybackslash}p{3.0cm}
    >{\raggedright\arraybackslash}p{4.1cm}
    >{\raggedright\arraybackslash}p{3.3cm}
    >{\raggedright\arraybackslash}p{3.4cm}
    >{\raggedright\arraybackslash}p{4.0cm}
    @{}
}
\caption{Comparison of the interpretation methods in the
configurations used in this study. For tree methods, construction
criteria refer to variable and split-point selection. For rule-based
methods, they refer to rule generation and selection.}
\label{tab:method_comparison}\\

\toprule
Method &
Information used &
Final representation &
Construction and selection criteria &
Stopping and complexity control &
Prediction components &
Final prediction \\
\midrule
\endfirsthead

\caption[]{Comparison of the interpretation methods
(continued).}\\
\toprule
Method &
Information used &
Final representation &
Construction and selection criteria &
Stopping and complexity control &
Prediction components &
Final prediction \\
\midrule
\endhead

\midrule
\multicolumn{7}{r}{Continued on next page}\\
\endfoot

\bottomrule
\endlastfoot

\multicolumn{7}{@{}l}{
    \textbf{Methods using the oracle coaching approach}
}\\
\addlinespace[3pt]

LM &
Variables and RF predictions &
Global linear regression model &
Least-squares fitting.
Forward selection can reduce the variable set &
Variable selection stops when the selection criterion
no longer improves &
Intercept and linear variable terms &
Intercept plus the sum of variable values multiplied
by their coefficients \\
\addlinespace

OC-CART &
Variables and RF predictions &
Single tree with constant terminal-node predictions &
Variable and split point selected jointly by maximizing
impurity reduction &
Node-size and other growth constraints.
No post-pruning in this study &
Terminal-node mean or majority class &
Prediction of the terminal node reached by the observation \\
\addlinespace

GUIDE &
Variables and RF predictions &
Single tree with local linear main effects models &
Variable selected using residual-sign association and
interaction tests.
Split point minimizes the combined residual sum of squares
of the child-node models &
Minimum improvement criterion, node-size constraints,
and a limit on tree-growth levels.
Variable-selection tests do not require significance
for continued growth &
Linear model within each terminal node &
Prediction from the local model in the terminal node
reached by the observation \\
\addlinespace

SLIM &
Variables and RF predictions &
Single tree with local main effects models &
Variable and split point selected jointly by minimizing
the combined residual sum of squares of the child-node models &
Node-size constraints and a limit on tree-growth levels
in this study.
A minimum improvement criterion also controls splitting &
Main effects model within each terminal node &
Prediction from the local model in the terminal node
reached by the observation \\
\addlinespace

MOB &
Variables and RF predictions &
Single tree with intercept-only node models &
Variable selected by parameter instability tests.
Split point minimizes the combined residual sum of squares
around the child-node means &
No significant instability after multiplicity correction,
or growth constraints &
Mean RF prediction within each terminal node &
Prediction of the terminal node reached by the observation \\
\addlinespace

CTree &
Variables and RF predictions &
Single tree with constant terminal-node predictions &
Variable selected by conditional independence tests.
Split point maximizes a standardized separation statistic &
No significant association after multiplicity correction,
or growth constraints &
Terminal-node mean or majority class &
Prediction of the terminal node reached by the observation \\
\addlinespace[6pt]

\multicolumn{7}{@{}l}{
    \textbf{Representative trees}
}\\
\addlinespace[3pt]

MRT &
Trees of the fitted RF &
One existing RF tree &
Selects the tree with the smallest average distance
to the other RF trees &
Inherits the selected tree's size.
No additional tree growth &
Predictions retained from the selected RF tree &
Prediction of the terminal node reached in the selected tree \\
\addlinespace

ART &
Fitted RF, variables, and observed outcomes &
New representative tree with constant terminal-node
predictions &
Selects the candidate split yielding the smallest
distance between the resulting tree and the RF &
Stops when distance increases or remains unchanged
without sufficient prediction improvement.
Node-size constraints also apply &
Mean observed outcome or majority observed class
within each terminal node &
Prediction of the terminal node reached by the observation \\
\addlinespace[6pt]

\multicolumn{7}{@{}l}{
    \textbf{Rules extracted from RF trees}
}\\
\addlinespace[3pt]

inTrees &
Trees of the fitted RF, variables, and observed outcomes &
Ordered rule list &
Extracts path rules, removes redundant conditions,
and selects rules using prediction error, coverage,
and length &
Rule simplification and selection.
An additional decision budget is applied in this study &
Prediction assigned to each rule using observed outcomes &
Prediction of the first matching rule.
Default prediction if no selected rule matches \\
\addlinespace[6pt]

\multicolumn{7}{@{}l}{
    \textbf{Rules learned from RF predictions}
}\\
\addlinespace[3pt]

SIRUS &
Variables and RF predictions &
Weighted collection of rules with an intercept &
Builds its own modified RF with quantile-based numerical
split points.
Selects frequent paths and removes linearly dependent rules &
Rule-frequency threshold and forest-stability criterion &
Each rule returns the mean RF prediction inside or outside
its conditions.
Non-negative weights fitted by ridge regression &
Intercept plus the weighted sum of all selected rule outputs \\
\addlinespace

NH &
Variables and RF predictions &
Weighted collection of overlapping rules &
Generates candidate rules from its own forest.
Joint quadratic optimization assigns non-negative weights.
Only rules with positive weights contribute &
Candidate-set size, rule length, and node-size constraints.
Weight optimization determines the contributing rules &
Mean RF prediction among training observations
satisfying each rule &
Weighted mean of all matching rule predictions,
normalized by the sum of their weights \\
\addlinespace

RuleFit &
Variables and RF predictions &
Regularized regression model containing rule indicators
and linear variable terms &
Generates path rules from new conditional inference trees.
LASSO jointly estimates coefficients and selects terms &
Tree number and depth, regularization strength,
and the additional decision budget used in this study &
Binary rule indicators, retained linear terms,
and an intercept &
Intercept plus weighted rule indicators and linear terms.
For classification, linear predictors are transformed
into class probabilities \\

\end{longtable}
\endgroup
\end{landscape}

\begin{landscape}
      \begin{table}[ht]
          \centering
          \small
          \begin{tabular}{
                  p{5.1cm}
                  p{1.75cm}
                  c
                  l
                  p{4cm}
                  p{2cm}
                  p{1.8cm}
                  p{1.8cm}}
              \toprule
              \textbf{Name} & \textbf{Abbreviation} & \textbf{Type} & \textbf{Publication} & \textbf{Implementation} & \textbf{Used
              version} & \textbf{Regression} & \textbf{Classification} \\
              \midrule
              Artificial Representative Tree & ART & S & \citep{laabsConstructionArtificialMost2024} & R package \emph{timbR} on GitHub
              \url{https://github.com/imbs-hl/timbR} & V3.3.2 & yes & yes \\ \addlinespace[1pt]
              Conditional Inference Tree & CTree & S & \citep{hothornUnbiasedRecursivePartitioning2006} & R package \emph{partykit} on CRAN & V1.2-26 &
              yes & yes \\ \addlinespace[1pt]
              Classification and Regression Tree & decision tree (reference model) & - & \citep{breimanClassificationRegressionTrees1984} & R package \emph{ranger} on CRAN &
              V0.17.0 & yes & yes \\ \addlinespace[1pt]
              Generalized, Unbiased Interaction Detection and Estimation & GUIDE & S & \citep{lohRegressionTressUnbiased2002} & R implementation
              on GitHub \url{https://github.com/slds-lmu/mbt_comparison} & Downloaded 18.01.2026 & yes & no \\ \addlinespace[1pt]
              Interpretable Trees & inTrees & R & \citep{dengInterpretingTreeEnsembles2019} & R package \emph{inTrees} in CRAN archive & V1.4 &
              yes & yes \\ \addlinespace[1pt]
              Model-Based Recursive Partitioning & MOB & S & \citep{zeileisModelBasedRecursivePartitioning2008} & R package \emph{partykit} on CRAN &
              V1.2-26 & yes & no \\ \addlinespace[1pt]
              Most Representative Tree & MRT & S & \citep{banerjeeIdentifyingRepresentativeTrees2012a} & R package \emph{timbR} on GitHub \url{https://
              github.com/imbs-hl/timbR} & V3.3.2 & yes & yes \\ \addlinespace[1pt]
              Node Harvest & NH & R & \citep{meinshausenNodeHarvest2010} & R package \emph{nodeHarvest} in CRAN archive & V0.7-3 & yes & no
              \\ \addlinespace[1pt]
              Oracle Coaching using CART & OC-CART & S & \citep{johanssonObtainingAccurateComprehensible2012} & R package \emph{ranger} on CRAN &
              V0.17.0 & yes & yes \\ \addlinespace[1pt]
              RuleFit & RuleFit & R & \citep{friedmanPredictiveLearningRule2008} & R package \emph{pre} on CRAN & V1.0.8 & yes & yes \\
              \addlinespace[1pt]
              Stable and Interpretable Rule Set & SIRUS & R & \citep{benardSIRUSStableInterpretable2021} & R package \emph{sirus} in CRAN archive &
              V0.3.3 & yes & no \\ \addlinespace[1pt]
              Surrogate Locally-Interpretable Models & SLIM & S & \citep{huSurrogateLocallyInterpretableModels2020} & R implementation on GitHub \url{https://
              github.com/slds-lmu/mbt_comparison} & Downloaded 18.01.2026 & yes & no \\ \addlinespace[1pt]
              Oracle Coaching using Linear Model & LM & S & \citep{molnarInterpretableMachineLearning2022} & R package \emph{stats} on CRAN & V4.3.0
              & yes & no \\
              \bottomrule
          \end{tabular}
          \caption{Overview of methods used in this study.
          S: surrogate model; R: rule extraction method.
          The decision tree serves as a prediction model fitted to the observed outcome
          and is therefore assigned neither category.}
          \label{tab:method_description}
      \end{table}
  \end{landscape}

\begin{landscape}
    \centering
    \scriptsize
    \setlength{\tabcolsep}{2pt}
    \renewcommand{\arraystretch}{1.08}
    \begin{longtable}{l>{\centering\arraybackslash}p{1.5cm}>{\centering\arraybackslash}p{1.5cm}>{\centering\arraybackslash}p{1.5cm}>{\centering\arraybackslash}p{1.5cm}llp{6.5cm}p{3cm}}
        \caption{Overview of the data sets used in the method comparison. The name is displayed in combination with the dimension of the data set after downloading the benchmark data (bench) or simulating it (sim). The columns "after QC" give the number after the preprocessing steps described in the text are applied as quality control (QC). We investigated two outcomes: classification (classif) with the number of classes in brackets and regression (regr). The last column describes which method was introduced on this data set. Details about the methods and the corresponding publications are listed in \Cref{tab:method_description}.}
        \label{tab:dataset_description}\\
        \toprule
        \textbf{Data set} & \shortstack{\textbf{N}\\\textbf{(original)}} & \shortstack{\textbf{N}\\\textbf{(after QC)}}& \shortstack{\textbf{P}\\\textbf{(original)}} &  \shortstack{\textbf{P}\\\textbf{(after QC)}} & \textbf{outcome} & \textbf{type} & \textbf{Source} & \textbf{Used in publication of} \\
        \midrule
        BreastCancer & 699 & 683 & 11 & 10 & classif (2) & bench & UCI: \url{https://doi.org/10.24432/C5HP4Z} & CTree \\
        \addlinespace[1pt]
        Digit recognition & 5200 & 5200 & 7 & 7 & classif (10) & sim & Simulation from \citep{breimanClassificationRegressionTrees1984} & decision tree \\
                \addlinespace[1pt]
        Digit recognition noisy& 5200 & 5200 & 24 & 24 & classif (10) & sim & Simulation from \citep{breimanClassificationRegressionTrees1984} & CART algorithm \\
        \addlinespace[1pt]
        GlaucomaMVF & 170 & 153 & 67 & 67 & classif (2) & bench & R package \emph{ipred} V0.9.15 & CTree \\
        \addlinespace[1pt]
        XOR-problem & 200 & 200 & 20 & 20 & classif (2) & sim & Simulation from \citep{dengInterpretingTreeEnsembles2019} & inTrees \\
        \addlinespace[1pt]
        Soybean & 683 & 562 & 36 & 36 & classif (19) & bench & R package \emph{mlbench} V2.1.7 & CTree \\
        \addlinespace[1pt]
        automobile & 205 & 159 & 26 & 25 & classif (6) & bench & UCI: \url{https://doi.org/10.24432/C5B01C} & inTrees \\
        \addlinespace[1pt]
        balance\_scale & 625 & 625 & 5 & 5 & classif (3) & bench & UCI: \url{https://doi.org/10.24432/C5488X} & OC-CART \\
        \addlinespace[1pt]
        breast\_cancer & 286 & 277 & 10 & 10 & classif (2) & bench & R package \emph{mlbench} V2.1.7 & CTree \\
        \addlinespace[1pt]
        breast\_cancer\_wisconsin\_diagnostic & 569 & 569 & 32 & 31 & classif (2) & bench & UCI: \url{https://doi.org/10.24432/C5DW2B} & OC-CART, inTrees, nodeHarvest \\
        \addlinespace[1pt]
        congressional\_voting\_records & 435 & 232 & 17 & 17 & classif (2) & bench & UCI: \url{https://doi.org/10.24432/C5C01P} & OC-CART \\
        \addlinespace[1pt]
        connectionist\_bench\_sonar\_mines\_vs\_rocks & 208 & 208 & 61 & 61 & classif (2) & bench & UCI: \url{https://doi.org/10.24432/C5T01Q} & OC-CART, CTree \\
        \addlinespace[1pt]
        contraceptive\_method\_choice & 1473 & 1473 & 10 & 10 & classif (3) & bench & UCI: \url{https://doi.org/10.24432/C59W2D} & OC-CART \\
        \addlinespace[1pt]
        credit\_approval & 690 & 653 & 16 & 16 & classif (2) & bench & UCI: \url{https://doi.org/10.24432/C5FS30} & inTrees \\
        \addlinespace[1pt]
        dermatology & 366 & 358 & 35 & 35 & classif (6) & bench & UCI: \url{https://doi.org/10.24432/C5FK5P} & OC-CART \\
        \addlinespace[1pt]
        diabetes & 768 & 768 & 9 & 9 & classif (2) & bench & R package \emph{mlbench} V2.1.7 & CTree, OC-CART, inTrees, nodeHarvest, SIRUS, MOB \\
        \addlinespace[1pt]
        ecoli & 336 & 336 & 9 & 8 & classif (8) & bench & UCI: \url{https://doi.org/10.24432/C5388M} & OC-CART \\
        \addlinespace[1pt]
        glass\_identification & 214 & 214 & 11 & 10 & classif (6) & bench & UCI: \url{https://doi.org/10.24432/C5WW2P} & OC-CART, inTrees, CTree \\
        \addlinespace[1pt]
        haberman\_survival & 306 & 306 & 4 & 4 & classif (2) & bench & UCI: \url{https://doi.org/10.24432/C5XK51} & OC-CART, SIRUS \\
        \addlinespace[1pt]
        heart\_disease\_c & 303 & 297 & 14 & 14 & classif (5) & bench & UCI: \url{https://doi.org/10.24432/C52P4X} & OC-CART, SIRUS \\
        \addlinespace[1pt]
        heart\_disease\_h & 294 & 261 & 14 & 11 & classif (5)  & bench & UCI: \url{https://doi.org/10.24432/C52P4X} & OC-CART, SIRUS \\
        \addlinespace[1pt]
        heart\_disease\_s & 123 & 55 & 14 & 11 & classif (5)  & bench & UCI: \url{https://doi.org/10.24432/C52P4X} & OC-CART \\
        \addlinespace[1pt]
        horse\_colic & 368 & 61 & 28 & 24 & classif (2)  & bench & UCI: \url{https://doi.org/10.24432/C58W23} & OC-CART, inTrees, SIRUS \\
        \addlinespace[1pt]
        image\_segmentation & 210 & 210 & 20 & 19 & classif (7)  & bench & UCI: \url{https://doi.org/10.24432/C5GP4N} & OC-CART \\
        \addlinespace[1pt]
        ionosphere & 351 & 351 & 35 & 34 & classif (2)  & bench & UCI: \url{https://doi.org/10.24432/C5W01B} & OC-CART, CTree, SIRUS \\
        \addlinespace[1pt]
        iris & 150 & 150 & 5 & 5 & classif (3)  & bench & UCI: \url{https://doi.org/10.24432/C56C76} & OC-CART, inTrees \\
        \addlinespace[1pt]
        liver\_disorders & 345 & 345 & 7 & 7 & classif (2) & bench & UCI: \url{https://doi.org/10.24432/C54G67} & OC-CART, SIRUS \\
        \addlinespace[1pt]
        lymphography & 148 & 148 & 19 & 19 & classif (4)  & bench & UCI: \url{https://doi.org/10.24432/C54598} & inTrees \\
        \addlinespace[1pt]
        parkinsons & 195 & 195 & 24 & 23 & classif (2)  & bench & UCI: \url{https://doi.org/10.24432/C59C74} & nodeHarvest \\
        \addlinespace[1pt]
        secom & 1567 & 461 & 591 & 441 & classif (2)  & bench & UCI: \url{https://doi.org/10.24432/C54305} & SIRUS \\
        \addlinespace[1pt]
        spambase & 4601 & 4601 & 58 & 58 & classif (2)  & bench & UCI: \url{https://doi.org/10.24432/C53G6X} & OC-CART \\
        \addlinespace[1pt]
        statlog\_australian\_credit\_approval & 690 & 690 & 15 & 15 & classif (2)  & bench & UCI: \url{https://doi.org/10.24432/C59012} & OC-CART, SIRUS, inTrees \\
        \addlinespace[1pt]
        statlog\_german\_credit\_data & 1000 & 1000 & 21 & 21 & classif (2)  & bench & UCI: \url{https://doi.org/10.24432/C5NC77} & OC-CART, SIRUS \\
        \addlinespace[1pt]
        statlog\_heart & 270 & 270 & 14 & 14 & classif (2)  & bench & UCI: \url{https://doi.org/10.24432/C57303} & inTrees, SIRUS \\
        \addlinespace[1pt]
        statlog\_vehicle\_silhouettes & 94 & 94 & 19 & 19 & classif (4)  & bench & UCI: \url{https://doi.org/10.24432/C5HG6N} & OC-CART, inTrees, CTree \\
        \addlinespace[1pt]
        teaching\_assistant\_evaluation & 151 & 151 & 6 & 6 & classif (3) & bench & UCI: \url{https://doi.org/10.24432/C55P6M} & OC-CART \\
        \addlinespace[1pt]
        thyroid\_disease & 3163 & 2000 & 26 & 20 & classif (3) & bench & UCI: \url{https://doi.org/10.24432/C5D010} & OC-CART \\
        \addlinespace[1pt]
        tic\_tac\_toe\_endgame & 958 & 958 & 10 & 10 & classif (2) & bench & UCI: \url{https://doi.org/10.24432/C5688J} & OC-CART, inTrees \\
        \addlinespace[1pt]
        waveform\_database\_generator\_version\_1 & 5000 & 5000 & 22 & 22 & classif (3) & bench & UCI: \url{https://doi.org/10.24432/C5688J} & OC-CART, inTrees, CART algorithm \\
        \addlinespace[1pt]
        waveform\_database\_generator\_version\_2 & 5000 & 5000 & 41 & 41 & classif (3) & bench & OpenML: ID 60 & CART algorithm \\
        \addlinespace[1pt]
        wine & 178 & 178 & 14 & 14 & classif (3) & bench & UCI: \url{https://doi.org/10.24432/C5PC7J} & OC-CART, inTrees, ART \\
        \addlinespace[1pt]
        zoo & 101 & 101 & 18 & 17 & classif (7) & bench & UCI: \url{https://doi.org/10.24432/C5R59V} & inTrees \\
        \addlinespace[1pt]
        Allstate\_Claims\_Severity & 188318 & 187027 & 132 & 127 & regr & bench & OpenML: ID 42571 & ART \\
        \addlinespace[1pt]
        Brazilian\_houses & 10692 & 10671 & 13 & 13 & regr & bench & OpenML: ID 42688 & ART \\
        \addlinespace[1pt]
        Journals & 180 & 180 & 9 & 9 & regr & bench & R package \emph{AER} V1.2.16 & MOB \\
        \addlinespace[1pt]        
        Buzzinsocialmedia\_Twitter & 583250 & 541125 & 78 & 78 & regr & bench & OpenML: ID 4549 & ART \\
        \addlinespace[1pt]
        Mercedes\_Benz\_Greener\_Manufacturing & 4209 & 4208 & 378 & 361 & regr & bench & OpenML: ID 42570 & ART \\
        \addlinespace[1pt]
        Moneyball & 1232 & 1232 & 15 & 10 & regr & bench & OpenML: ID 41021 & ART \\
        \addlinespace[1pt]
        Ozone & 366 & 203 & 13 & 11 & regr & bench & R package \emph{mlbench} V2.1.7 & CTree, CART algorithm, nodeHarvest \\
        \addlinespace[1pt]
        Servo & 167 & 167 & 5 & 5 & regr & bench & R package \emph{mlbench} V2.1.7 & CTree \\
        \addlinespace[1pt]
        Simulation Few large effects & 2000 & 2000 & 100 & 100 & regr & sim & Simulation from \citep{laabsConstructionArtificialMost2024} & ART \\
        \addlinespace[1pt]
        Simulation Many small effects & 2000 & 2000 & 100 & 100 & regr & sim & Simulation from \citep{laabsConstructionArtificialMost2024}  & ART \\
        \addlinespace[1pt]
        Simulation Correlated variables & 2000 & 2000 & 100 & 100 & regr & sim & Simulation from \citep{laabsConstructionArtificialMost2024}  & ART \\
        \addlinespace[1pt]
        Simulation Interaction effects & 2000 & 2000 & 100 & 100 & regr & sim & Simulation from \citep{laabsConstructionArtificialMost2024}  & ART \\
        \addlinespace[1pt]
        Simulation Continuous variables & 2000 & 2000 & 100 & 100 & regr & sim & Simulation from \citep{laabsConstructionArtificialMost2024}  & ART \\
        \addlinespace[1pt]
        Simulation Subgroups & 200 & 200 & 10 & 10 & regr & sim & Simulation from \citep{breimanClassificationRegressionTrees1984} & CART algorithm \\
        \addlinespace[1pt]
        Simulation Nonlinear effects & 2000 & 2000 & 2 & 2 & regr & sim & Simulation from \citep{meinshausenNodeHarvest2010}  & nodeHarvest \\
        \addlinespace[1pt]
        Simulation Exponential effects & 55000 & 55000 & 100 & 100 & regr & sim & Simulation from \citep{friedmanPredictiveLearningRule2008}  & RuleFit \\
        \addlinespace[1pt]
        abalone & 4177 & 4170 & 9 & 9 & regr & bench & R package \emph{mlbench} V2.1.7 & nodeHarvest, ART \\
        \addlinespace[1pt]
        bike\_sharing\_dataset & 17379 & 17379 & 17 & 15 & regr & bench & UCI: \url{https://doi.org/10.24432/C5W894} & SLIM \\
        \addlinespace[1pt]
        black\_friday & 166821 & 166821 & 10 & 10 & regr & bench & OpenML: ID 41540 & ART \\
        \addlinespace[1pt]
        bone & 485 & 485 & 4 & 3 & regr & bench & \url{https://hastie.su.domains/ElemStatLearn/datasets} \citep{bachrachBoneMineralAcquisition1999} & nodeHarvest \\
        \addlinespace[1pt]
        boston\_housing & 506 & 506 & 14 & 14 & regr & bench & R package \emph{mlbench} V2.1.7 & CTree, CART algorithm, nodeHarvest, RuleFit, MOB, ART \\
        \addlinespace[1pt]
        crime & 1994 & 1993 & 128 & 103 & regr & bench & UCI: \url{https://doi.org/10.24432/C53W3X} & nodeHarvest, ART \\
        \addlinespace[1pt]
        diamonds & 53940 & 53940 & 10 & 10 & regr & bench & OpenML: ID 42225 & ART \\
        \addlinespace[1pt]
        elevators & 16599 & 16361 & 19 & 19 & regr & bench & OpenML: ID 216 & ART \\
        \addlinespace[1pt]
        galaxy & 323 & 323 & 6 & 5 & regr & bench & \url{https://hastie.su.domains/ElemStatLearn/datasets} & nodeHarvest \\
        \addlinespace[1pt]
        house\_16H & 22784 & 22307 & 17 & 17 & regr & bench & OpenML: ID 574 & ART \\
        \addlinespace[1pt]
        house\_sales & 21613 & 21422 & 24 & 22 & regr & bench & OpenML: ID 42731 & ART \\
        \addlinespace[1pt]
        lungcap & 654 & 654 & 5 & 5 & regr & bench & R package \emph{GLMsData} V1.4 & GUIDE \\
        \addlinespace[1pt]
        machine & 209 & 202 & 10 & 8 & regr & bench & UCI: \url{https://doi.org/10.24432/C5830D} & nodeHarvest \\
        \addlinespace[1pt]
        mpg & 398 & 392 & 9 & 8 & regr & bench & UCI: \url{https://doi.org/10.24432/C5859H} & nodeHarvest \\
        \addlinespace[1pt]
        nyc-taxi-green-dec-2016 & 581835 & 577601 & 19 & 16 & regr & bench & OpenML: ID 42729 & ART \\
        \addlinespace[1pt]
        pol & 15000 & 15000 & 49 & 27 & regr & bench & OpenML: ID 201 & ART \\
        \addlinespace[1pt]
        prostate & 97 & 97 & 10 & 9 & regr & bench & \url{https://hastie.su.domains/ElemStatLearn/datasets} & nodeHarvest \\
        \addlinespace[1pt]
        socmob & 1156 & 1087 & 6 & 4 & regr & bench & OpenML: ID 541 & ART \\
        \addlinespace[1pt]
        space\_ga & 3107 & 3106 & 7 & 7 & regr & bench & OpenML: ID 507 & ART \\
        \addlinespace[1pt]
        tecator & 240 & 240 & 126 & 125 & regr & bench & OpenML: ID 505 & ART \\
        \bottomrule
    \end{longtable}
\end{landscape}

\begin{landscape}
    \centering
    \small
    \setlength{\tabcolsep}{4pt}
    \renewcommand{\arraystretch}{1.1}
    \begin{longtable}{lllp{18cm}}
        \caption{Failed simulation repetitions and benchmark cross-validation folds by method, outcome, source, data set, and recorded cause.}\label{tab:method_failures}\\
        \toprule
        \textbf{Method} & \textbf{Outcome} & \textbf{Source} & \textbf{Details} \\
        \midrule
        \endfirsthead
        \multicolumn{4}{l}{\tablename~\thetable{} continued}\\
        \toprule
        \textbf{Method} & \textbf{Outcome} & \textbf{Source} & \textbf{Details} \\
        \midrule
        \endhead
        \midrule
        \multicolumn{4}{r}{Continued on next page}\\
        \endfoot
        \bottomrule
        \endlastfoot
        ART & regr & bench & runtime above 24 h for data sets Allstate (10 times), Buzzinsocialmedia\_Twitter (10 times), nyc-taxi (7 times) \\ \addlinespace[1pt]
        ART & classif & bench & no variables selected for data set heart\_disease\_s (4 times) \\ \addlinespace[1pt]
        NH & regr & bench & constraints are inconsistent, no solution for data set bike\_sharing (8 times); negative length vectors are not allowed for data sets Allstate (10 times), Brazilian\_houses (10 times), Journals (10 times), Mercedes\_Benz (10 times), Moneyball (10 times), Ozone (10 times), Servo (10 times), abalone (10 times), black\_friday (10 times), nyc-taxi (10 times), socmob (10 times); quadratic function is not positive definite for data sets Buzzinsocialmedia\_Twitter (10 times), diamonds (10 times), house\_16H (10 times), house\_sales (10 times) \\ \addlinespace[1pt]
        SLIM & regr & bench & no applicable method for `predict' applied to an object of class for data sets Journals (4 times), Ozone (2 times), machine (3 times), prostate (7 times) \\ \addlinespace[1pt]
        ruleFit & regr & bench & runtime above 24 h for data sets Buzzinsocialmedia\_Twitter (10 times), nyc-taxi (10 times) \\ \addlinespace[1pt]
        ruleFit & classif & sim & multinomial or binomial class has 1 or 0 observations for data set Digits recognition (5 times) \\ \addlinespace[1pt]
        ruleFit & classif & bench & factor has new levels for data sets automobile (1 time), breast\_cancer (8 times), credit\_approval (1 time); multinomial or binomial class has 1 or 0 observations for data sets heart\_disease\_c (10 times), heart\_disease\_s (2 times), tic\_tac\_toe\_endgame (10 times) \\
    \end{longtable}
\end{landscape}


\FloatBarrier
\subsection*{Supplementary figures: Simulation study}
\begin{figure}
    \centering
    \includegraphics[width=\linewidth]{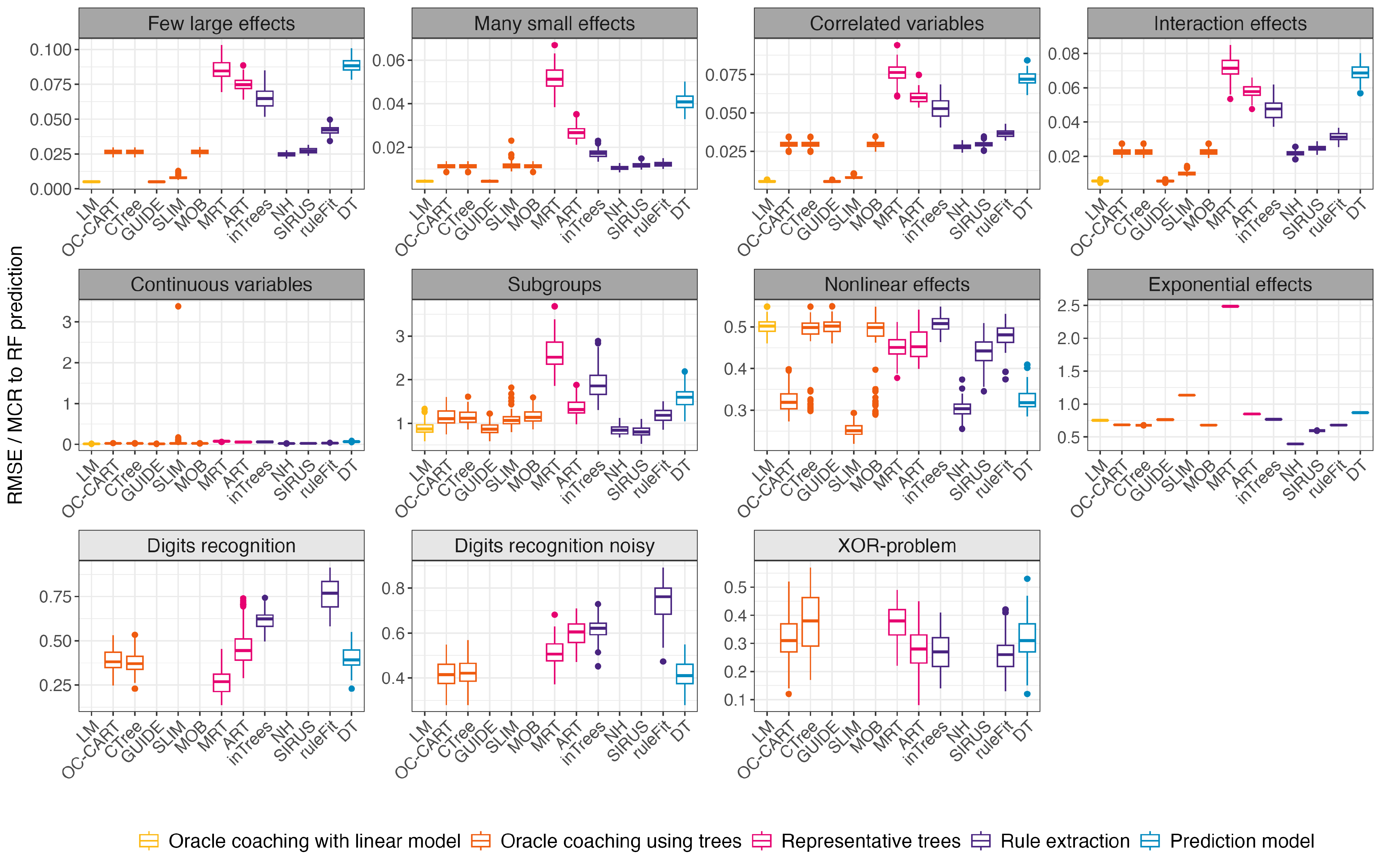}
    \caption{Root mean squared error (RMSE) for regression and misclassification rate (MCR) for classification between the predictions of each method and the corresponding RF across all simulated test data sets. Methods are displayed on the x-axis and colored by method type: rule extraction method, surrogate model, or prediction model (decision tree; DT). Facet headers indicate the outcome type, with light gray denoting classification and dark gray denoting regression.}
    \label{app:fig:sim_dev_rf_pred}
\end{figure}

\begin{figure}
    \centering
    \includegraphics[width=\linewidth]{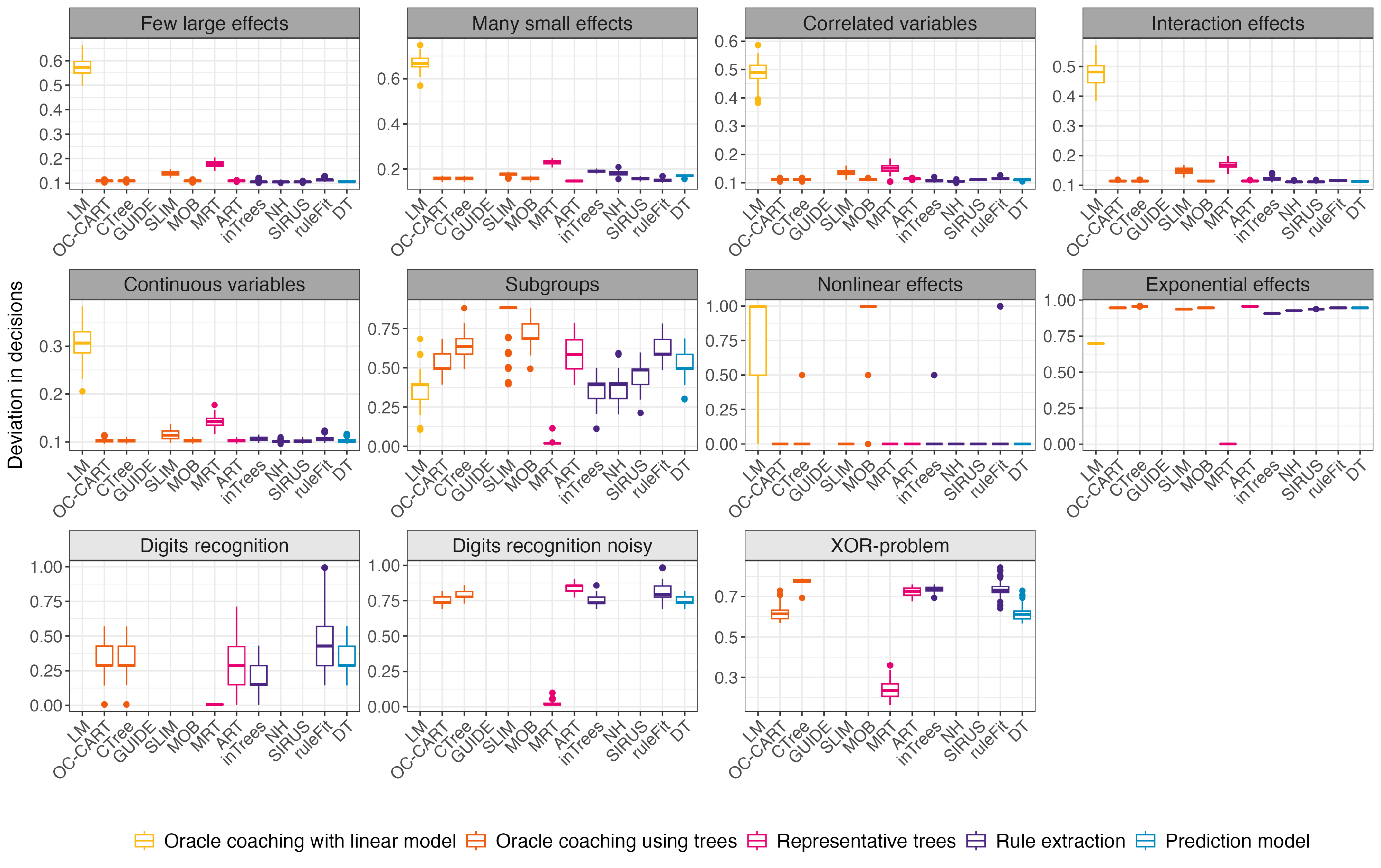}
    \caption{Average proportion of variables whose use in model decisions differed between each method and the individual trees of the corresponding RF across all simulated data sets. Methods are displayed on the x-axis and colored by method type: rule extraction method, surrogate model, or prediction model (decision tree; DT). Facet headers indicate the outcome type, with light gray denoting classification and dark gray denoting regression.}
    \label{app:fig:sim_dist_sv}
\end{figure}

\begin{figure}
    \centering
    \includegraphics[width=\linewidth]{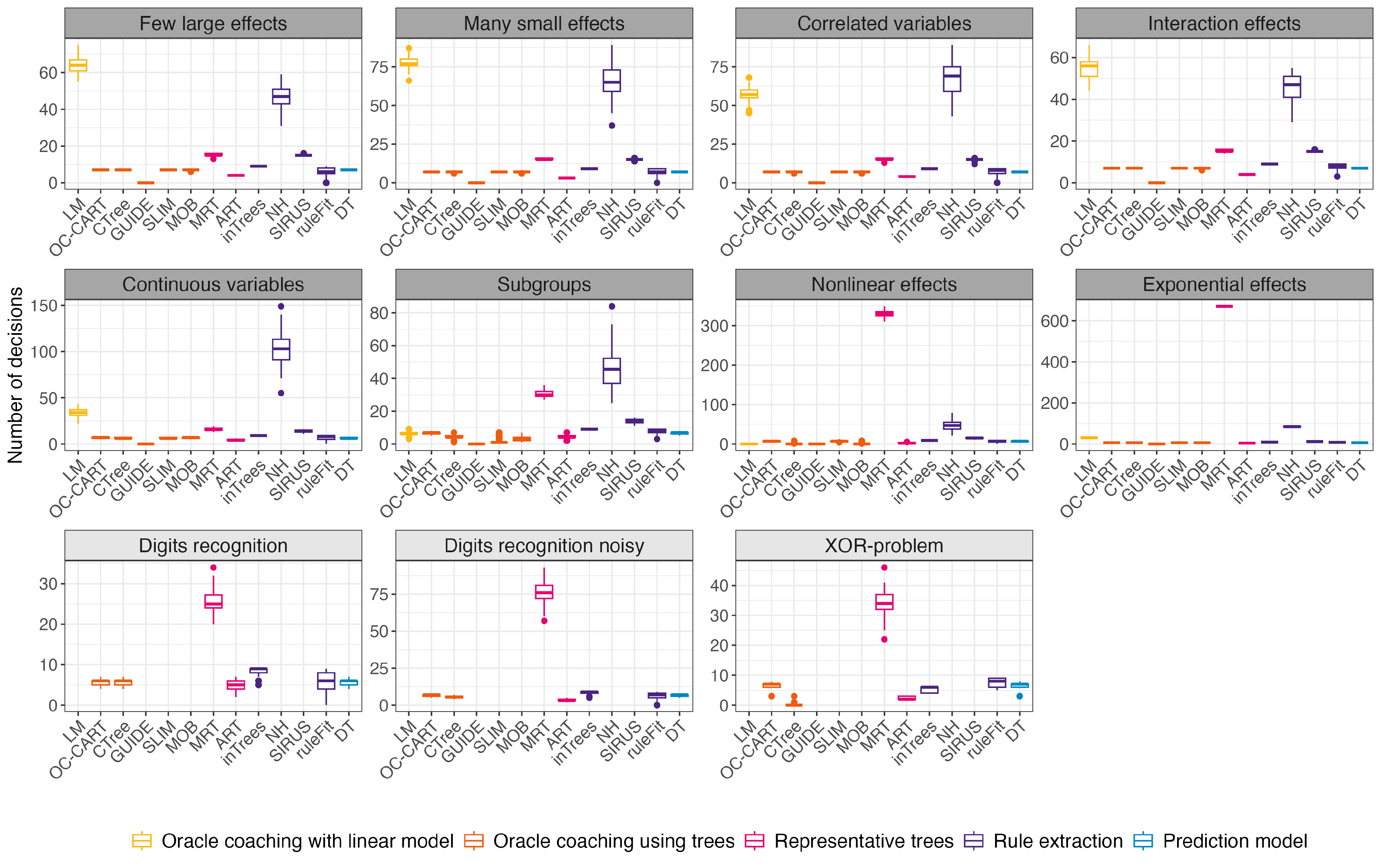}
    \caption{Number of decisions of each method across all simulated data sets. Methods are displayed on the x-axis and colored by method type: rule extraction method, surrogate model, or prediction model (decision tree; DT). Facet headers indicate the outcome type, with light gray denoting classification and dark gray denoting regression. Lower values indicate greater interpretability.}
    \label{app:fig:sim_number_decisions}
\end{figure}

\begin{figure}
    \centering
    \includegraphics[width=\linewidth]{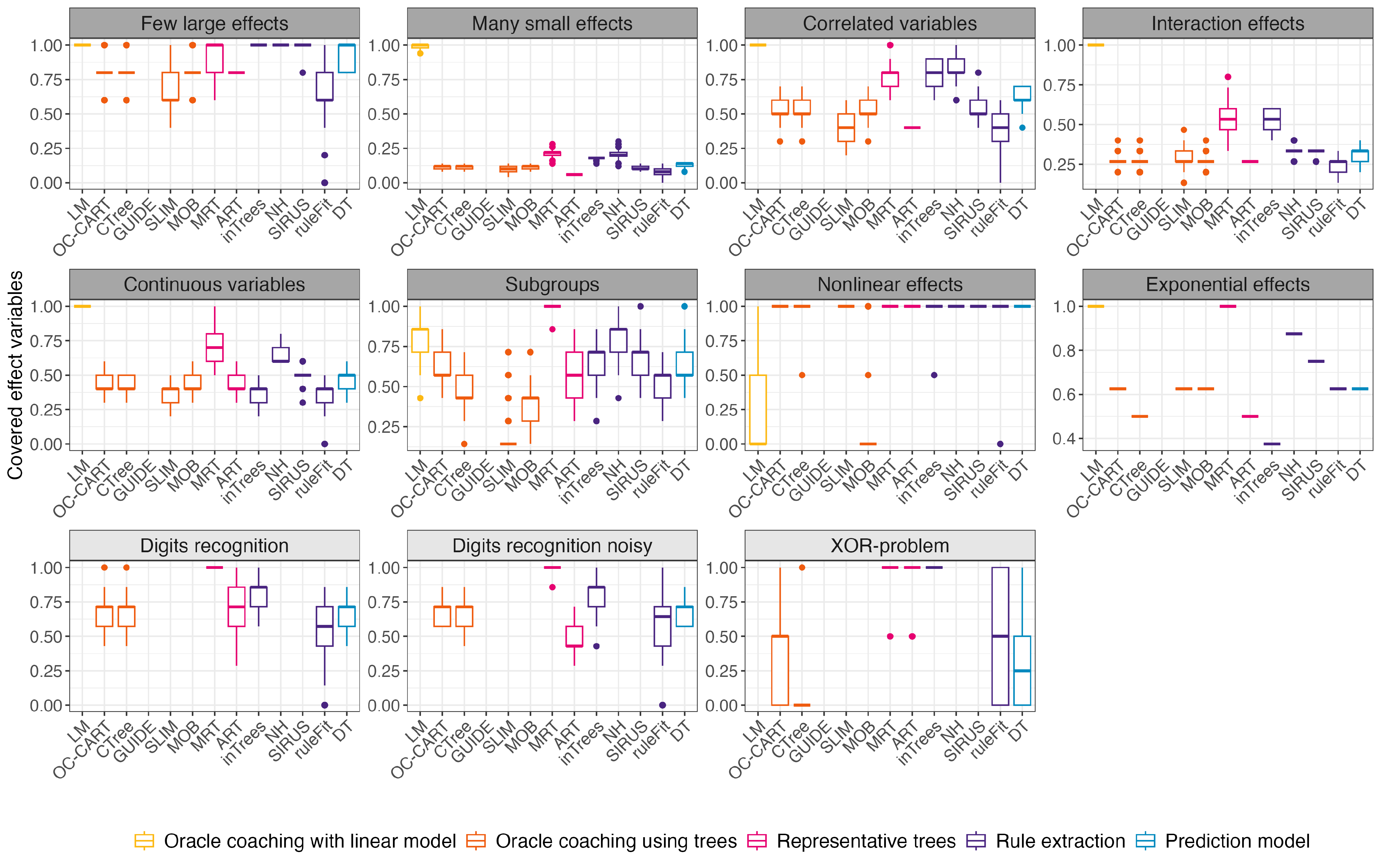}
    \caption{Covered effect vars in the decisions of each method across all simulated data sets. Methods are displayed on the x-axis and colored by method type: rule extraction method, surrogate model, or prediction model (decision tree; DT). Facet headers indicate the outcome type, with light gray denoting classification and dark gray denoting regression.}
    \label{app:fig:sim_covered_effect_vars}
\end{figure}

\begin{figure}
    \centering
    \includegraphics[width=\linewidth]{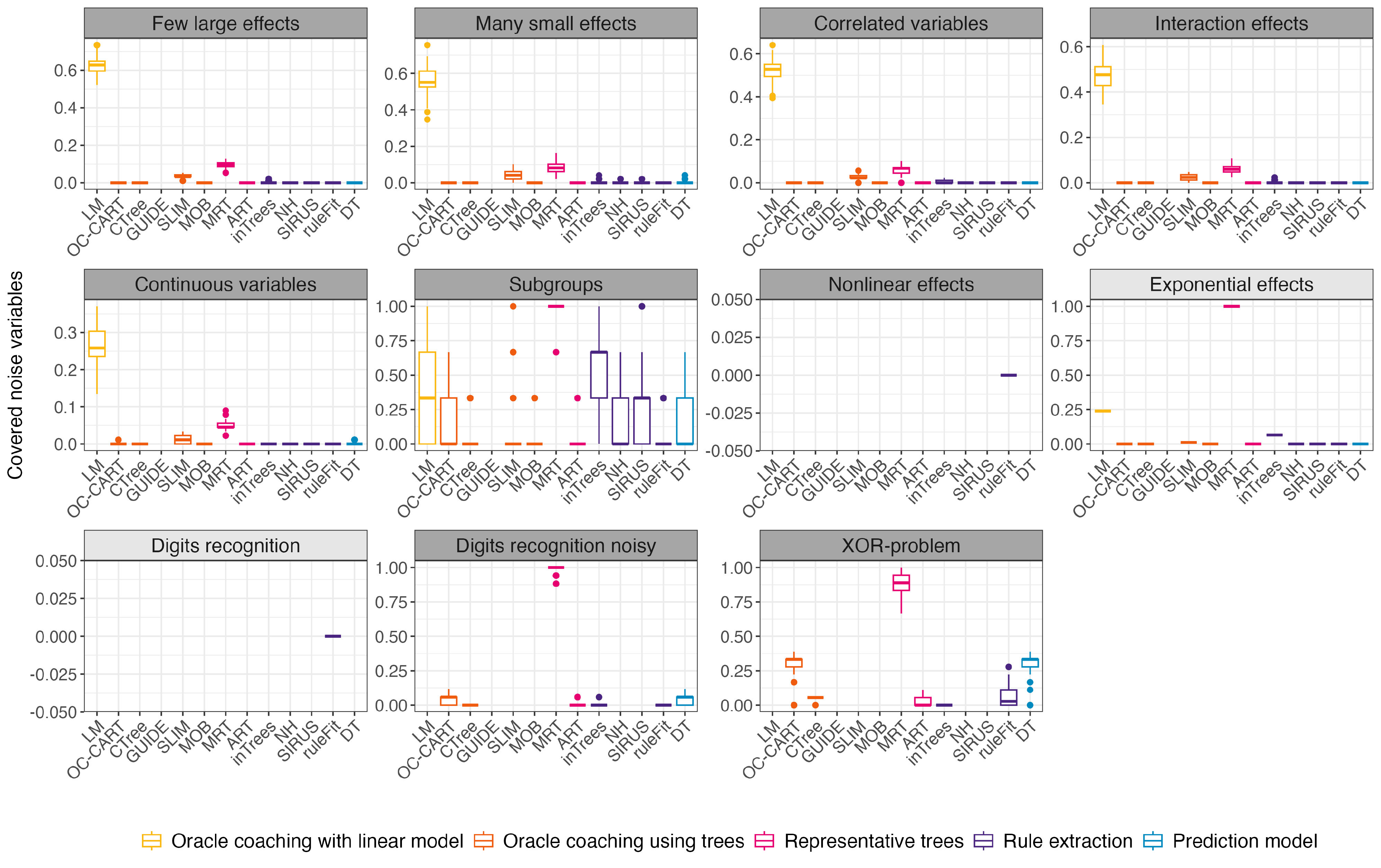}
    \caption{Covered noise vars in the decisions of each method across all simulated data sets. Methods are displayed on the x-axis and colored by method type: rule extraction method, surrogate model, or prediction model (decision tree; DT). Facet headers indicate the outcome type, with light gray denoting classification and dark gray denoting regression.}
    \label{app:fig:sim_covered_noise_vars}
\end{figure}

\begin{figure}
    \centering
    \includegraphics[width=\linewidth]{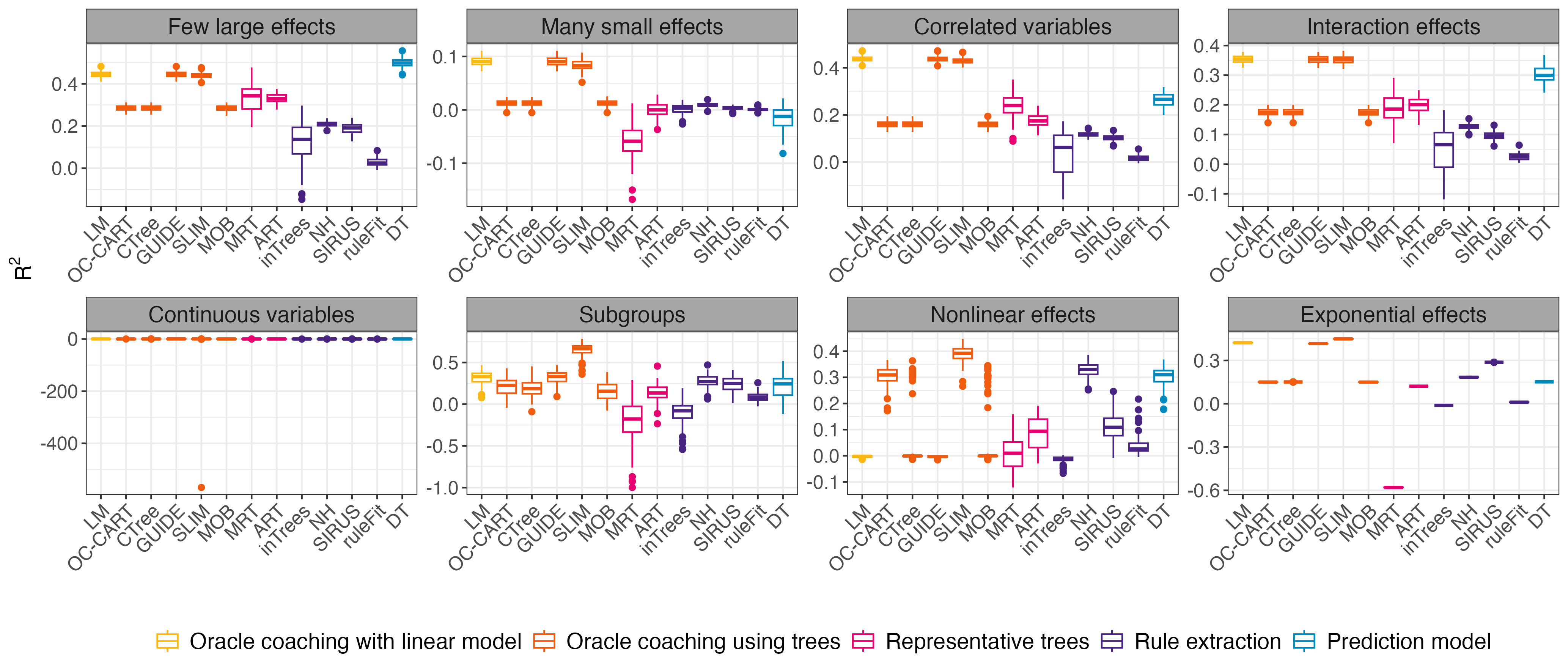}
    \caption{$R^2$ of each regression method across all simulated data sets. Methods are displayed on the x-axis and colored by method type: rule extraction method, surrogate model, or prediction model. (decision tree; DT)}
    \label{app:fig:sim_r_squared}
\end{figure}

\begin{figure}
    \centering
    \includegraphics[width=\linewidth]{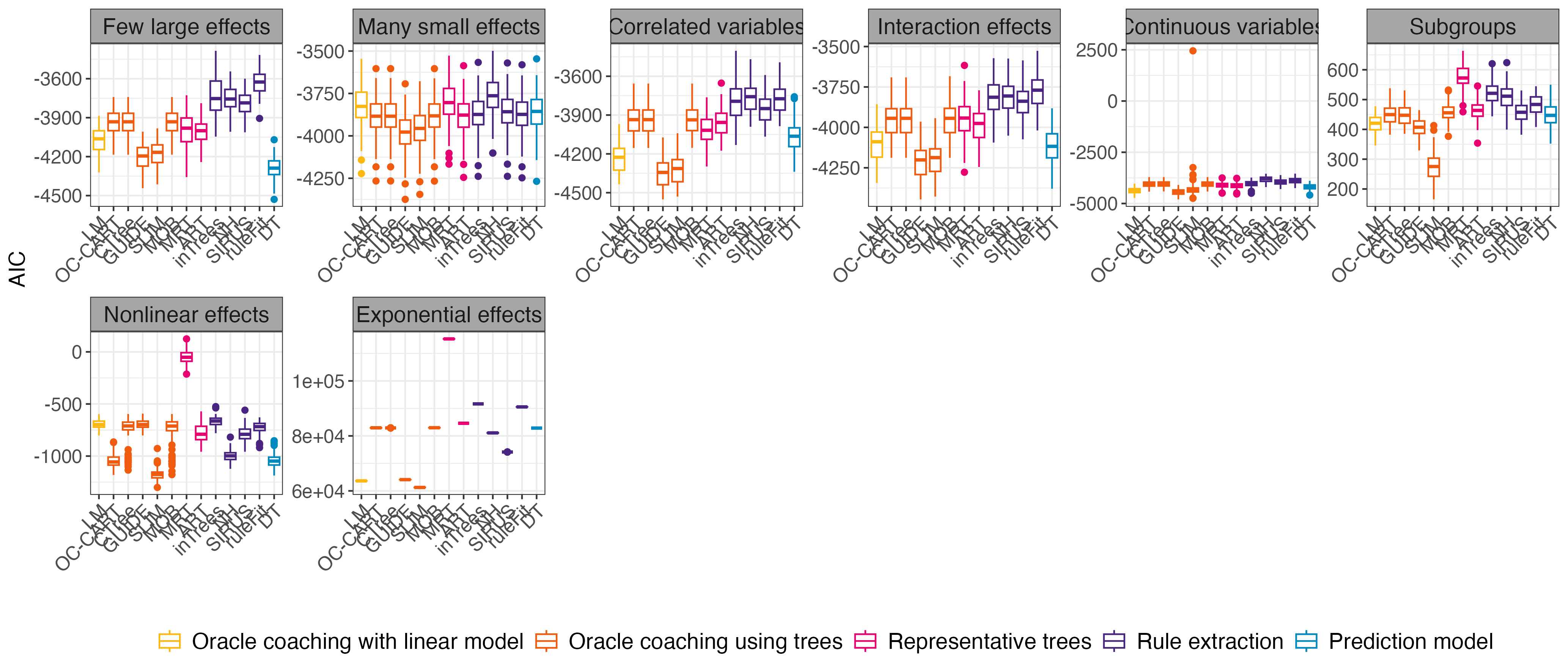}
    \caption{AIC of each regression method across all simulated data sets. Methods are displayed on the x-axis and colored by method type: rule extraction method, surrogate model, or prediction model (decision tree; DT).}
    \label{app:fig:sim_aic}
\end{figure}

\begin{figure}
    \centering
    \includegraphics[width=\linewidth]{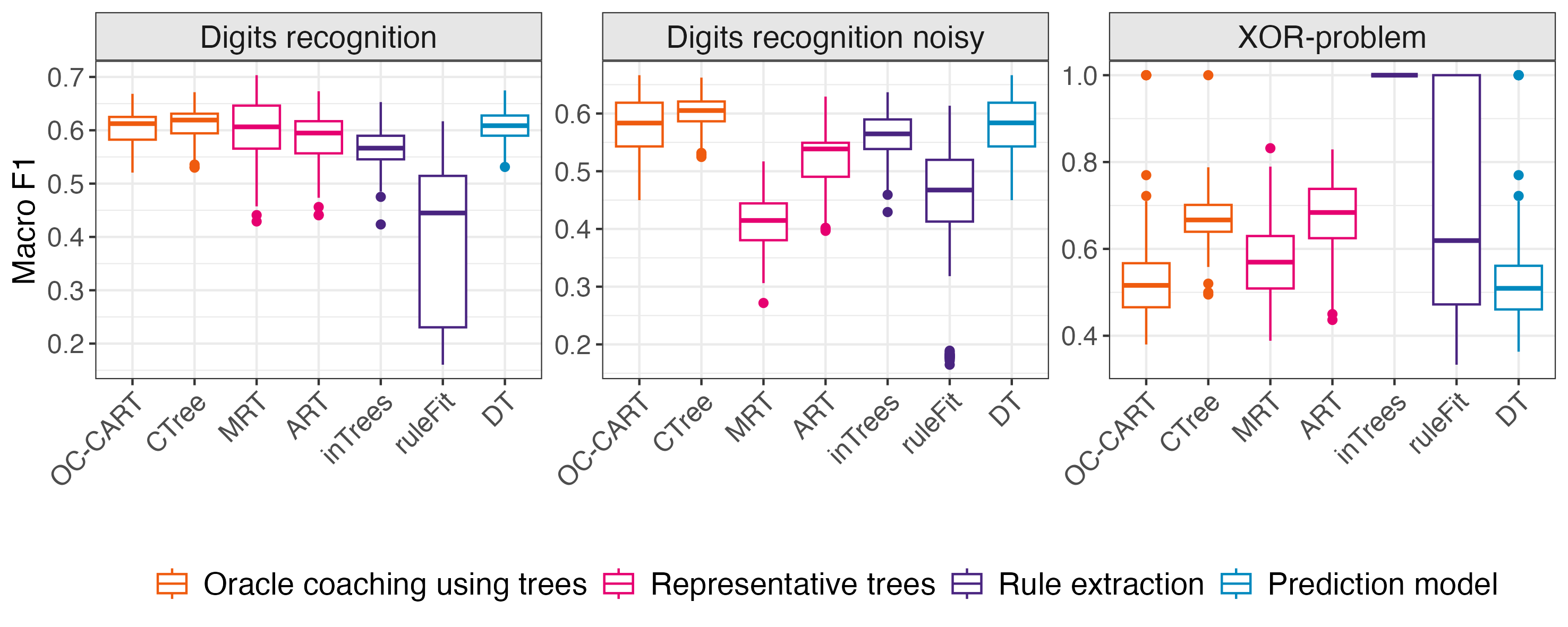}
    \caption{Macro F1 score of each classification method across all simulated data sets. Methods are displayed on the x-axis and colored by method type: rule extraction method, surrogate model, or prediction model (decision tree; DT).}
    \label{app:fig:sim_macro_f1}
\end{figure}

\begin{figure}
    \centering
    \includegraphics[width=\linewidth]{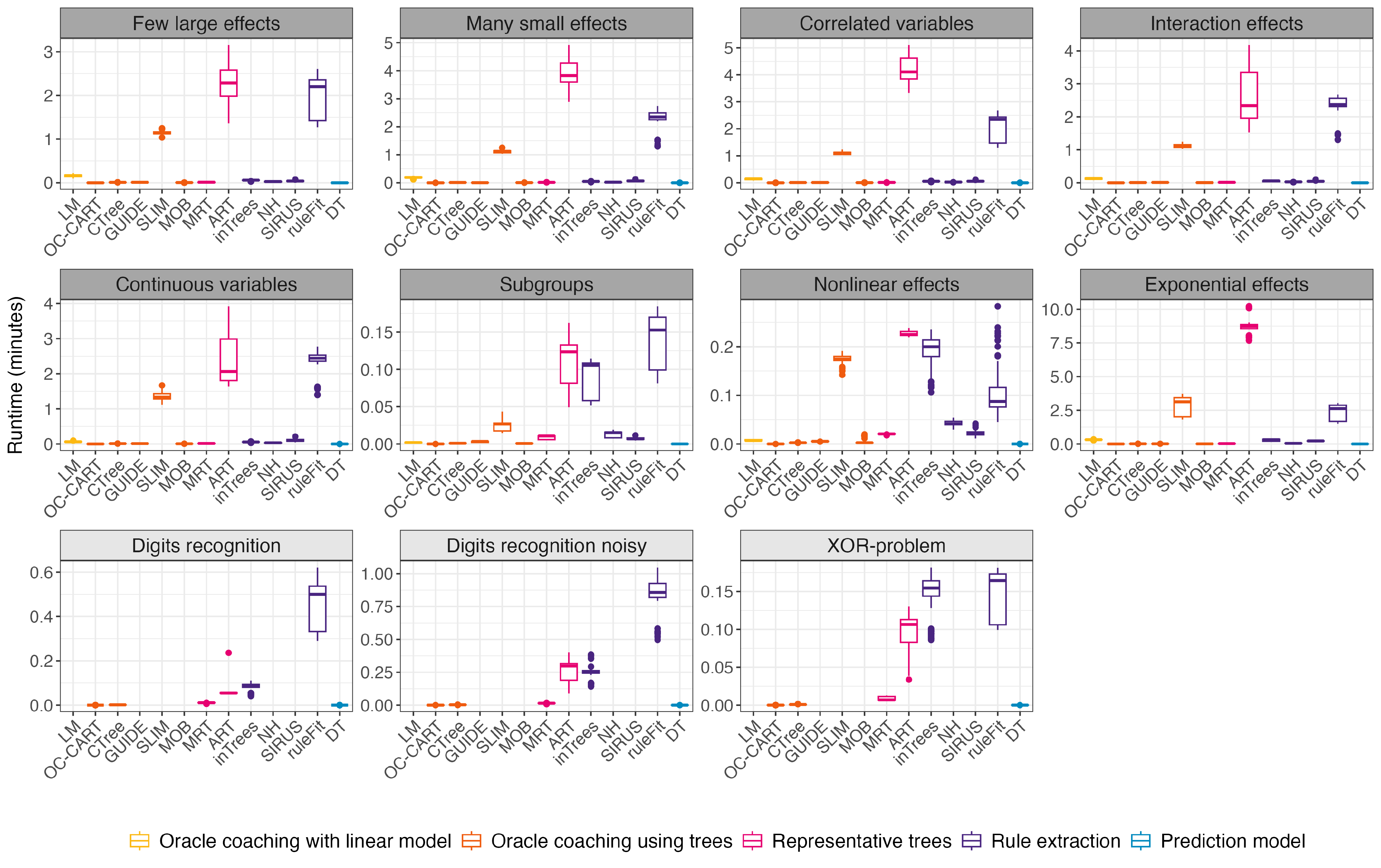}
    \caption{CPU runtime in minutes of each method across all simulated data sets. Methods are displayed on the x-axis and colored by method type: rule extraction method, surrogate model, or prediction model (decision tree; DT). Facet headers indicate the outcome type, with light gray denoting classification and dark gray denoting regression.}
    \label{app:fig:sim_runtime}
\end{figure}

\FloatBarrier
\subsection*{Supplementary figures: Benchmark experiment study}
\begin{figure}
    \centering
    \includegraphics[width=\linewidth]{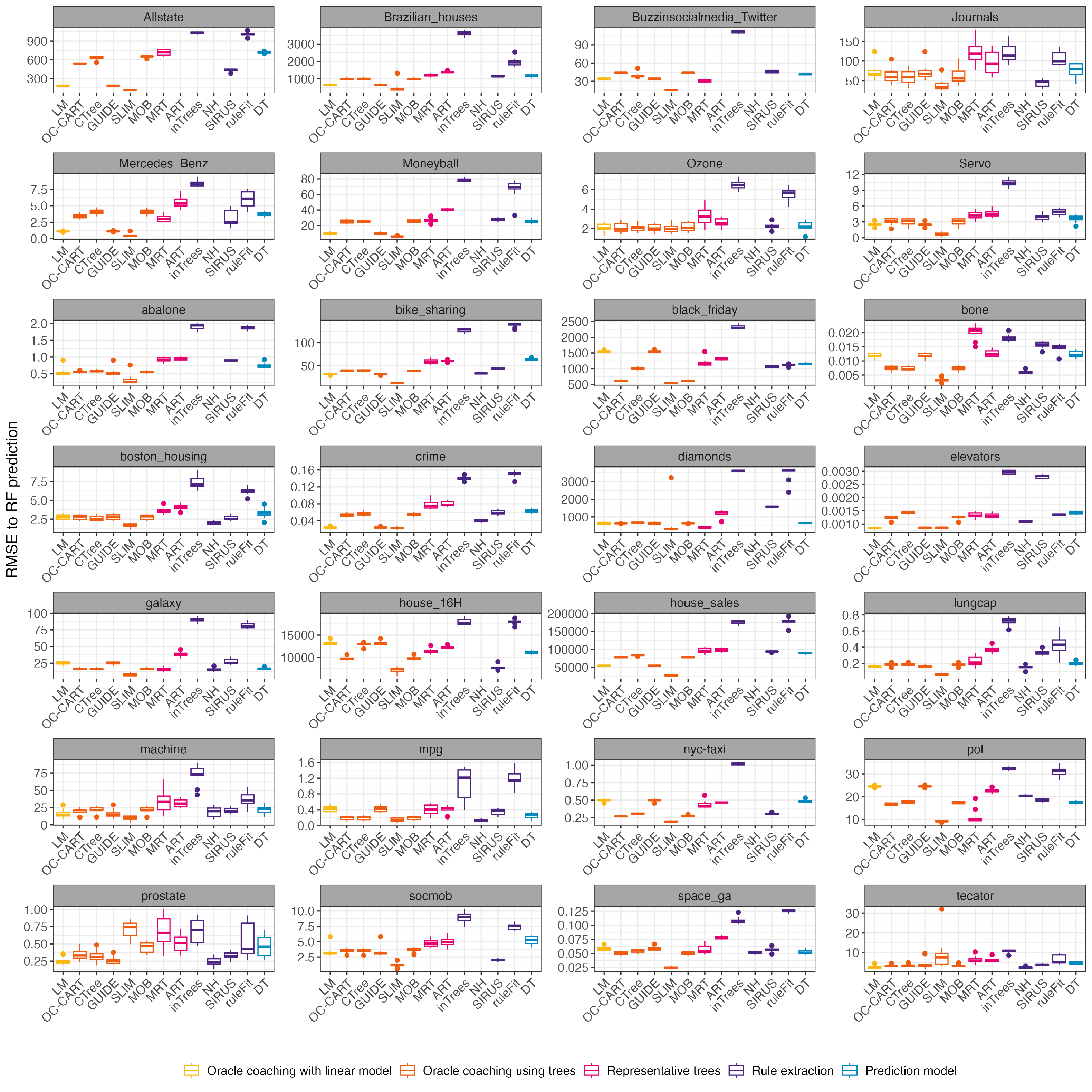}
    \caption{Root mean squared error (RMSE) for regression between the predictions of each method and the corresponding RF across all benchmark test data sets. Methods are displayed on the x-axis and colored by method type: rule extraction method, surrogate model, or prediction model (decision tree; DT).}
    \label{app:fig:bench_dev_rf_pred_rmse}
\end{figure}

\begin{figure}
    \centering
    \includegraphics[width=\linewidth]{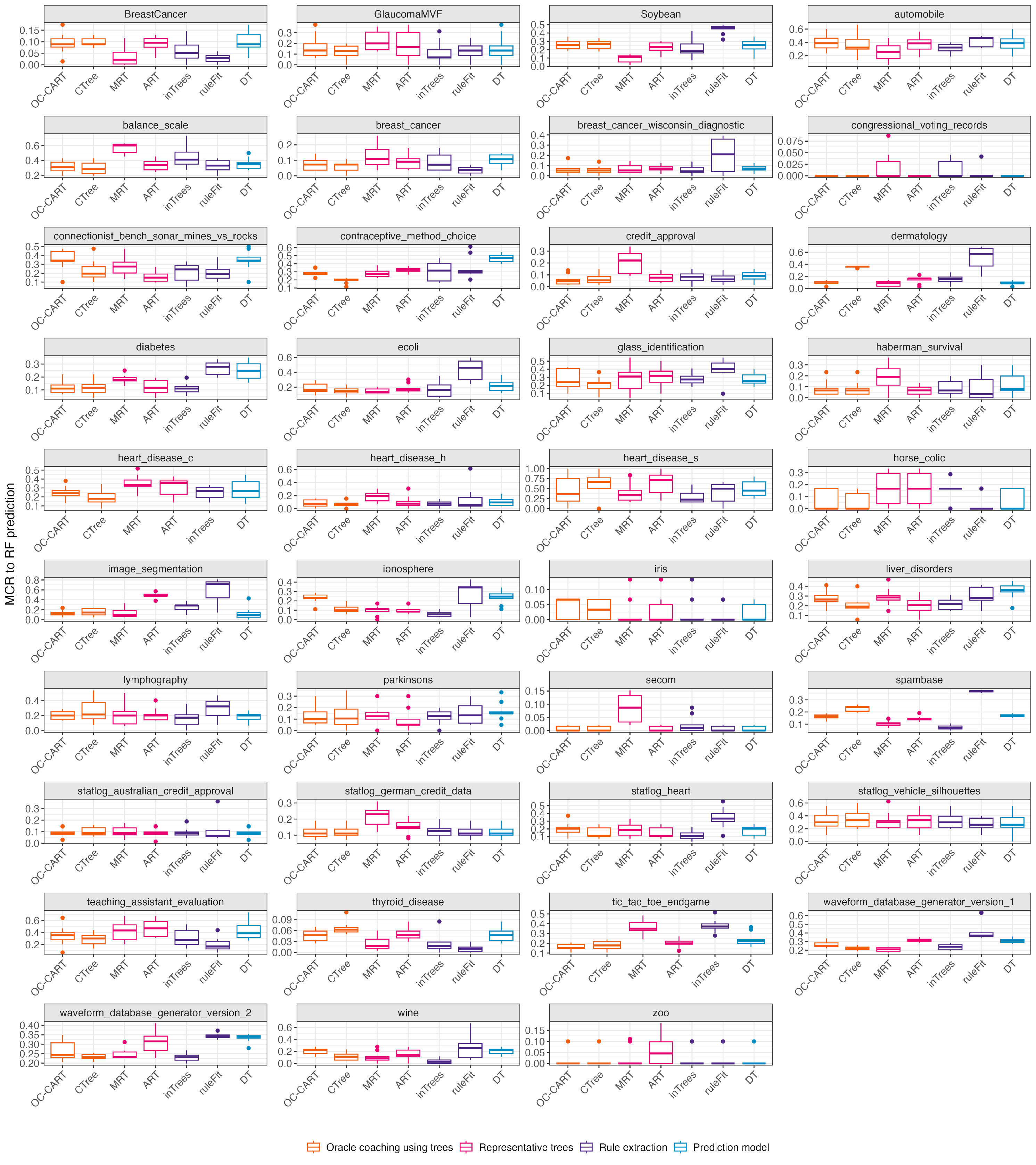}
    \caption{Misclassification rate (MCR) for classification between the predictions of each method and the corresponding RF across all benchmark test data sets. Methods are displayed on the x-axis and colored by method type: rule extraction method, surrogate model, or prediction model (decision tree; DT).}
    \label{app:fig:bench_dev_rf_pred_mcr}
\end{figure}

\begin{figure}
    \centering
    \includegraphics[width=\linewidth]{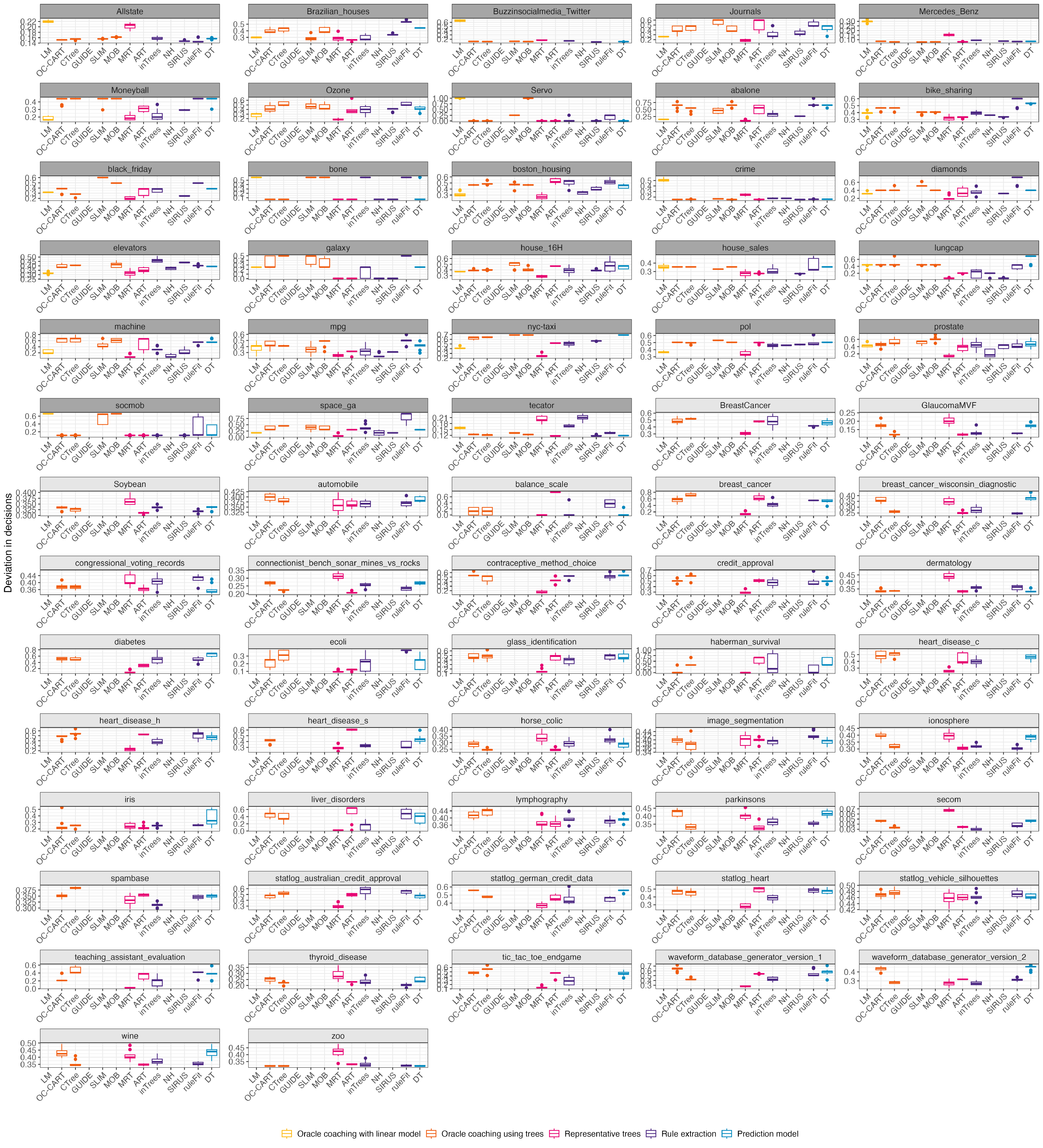}
    \caption{Average proportion of variables whose use in model decisions differed between each method and the individual trees of the corresponding RF across all benchmark data sets. Methods are displayed on the x-axis and colored by method type: rule extraction method, surrogate model, or prediction model (decision tree; DT).}
    \label{app:fig:bench_dist_sv}
\end{figure}

\begin{figure}
    \centering
    \includegraphics[width=\linewidth]{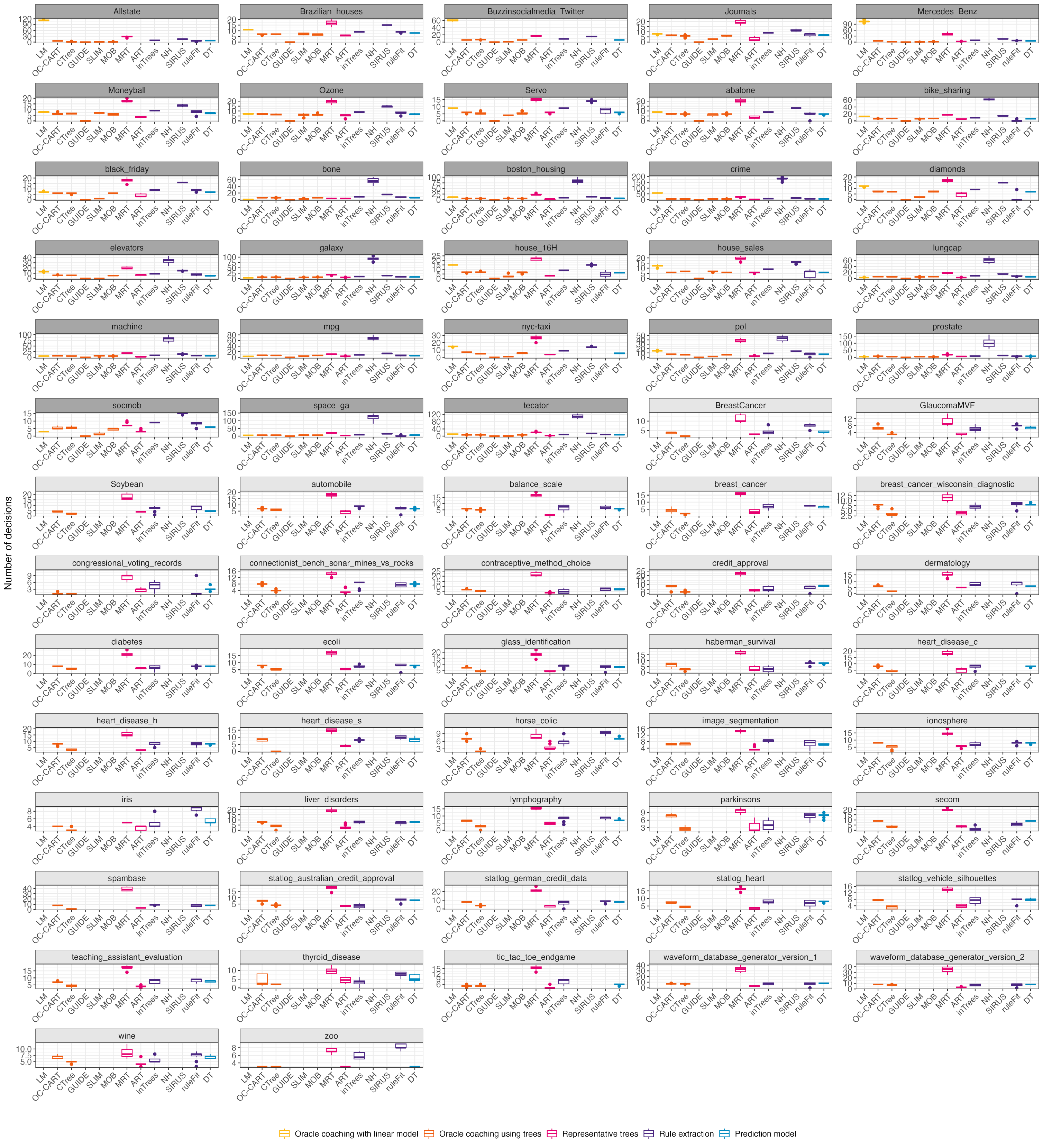}
    \caption{Number of decisions of each method across all benchmark data sets. Methods are displayed on the x-axis and colored by method type: rule extraction method, surrogate model, or prediction model (decision tree; DT). Facet headers indicate the outcome type, with light gray denoting classification and dark gray denoting regression. Lower values indicate greater interpretability.}
    \label{app:fig:bench_number_decisions}
\end{figure}

\begin{figure}
    \centering
    \includegraphics[width=\linewidth]{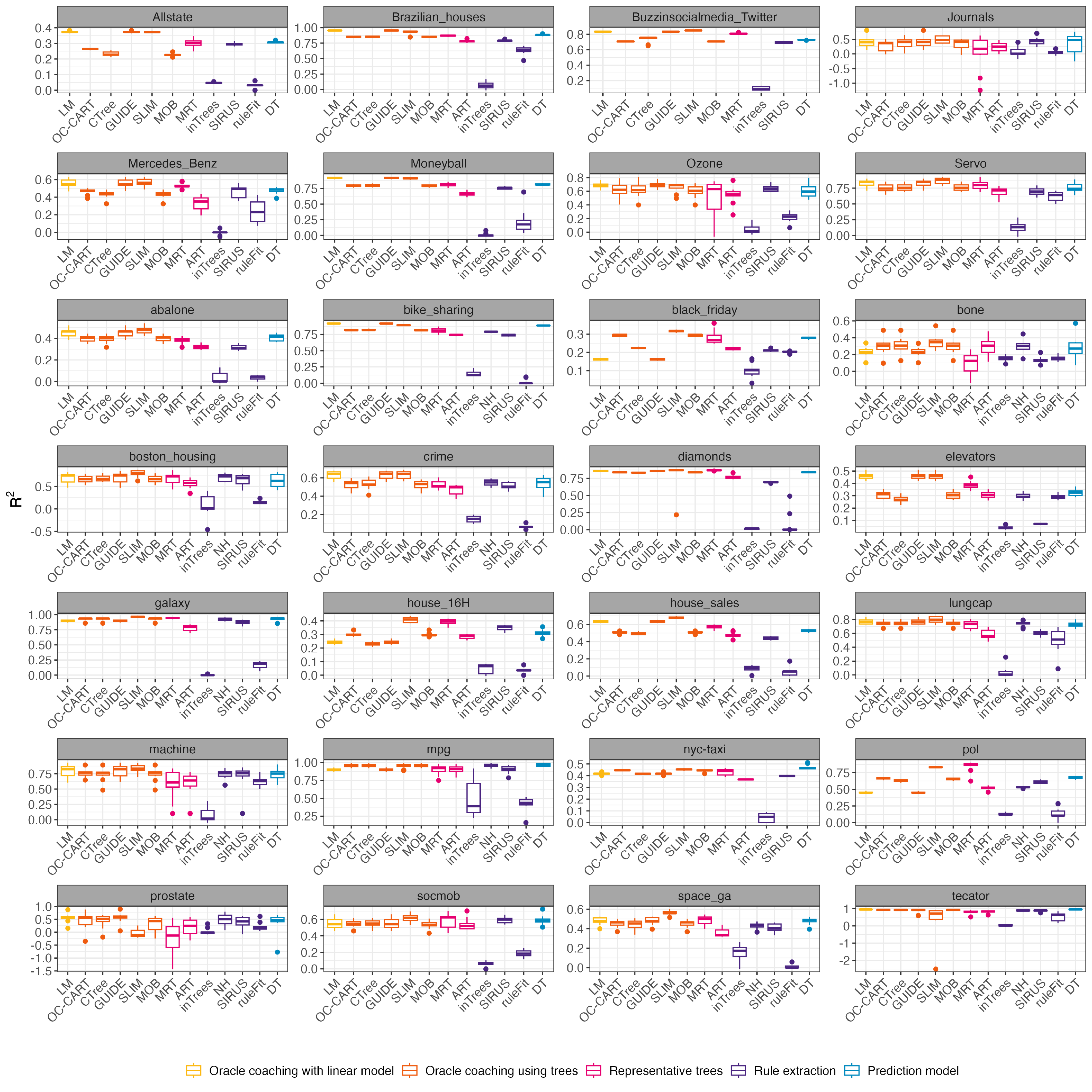}
    \caption{$R^2$ of each regression method across all benchmark data sets. Methods are displayed on the x-axis and colored by method type: rule extraction method, surrogate model, or prediction model (decision tree; DT).}
    \label{app:fig:bench_r_squared}
\end{figure}

\begin{figure}
    \centering
    \includegraphics[width=\linewidth]{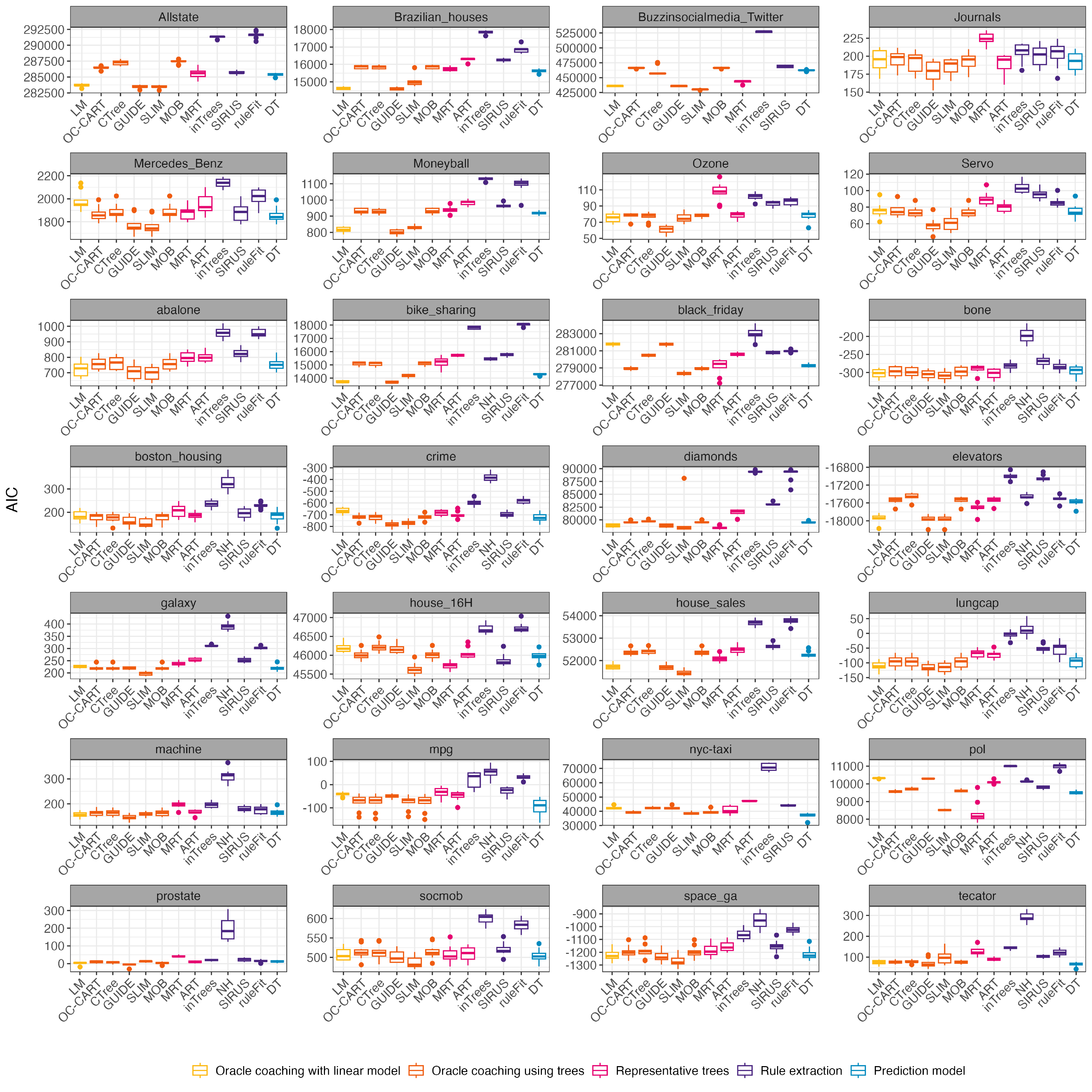}
    \caption{AIC of each regression method across all benchmark data sets. Methods are displayed on the x-axis and colored by method type: rule extraction method, surrogate model, or prediction model (decision tree; DT).}
    \label{app:fig:bench_aic}
\end{figure}

\begin{figure}
    \centering
    \includegraphics[width=\linewidth]{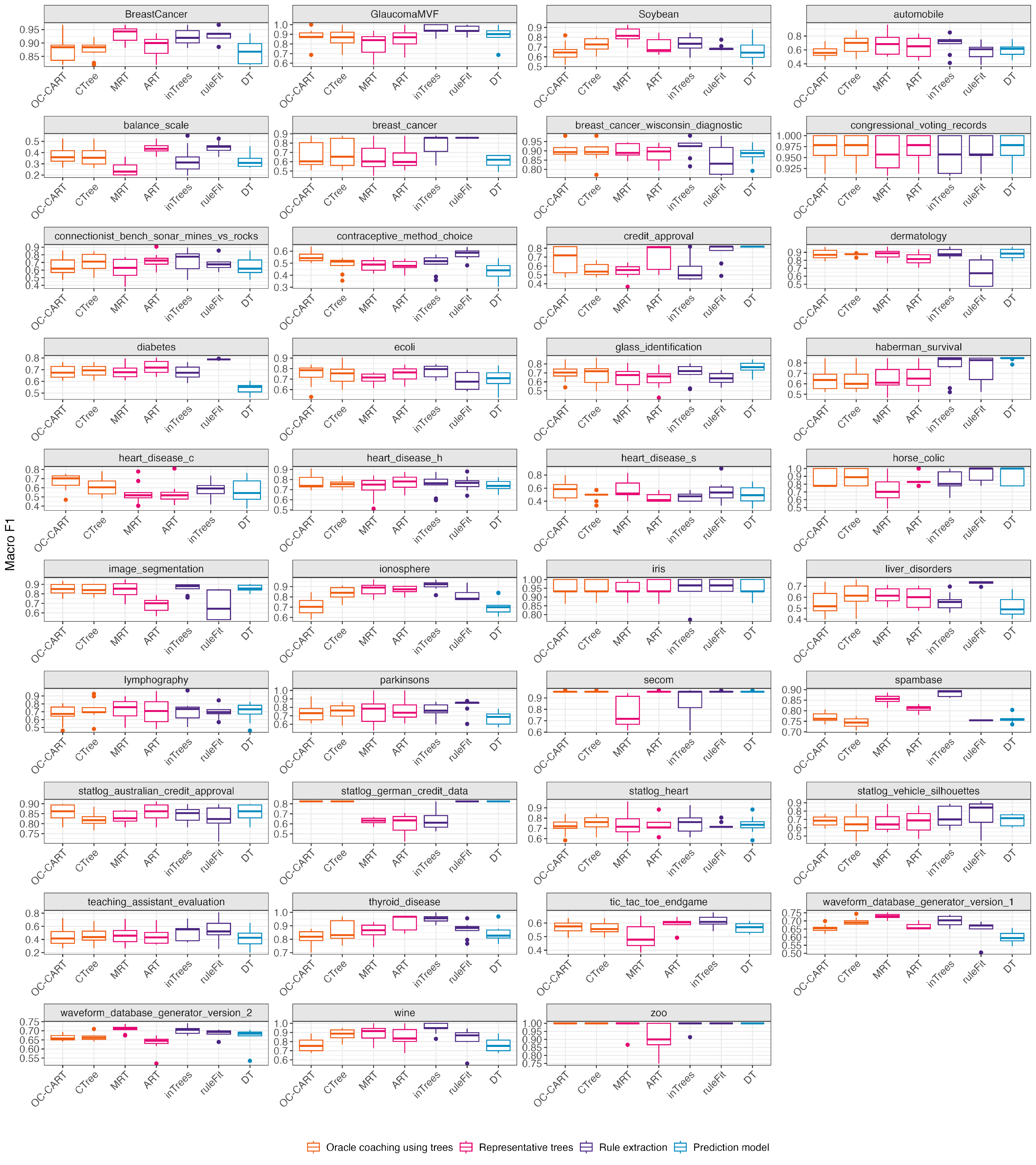}
    \caption{Macro F1 score of each classification method across all benchmark data sets. Methods are displayed on the x-axis and colored by method type: rule extraction method, surrogate model, or prediction model (decision tree; DT).}
    \label{app:fig:bench_macro_f1}
\end{figure}

\begin{figure}
    \centering
    \includegraphics[width=\linewidth]{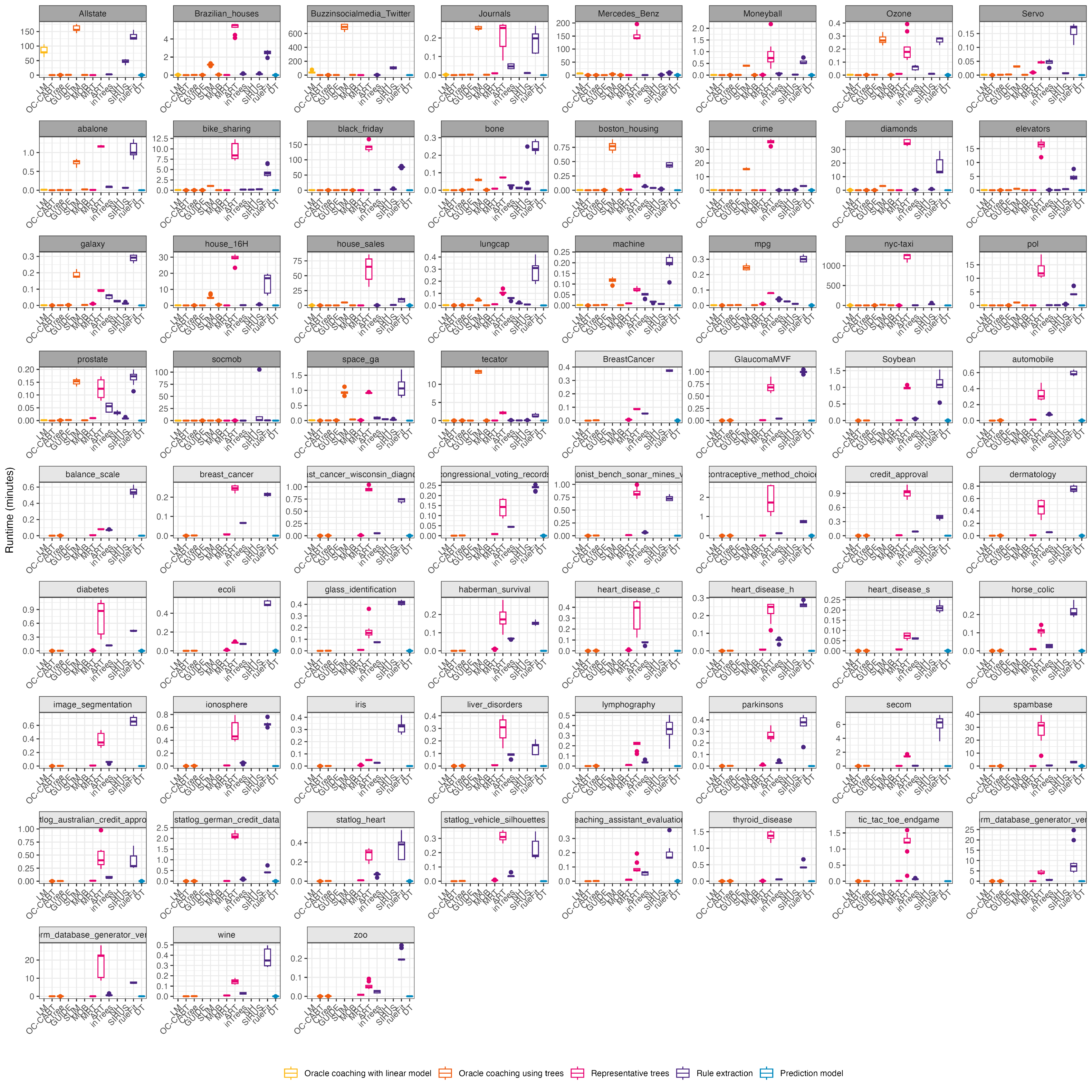}
    \caption{CPU runtime in minutes of each method across all benchmark data sets. Methods are displayed on the x-axis and colored by method type: rule extraction method, surrogate model, or prediction model (decision tree; DT). Facet headers indicate the outcome type, with light gray denoting classification and dark gray denoting regression.}
    \label{app:fig:bench_runtime}
\end{figure}

\end{document}